\documentclass[a4paper,10pt]{article}
\usepackage[T1]{fontenc}
\usepackage[utf8]{inputenc} 
\usepackage[italian,english]{babel}
\usepackage[autostyle,italian=guillemets]{csquotes}
\usepackage{multicol}
\usepackage{microtype}
\usepackage{guit}
\usepackage{booktabs}
\usepackage{graphicx}
\usepackage{mathptmx}
\usepackage{amsfonts}
\usepackage{amssymb}
\usepackage{amsmath}
\usepackage{ifpdf}
\usepackage{dcolumn}
\usepackage{longtable} 
\usepackage{hyperref}
\usepackage{textcomp}
\usepackage{tabularx,array}
\usepackage[normalem]{ulem} 
\usepackage{placeins}
\usepackage{epstopdf}
\usepackage{rotating}
\makeatletter
\DeclareUnicodeCharacter{00A0}{~}       
\DeclareUnicodeCharacter{201C}{``}      
\DeclareUnicodeCharacter{201D}{''}      
\DeclareUnicodeCharacter{2014}{---}     
\DeclareUnicodeCharacter{2013}{--}      
\DeclareUnicodeCharacter{200B}{}        
\DeclareUnicodeCharacter{00B0}{\textdegree} 
\makeatother

\title{New Astronomical, Meteorological and Geological Study of the "Torre Spaccata District" - VII district of the Municipality of Rome}

\author{
	Diego Tasselli$^{1,4}$ \\
	\small \texttt{diego.tasselli@tscorporation.it}
	\and
	Stefano Ricci$^{2,4}$ \\
	\small \texttt{stefano.ricci@tscorporation.it}
	\and
	Pamela Bianchi$^{3,4}$ \\
	\small \texttt{pamela.bianchi@tscorporation.it}
}

\date{
	\small
	$^{1}$TS Corporation - Astronomy and Astrophysics Department, Via Rugantino 71, 00169 Roma RM, Italy\\
	$^{2}$TS Corporation - Meteorological and Climatic Change Department, Via Giordano Bruno 3, 73057 Taviano LE, Italy\\
	$^{3}$TS Corporation - Geology and Geophysics Department, Via Reclusorio 38, 06084 Foligno PG, Italy\\
	$^{4}$Halminak S.r.l., Parent Company and Strategic Coordination Node, Via Rugantino 71, 00169 Roma RM, Italy
}

\begin{document}
 
\maketitle
 
\begin{abstract}
	This study, part of the NGICS (New Italian City Geological Study) project, analyzes 25 years of multi-disciplinary data to assess long-term environmental, climatic, and geological trends in Torrespaccata, an urban district within Rome's Seventh Municipality (VII Administrative District). Specifically, it investigates the temporal evolution of Heating Degree Days (HDD) by comparing the historical stationary climate period (1961–1989) with the recent multidecadal interval (1990–2014). Statistical analysis reveals a clear climatic transition: while the pre-1989 period showed a stationary trend ($m=-1.14$ HDD/year), the modern era exhibits a sharp decline in heating demand, characterized by a negative trend of $-11.52$ HDD/year ($R^{2}=0.511$). This significant reduction of 80.14 HDD between the two periods is robustly validated via Welch's t-test ($t=3.045$, $p=0.0039$). Concurrently, interannual variability increased by 29.8\%, driven by the synergistic feedback between global warming and the Urban Heat Island (UHI) effect. Furthermore, the correlation between climatic anomalies and local geological events is evaluated using robust statistical imputation to mitigate uncertainties caused by data gaps. These findings provide critical insights into long-term anthropic and natural changes within the urban ecosystem.
\end{abstract}

\vfill
\noindent {\footnotesize This paper was prepared using the \LaTeX\ typesetting system.}
\begin{multicols}%
{2}
\normalsize
\section{Introduction} 
In this work, we present an analysis of the astronomical, geological, meteorological, and climatic data collected in Torrespaccata (Municipio VII of Rome). Based on measurements obtained from local astronomical, meteorological, and geological records~\cite{datimeteo:ref2013, eumetsat:ref2013}, these data are compared to verify localized variations in astronomical, climatological, and geological conditions. 
\subsection{Location}
Torrespaccata is an urban district administratively included within Municipio VII (formerly Municipio VI until January 2021) of the Municipality of Rome, situated in the eastern suburban sector of the metropolitan area.

\textbf{Geomorphologically}, the territory is characterized by a low-amplitude ignimbritic plateau or paleovolcanic flatland, with average elevations ranging between 50 and 70 meters above sea level. This landscape is a direct product of Quaternary explosive activity from the nearby Colli Albani Volcanic District (specifically the Vulcano Laziale phases), which deposited thick successions of pyroclastic lithotypes over older marine sedimentary substrates.

\textbf{Hydrographically}, the district does not border the Tiber River; instead, it is located within the hydrographic sub-basin of the Aniene River, the primary left-bank tributary of the Tiber. The area was historically drained by a minor, low-hierarchy natural drainage network consisting of small torrential channels and seasonal creeks. The primary collector of this quadrant is the Tor Tre Teste stream (historically known as the "Marrana di Tor Sapienza"), which flows northward toward the Aniene valley. Due to intensive post-war urban development, almost the entirety of this minor network has been rectified, culverted, or integrated into the municipal mixed sewerage system. A significant exception is the non-urbanized area known as the "Pratone di Torre Spaccata", which preserves the pristine morphology of a gently incised plateau and acts as a localized natural retention basin during extreme run-off events.
\begin{center}
	\footnotesize
	\begin{tabular}{ccc}
		\hline
		\rule{0pt}{3ex} \textbf{Latitude} & \textbf{Longitude} & \textbf{Elevation} \\[0.5ex]
		\hline
		\rule{0pt}{3ex} $41^\circ 54^\prime 29.34^{\prime\prime}$~N & $12^\circ 28^\prime 54.48^{\prime\prime}$~E & 60~m a.s.l. \\[0.5ex]
		\hline
	\end{tabular}
	
	\vspace{2mm}
	\textit{Geographic Data of Torrespaccata.}
\end{center}

The geological map of the area is shown in Section 3.5. $^{\scriptsize \cite{mappageologica:2013}}$ \\ \\
\begin{tabular}{lp{0.28\textwidth}}
\hline
\footnotesize \textbf{Seismic Zone} & \footnotesize \textbf{Description}  \cr
\hline 
\footnotesize 2B & \footnotesize Area with medium seismic hazard where strong earthquakes can occur. Sub-area 2B indicates a value of ag <0.20g. \cr
\hline 
\end{tabular} \\ \\
{\footnotesize \textit {Seismic Data of Torrespaccata}}
\\ \\ 
The following table identifies the climatic data assigned by the President of the Republic's Decree n. 412 of 26 August 1993.\\ \\
\begin{tabular}{c|c}
\hline
\footnotesize \textbf{Climatic Zone} & \footnotesize \textbf{Day Degrees} \cr
\hline 
\footnotesize D & \footnotesize 1.415 \cr
\hline 
\end{tabular} \\ \\ {\footnotesize {\textit {Climatic Parameter of Torrespaccata}}} 
\section{Meteo-Climatic Parameters}
In this section, we describe the atmospheric variables—including air temperatures ($T$), Dew Point, Relative Humidity, surface atmospheric pressure, diurnal thermal ranges, and the annual frequency of precipitation, fog, and thunder events—obtained through a rigorous statistical analysis of local micrometeorological records. The historical baseline parameters investigated in this study are synthesized in Table 1. It should be noted that all climatological values considered represent the 25-year operational average calculated for the multidecadal midrange interval spanning from 1990 to 2014.
\begin{table*}[t]
	\centering
	\caption{Baseline Meteorological and Climatological Parameters (1990–2014).}
	\label{tab:meteo_parameters}
	\vspace{0.5em}
	\begin{tabular}{lrrr}
		\hline
		\textbf{Climatic Parameter} & \textbf{Unit} & \textbf{Value} & \textbf{Record Baseline} \\ \hline
		&       &       &  \\
		Average annual temperature & $^\circ\text{C}$ & 15.81 &  \\
		Absolute maximum monthly mean temperature record & $^\circ\text{C}$ & 27.94 & Aug-03 \\
		Absolute minimum monthly mean temperature record & $^\circ\text{C}$ & 2.98  & Feb-12 \\
		Annual diurnal thermal range & $^\circ\text{C}$ & 9.83  &  \\
		Months with average temperature $> 20^\circ\text{C}$ & Count & 97    &  \\
		Total cumulative rainfall (1990–2014) & mm    & 32614.60 &  \\
		Total rain days over the study period & Count & 1711  &  \\
		Total fog days over the study period & Count & 254   &  \\
		Total storm / thunder days over the study period & Count &       & 140 \\
		Total isolated rain days over the study period & Count &       & 1069 \\
		Total rain/storm days over the study period & Count &       & 47 \\
		Total rain/snow days over the study period & Count &       & 10 \\
		Total rain/fog days over the study period & Count &       & 38 \\
		Total rain/thunder/fog days over the study period & Count &       & 23 \\
		Total snow days over the study period & Count &       & 11 \\
		Maximum wind speed peak & km/h  & 50.73 & Mar-01 \\
		Minimum wind speed baseline & km/h  & 0.00  & Feb-07 \\
		Maximum monthly rainfall record & mm    & 970.00 & Oct-92 \\
		Minimum monthly rainfall record & mm    & 0.00  & Mar-94 \\ \hline
	\end{tabular}%
\end{table*}
\subsection{Solar Radiation of the Territory} 
In the study area, the values of solar radiation are increasing. In fact, the solar irradiation data for the territory taken from the parameters and the data prepared by the European Union, demonstrate the trend of irradiation for the municipality, visible in next table: 

\begin{tabular}{|c|c|c|c|}
	\hline
	\textbf{Month} & \multicolumn{1}{c|}{\textbf{DNI}} & \multicolumn{1}{c|}{\textbf{Hh}} & \multicolumn{1}{c|}{\textbf{DL}} \\
	\hline
	Jan   & 12.05 & 6.14  & 12   \\ 
	Feb   & 14.32 & 8.85  & 11   \\ 
	Mar   & 19.44 & 13.50 & 9    \\ 
	Apr   & 21.54 & 16.79 & 6    \\ 
	May   & 26.24 & 21.00 & 1.5  \\ 
	Jun   & 28.69 & 23.08 & -    \\ 
	Jul   & 29.11 & 23.11 & -    \\ 
	Aug   & 24.99 & 19.51 & -    \\ 
	Sep   & 20.69 & 15.19 & -    \\ 
	Oct   & 15.27 & 10.16 & 3    \\ 
	Nov   & 11.67 & 6.61  & 7.5  \\ 
	Dec   & 9.67  & 4.93  & 11   \\ \hline
\end{tabular}
\vspace{0.2cm}
{\footnotesize \textit{DNI:Direct Normal Irradiance (Wh/m$^2$/day);Hh:Radiation on Horizontal Plane (Wh/m$^2$/day);DL:Average monthly day length expressed in hours.}}
\vspace{0.3cm}

The meteorological data and extrapolated charts for the investigated territory indicate a global solar radiation range between 1100 and 1200~kWh/1kWp, in strict agreement with the solar atlas prepared by the European Union~\cite{UE:2013} and shown in Figure 17. This monthly distribution is chronologically illustrated in Figures 18, 19, and 20, which match the dataset provided in the table above. Based on these monthly variations, the processed data yields an annual mean daily horizontal irradiance ($H_h$) equal to 5.92 kWh/m$^2$/day, in strict agreement with the solar atlas prepared by the European Union \cite{UE:2013}.
\subsection{Temperature} \normalsize {In this section we describe air temperatures (T) obtained by an accurate analysis of the meteorological data. The study shows a non-constant average in the temperatures recorded in the period between 1990 and 2014, alternating cold and hot periods, as shown in the following graph.} \\
\includegraphics[width=0.49\textwidth{}]{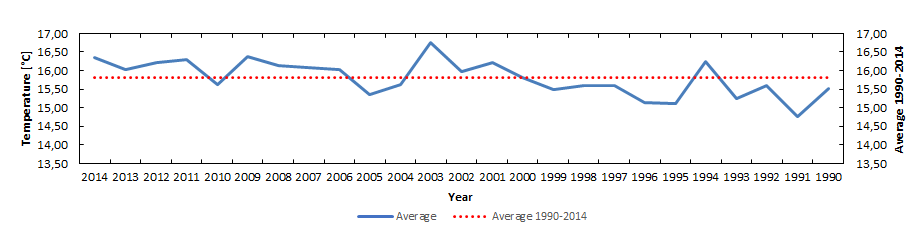} 
{\scriptsize \textit{\textit{{\footnotesize Relationship between annual temperature and the 1990–2014 average}}}}. \\
 \\The values are calculated across the entire measurement period (1990–2014), drawing on data from the historical records published by "Centro Epson Meteo"$^{\scriptsize \cite{datimeteo:ref2013}}$ for the period 1990-2014. \\ \\ \\ \includegraphics[width=0.49\textwidth{}]{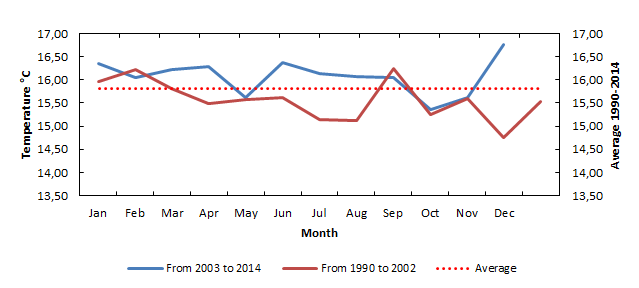} 
 {\scriptsize {\textit{{\footnotesize Variation of Month Temperature about year 1990-2002 and 2003-2014}}}} \\ \\
 The graphs shown in Figures 3 and 4 show the performance of the maximum and minimum temperatures proportional to the average of the period of study (1990-2014). The study shows the following trends:
\begin{itemize}
\item Minimum temperatures: the lowest monthly mean temperature record of the entire series was registered in February 2012 with a value of 2.98$^\circ$C;
\item Maximum temperatures: the absolute maximum monthly mean temperature record peaked at 27.94$^\circ$C in August 2003.
\end{itemize} 
Table 6 details the historical mean temperatures for the winter quarter (January--February--March) and the summer quarter (June--July--August). Analyzing these seasonal quarters, key macroclimatic anomalies and absolute historical records emerge within the 25-year series:
\begin{itemize}
	\item An above-average increase in temperatures during the second decade of the series (2003--2014), characterized by a sharp upward shift initiated in 2003, in strict agreement with the interannual anomalies plotted in Figure 12;
	\item A massive macroclimatic anomaly during the summer of 2003, which recorded unprecedented seasonal heatwaves with a June mean of $26.10 \ \text{}^{\circ}\text{C}$ and an absolute historical maximum average of $27.94 \ \text{}^{\circ}\text{C}$ in August, as structurally highlighted in Table 6;
	\item The most significant temperature divergence between the winter and summer core months was registered in 2003, reaching a seasonal delta ($\Delta T$) of $22.71 \ \text{}^{\circ}\text{C}$ between February and August.
\end{itemize}
\subsubsection{Average temperature fluctuation 1990-2014}
Analyzing the temperature data for the 1990–2014 period, Table 8 and the following graph show a trend and from the following graph, a trend characterized by strong ascending and descending steps, with the higher $\Delta$ in the summer quarter (June-August), the maximum value of which occurs in August ($11.84^\circ C$), while the lowest value was found in December ($7.67^\circ °C$).\\ 
\includegraphics[width=0.50\textwidth{}]{Variation_of_Average_Temperature_for_Month.png}
\\ 
\subsection{Climatic Analysis of Effective Heating Degree Days (HDD) over the 1990–2014 Period} 
The climatological calculation of effective Heating Degree Days (HDD) for the city of Rome over the 1990–2014 period reveals a real multiyear mean total of 1,362 HDD, marking a significant contraction compared to historical baseline regulatory data, as reported in Table 9.

\subsubsection{Analysis and Discussion of Variations} 
A comparative analysis between the normative values established by the D.P.R. 412/1993 for the city of Rome (Climate Zone D, equal to 1,415 HDD) and the processing of real datasets measured over the 25-year interval (1990–2014) highlights macroclimatic dynamics of primary interest for energy and environmental purposes.

\subsubsection{Global Warming Effects and Winter HDD Contraction} 
On an annual scale, an overall decrease of $3.7\%$ in actual thermal demand is observed compared to conventional design baselines. This contraction is predominantly concentrated within the core months of the heating season: November ($-16.0\%$), January ($-8.0\%$), and February ($-8.6\%$). This trend is strictly consistent with global climate change models and the Urban Heat Island (UHI) effect typical of the Rome metropolitan area, which elevates nocturnal minimum temperatures, thereby reducing the thermal integration required by space heating systems.

\subsubsection{Seasonal Shift and Extension of the Potential Operational Period} 
A mirroring phenomenon of opposite sign emerges during the months of April, May, September, and October. While the current regulation categorically excludes HDD calculations outside the rigid November 1st – April 15th window for Zone D, real microclimatic datasets show winter tails and advanced autumns with non-zero thermal contributions (e.g., 38 HDD in October and 12 HDD in May). The positive deviation observed in April ($+41.8\%$) indicates that, although the mean monthly temperature is rising, strong daily meteorological variability accumulates a number of effective heating degree days higher than the rigid legal partitioning (which is limited to only 15 days).

Statistical computation of the historical time series for Rome (1990–2014) highlights a negative linear trend equal to $-11.52$ HDD/year and an interannual standard deviation ($\sigma$) of $107.04$ HDD. These parameters unequivocally quantify the velocity of local warming over the analyzed 25-year series.
\\	
\subsection{Dew point} 
In this section we describe the dew point temperature ($T_{\text{dew}}$) obtained from an accurate analysis of the data referring to the municipal area over the 1990--2014 climatological interval. The historical average identified in the study period is equal to $10.73 \ \text{}^{\circ}\text{C}$, as summarized in Table 1. On an annual scale, out of the 25 total surveyed years, 14 years plot systematically above this climatological baseline, as shown in Figure 10 at the end of this article. 

Concurrently, the empirical dataset highlights a clear seasonal distribution at a monthly downscaled level. There are exactly 6 months with values shifting above the multiyear average, clustered precisely within the warm period spanning from May to October (as illustrated in Table 15). 

Within this seasonal framework, the historical minimum monthly averages are concentrated in the winter core quarter, with February exhibiting the lowest climatological monthly mean of $3.98 \ \text{}^{\circ}\text{C}$, while single-year winter anomalies reached historical minimums down to $-1.0\text{}^{\circ}\text{C}$ in February 2003 whereas atmospheric water vapor accumulation peaks severely during August, reaching a macroclimatic maximum monthly mean of $17.90 \ \text{}^{\circ}\text{C}$. The comprehensive, year-by-year chronological distribution of these micrometeorological parameters is fully structured and itemized within Table~\ref{tab:dew_point_matrix_clean} in the Data Appendix at the end of this document: \\ 
\includegraphics[width=0.49\textwidth{}]{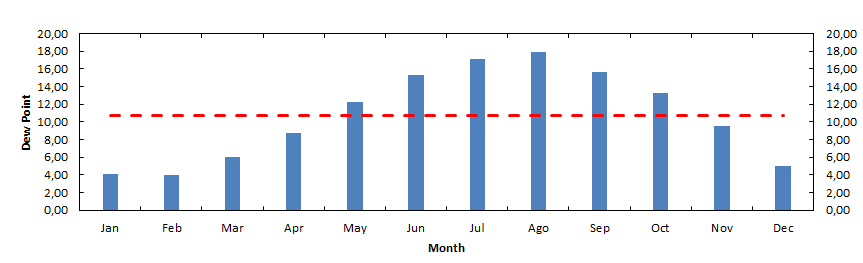} {\scriptsize \bf \textit{\textit{{\footnotesize Dew Point for months from 1990 to 2014}}}}. 
\subsection{Humidity} 
As regards relative humidity ($RH$), the long-term climatological analysis over the 1990--2014 baseline series exhibits a historical multiyear mean value of $73.94\%$, as structurally confirmed in Table 1. Temporally, the normalized anomalies are evaluated according to the following mathematical formulation:
\begin{equation}
	RH_{\text{ratio}} = \frac{AV}{AM}
\end{equation} 
where $AV$ denotes the single annual relative humidity value and $AM$ represents the historical multiyear mean average calculated for the 1990--2014 period. 

The downscaled seasonal fluctuations are graphically evaluated at a monthly level in Figure 8. The multiyear monthly averages highlight a substantial shift above the climatological mean from September to April, driven by low winter insolation and stable boundary layer pooling. Conversely, after a dry summer quarter characterized by intense solar radiation and enhanced convective boundary layer mixing, minimum values track down to a regional seasonal low of $63.59\%$ in July. 

On an annual scale (as mapped in Figure 8 at the end of this document), the dataset shows 11 years plotted above the long-term baseline, with the absolute saturation maximum reached in 2002 ($80\%$). Symmetrically, the remaining 12 years drop below the average, tracking an absolute historical minimum drought in 2003 ($67\%$), matching extreme pan-European heatwave conditions. The complete chronological breakdown is reported within Table~\ref{tab:humidity_matrix_clean} in the Data Appendix. \\ \\
\includegraphics[width=0.49\textwidth{}]{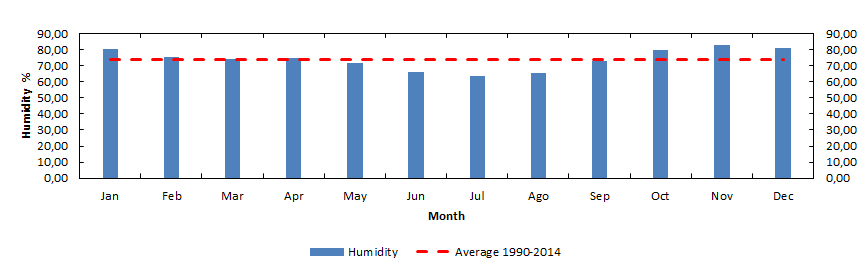} {\scriptsize \bf \textit{\textit{{\footnotesize Humidity Months from 1990 to 2014}}}}. \\ \\
The following graph, which illustrates the monthly values for the 1990–2014 study period, shows values above the average in the period between January and April, and after a below average summer quarter, return to rise above average again in the period September-December. \\ The annual study (figure 8 at the end of the document), highlights 11 years above the average, with the maximum value reached in 2002, and the remaining 12 years with values below the average, the lowest value of which is highlighted in 2003.
\subsection{Pressure} 
The analysis of the barometric data reports a stable historical mean surface pressure over the examined 1990--2014 study period, equal to $1013 \ \text{hPa}$, as summarized in Table 1. The seasonal variations are structured at a monthly downscaled level in the cross-sectional graph of Figure 9. Climatological tracking highlights that January, February, March, September, October, and November exhibit average monthly values shifting systematically above this regional baseline. Conversely, the remaining months are structurally positioned below the mean threshold, with May tracking down to the lowest microclimatic maximum of $1009 \ \text{hPa}$ due to spring convective mixing profiles: \\ \\
\includegraphics[width=0.50\textwidth{}]{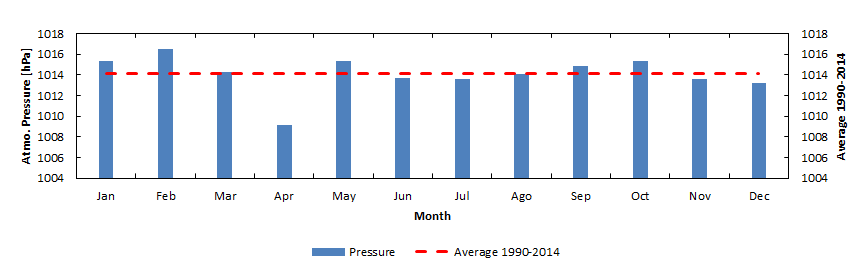} {\scriptsize \bf \textit{\textit{{\footnotesize Pressure Months from 1990 to 2014}}}}. \\ \\
Analyzing the processed dataset on an annual basis, as graphically represented in Figure 8 at the end of this document, we observe distinct barometric eras:
\begin{itemize}
	\item A persistent high-pressure era characterized by progressive positive deviations from the multiyear mean during the mid-stationarity window spanning from 1996 to 2008, driven by persistent sub-tropical anticyclonic ridges;
	\item A dynamic lower-pressure era featuring enhanced barometric instability and significant negative deviations from the long-term baseline concentrated in the years 1991, 1995, 2009, 2010, and 2012, matching regional cyclonic passage maximums as detailed in the chronological matrix of Table~\ref{tab:pressure_matrix_clean}.
\end{itemize}
\subsection{Day time and night time variation} 
An accurate analysis of the daily $\Delta T$ variation of the temperature values between day and night is highlighted in this section and reported in Table 3.
The calculated value of $9,83^\circ C$ relating to the years 1990-2014 clearly evident in the following graph highlights 4 periods that are quite distinct from each other in the values of the temperatures found and precisely:
\begin{itemize}
\item an above average presence over the years: 1993, 1996, 1997, 1998, 1999, 2000, 2002, 2003, 2004, 2006, 2007, 2008 and 2011;
\item values below the average 1990, 1991, 1994, 1995, 2005, 2009, 2010, 2012, 2013 and 2014.
\end{itemize}  
\includegraphics[width=0.50\textwidth{}]{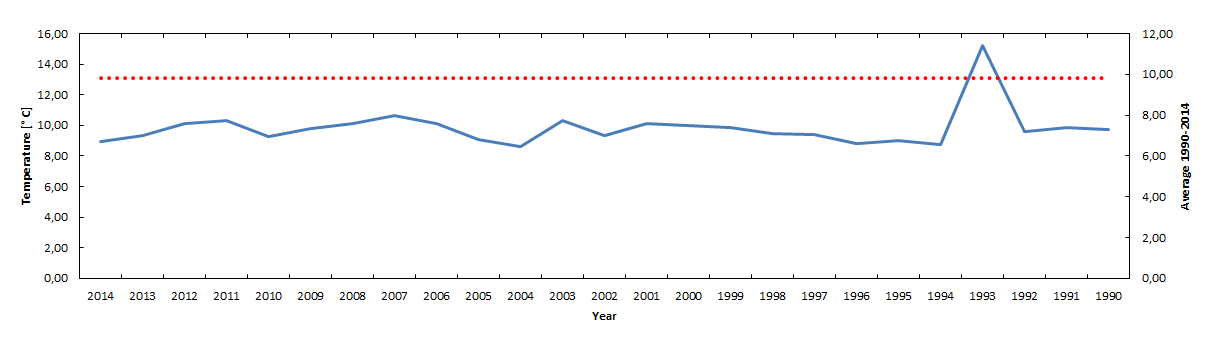}
{\scriptsize {\textit{{\footnotesize Day time and night time variations (solid line) and Average Day time and Night time variations 1990-2014 (dotted line)}}}}
\subsection{Rain} The analysis of the pluviometric data in the study period showed an average annual rainfall of $1304.58 \ \text{mm}$ in the municipal area, and an accumulation value for the period 1990-2014 equal to $32614.60 \ \text{mm}$. Furthermore, and an increase in intense, short-term rainfall events emerged, as clearly shown in Figure 18 at the end of the document, which shows the rainfall trends for each month of the year referring to the values between 1990 and 2014. 
\includegraphics[width=0.49\textwidth{}]{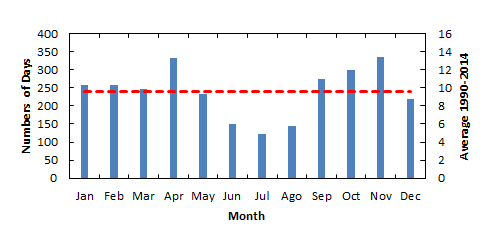} 
{\scriptsize {\textit{{\footnotesize Days Rain for Month from 1990 to 2014}}}}. \\ \\ These lead to an increase in hydro-geological risk as well as flash-floods, defined as the rapid flooding of a morphologically limited area, due to the rapid “saturation” of the surface soil which is no longer able to absorb rain. Such episodes of intense precipitation can also determine phenomena of surface runoff of rainwater with a possible increase in flooding, but also risk of water pollution (pollutants from agricultural and road runoff). \\ The study period showed an abnormal increase in rainfall.The pie chart below shows the percentage contribution of each single year's rainfall relative to the total cumulative precipitation measured over the 1990--2014 baseline period, according to the following formulation:
\begin{equation}
	P_{\text{year}} = \frac{a}{b} \times 100
\end{equation}
where $a$ denotes the specific annual rainfall value (mm) and $b$ represents the total cumulative precipitation of the multi-decadal series (32614.60 mm).
 \\ \includegraphics[width=0.49\textwidth{}]{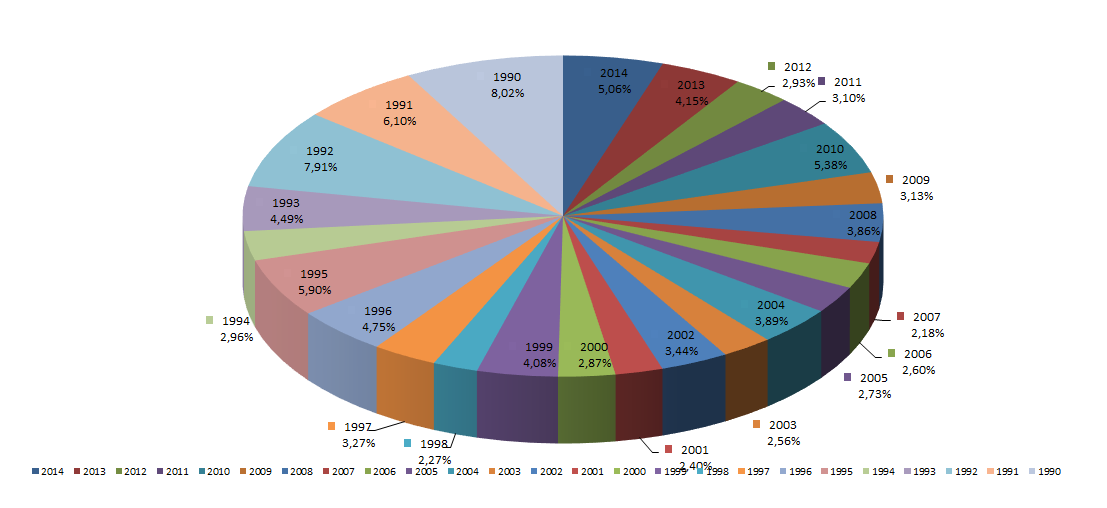}
{\scriptsize \textit{\% of Rain for this study}}. \\
In Figures 6 at the end of the document, we can see the total rainfall/year relating to this study, the values of which are shown in table 4. The next pie chart highlights the type of phenomena found for this study.
\includegraphics[width=0.5\textwidth{}]{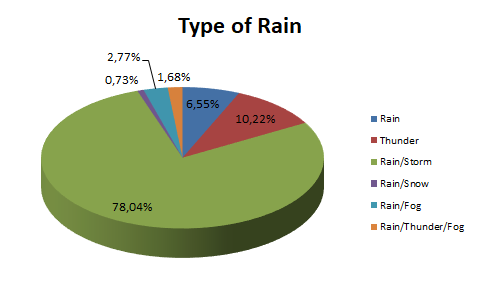}
{\scriptsize \textit{Type of Rain for this study}}. \\ \\ The statistical analysis of the 25-year multi-decadal series highlights the following cumulative environmental and macroclimatic metrics for the entire study period (1990–2014):
\begin{itemize}
	\item Total precipitation events: 1711 rain days (averaging 68.4 days/year) and 1,069 rain/storm days;
	\item Atmospheric obscuration and electrical events: 254 fog days (averaging 10.2 days/year) and 140 thunder days;
	\item Mixed and solid winter precipitation: 38 rain/fog days, 23 rain/thunder/fog days, 10 rain/snow days, and a low-frequency volume of 11 snow days (stabilizing at an operational mean of 0.44 days/year).
\end{itemize}

monthly:
\begin{itemize}
\item the month of October being the wettest.
\end{itemize}
in terms of number of rainy days per year:
\begin{itemize}
	\item 10 years out of 25 in total in which the cumulative annual values exceeded the climatological baseline average of $1304.58 \ \text{mm}$, with the historical maximum precipitation concentrated in the year 1990;
	\item 15 years out of 25 total in which the cumulative annual values were below the historical average, with the absolute minimum annual rainfall recorded during the dry year of 2007 ($709.50 \ \text{mm}$), in strict agreement with Table 4.
\end{itemize}
The comprehensive chronological breakdown of the monthly and annual precipitation metrics is fully itemized within Table~\ref{tab:rain_regimes_clean} in the Data Appendix at the end of this document.
\subsection{Snow}
Within the municipal area of Rome, snowfall represents a low-frequency but meteorologically significant climatological event. Statistical elaboration of the 1990--2014 dataset tracking the Torrespaccata sector isolates a total volume of $11 \ \text{days}$ of snow over the 25-year baseline period, resulting in a historical operational mean of $0.44 \ \text{days/year}$. Seasonally, occurrences are strictly confined to the winter window, as itemized in Table~\ref{tab:snow_days_clean}. The monthly distribution shows a relative maximum clustering in February with 4 total days, followed by December (3 days), January (2 days), and March (2 days), as visually summarized in the bar chart of Figure 21.\\
\includegraphics[width=0.49\textwidth{}]{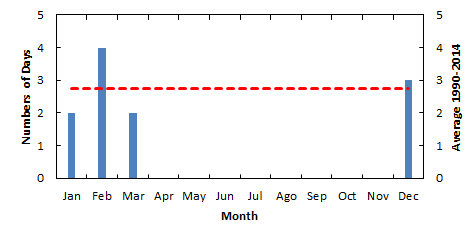}
\\
{\scriptsize \textit{Snow month days from 1990 to 2014}}
In terms of physical accumulation layer, the most severe single-event ground deposition occurred during February 2012, recording a localized snow blanket thickness of $11 \ \text{cm}$. Conversely, the maximum multi-event seasonal accumulation was observed during the winter of 1994, tracking a cumulative total profile thickness of $20 \ \text{cm}$ across three distinct days, as summarized in the baseline synthesis of Table 1.
 
\subsection{Fog} 
The statistical distribution of fog occurrences over the 1990-2014 climatological period shows a steady multi-decadal declining trend, tracking a cumulative volume of 254 fog days. This ongoing reduction is strictly associated with the progressive rise in urban winter night-time minimum temperatures, which limits boundary layer cooling below the saturation threshold. 

Analyzing the downscaled dataset on a monthly scale, the long-term historical average stabilizes at 2 days per month. Seasonally, fog phenomena are strongly clustered within the cold autumn and winter periods: January represents the absolute macroclimatic maximum with a cumulative count of 98 days, followed symmetrically by November (37 days) and December (30 days). Conversely, convective mixing and severe solar radiation suppress fog formation during the summer quarter, restricting occurrences to historical minimums in June (8 days) and July (only 4 days across the 25-year series). 

On an annual scale, the dataset exhibits significant non-stationary steps, marking a progressive drop in boundary layer stability during the last decade. The historical maximum was recorded during the early stationary period in 1993 with a peak of 15 fog days, whereas the absolute minimum was reached in 2014 with a record low of only 6 days. Furthermore, local micrometeorological observations confirm that winter radiation fogs (highly localized "humidification mists") are transient phenomena restricted to late-night hours over unpaved open soils, such as the non-urbanized sectors of the ``Pratone di Torrespaccata'', rapidly dissipating during early morning hours due to mechanical boundary layer mixing.
\includegraphics[width=0.49\textwidth{}]{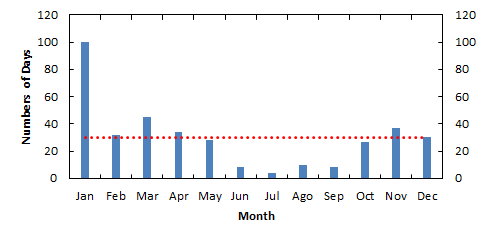}
\\
{\scriptsize \textit{Foggy month days from 1990 to 2014}}
\subsection{Wind speed} 
The anemometric regime of the urban quadrant under examination is structurally governed by local planetary boundary layer dynamics and regional coastal thermo-baric gradients. Climatological baseline data calculated for the 1990--2014 series indicates that the atmospheric circulation over the Torrespaccata district is dominated by a classic Mediterranean breeze cycle, highly modulated by the surrounding urban canopy architecture. 

Thermodynamic tracking highlights two symmetrical, anti-diurnal air mass circulation phases:
\begin{itemize}
	\item During night-time hours, the atmospheric circulation is primarily driven by a weak, low-velocity catabatic drainage flow descending from the nearby Colli Albani volcanic complex toward the Tiber-Aniene valley depression, maintaining stable and stratified boundary layer conditions;
	\item During daytime hours, this pattern is completely reversed by the inland penetration of the regional maritime sea breeze (the local ``Ponentino''), blowing from the Tyrrhenian coast toward the interior plains, which mitigates surface air temperatures while systematically increasing absolute moisture levels across the urban canopy layer.
\end{itemize}
The regional ventilation parameters are structurally integrated with the national anemometric database made available by the Wind Atlas of Italy \cite{atlantevento:2013}, evaluating the historical air mass velocity calculated at a reference altitude of 25 meters above ground level, as visually mapped in Figure 11. 

Statistical elaboration of the long-term historical series demonstrates a stable boundary layer structure, characterized by seasonal extreme events. The absolute maximum historical wind speed reached a peak of $86.00 \ \text{km/h}$ during a severe convective frontal passage in December 2001, while the highest wind speed baseline for the spring quarter occurred in March 2001 with ($50.73 \ \text{km/h}$) Conversely, the lowest calm periods (\(0.00\text{ km/h}\)) are statistically clustered around February 2007, as summarized in Table 1. 

The maximum monthly values also exhibit a persistent below-average trend, stabilizing at a suppressed operational peak of $33.84 \ \text{km/h}$ specifically during the month of June, as cross-sectionally plotted in Figure 11. Daily anemometric records - including the historical monthly means, absolute minima, maxima, and the calculated baseline index of $19.18 \ \text{km/h}$ for the multi-decadal midrange velocity - are fully cross-examined and structured within the comprehensive chronological framework of Table 8. Seasonal and diurnal variations are visually raccorded in the multi-panel comparative bar charts of Figure 11, Figure 12, and Figure 23 at the end of this document. \\ \\ 
\includegraphics[width=0.49\textwidth{}]{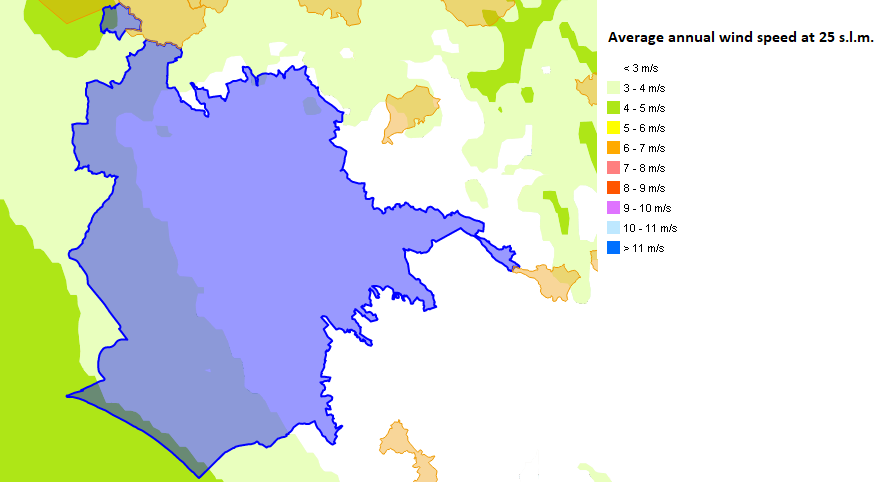}
{\scriptsize \textit{Wind Speed Atlas of Rome - RM \cite{atlantevento:2013}}}
\subsection{Blackbird days}
The ``Days of the Blackbird'' (traditionally corresponding to the final three days of January: 29, 30, and 31) are investigated within this framework to evaluate localized winter core thermal anomalies. Climatological statistics demonstrate that while regional folklore identifies this winter quarter phase as the absolute coldest interval of the year, atmospheric circulation over central Italy frequently exhibits a distinct warming response after the first decadal transition of January. 

To structurally quantify the thermal deviations across the 1990--2014 series, the interannual temperature delta ($\Delta T_{\text{blackbird}}$, $^{\circ}\text{C}$) is formulated according to the following mathematical framework:
\begin{equation}
	\Delta T_{\text{blackbird}} = \overline{T}_{\text{jan\_core}} - \mu_{\text{jan\_clim}}
\end{equation}
where $\overline{T}_{\text{jan\_core}}$ represents the specific mean air temperature calculated exclusively across the three-day target interval for each single year, and $\mu_{\text{jan\_clim}}$ denotes the historical multiyear baseline monthly average computed for January ($7.56 \ \text{}^{\circ}\text{C}$). 

The computed annual anomalies are structured within Table~\ref{tab:blackbird_days_clean} and visually cross-examined in the historical baseline variations plotted in Figure 17 at the end of this document.

\section{Geological, Hydrogeological and Earthquake Data} 
In this section we describe the Geological, Hydrogeological and Earthquake data for this study. \\ \\
\includegraphics[width=0.49\textwidth{}]{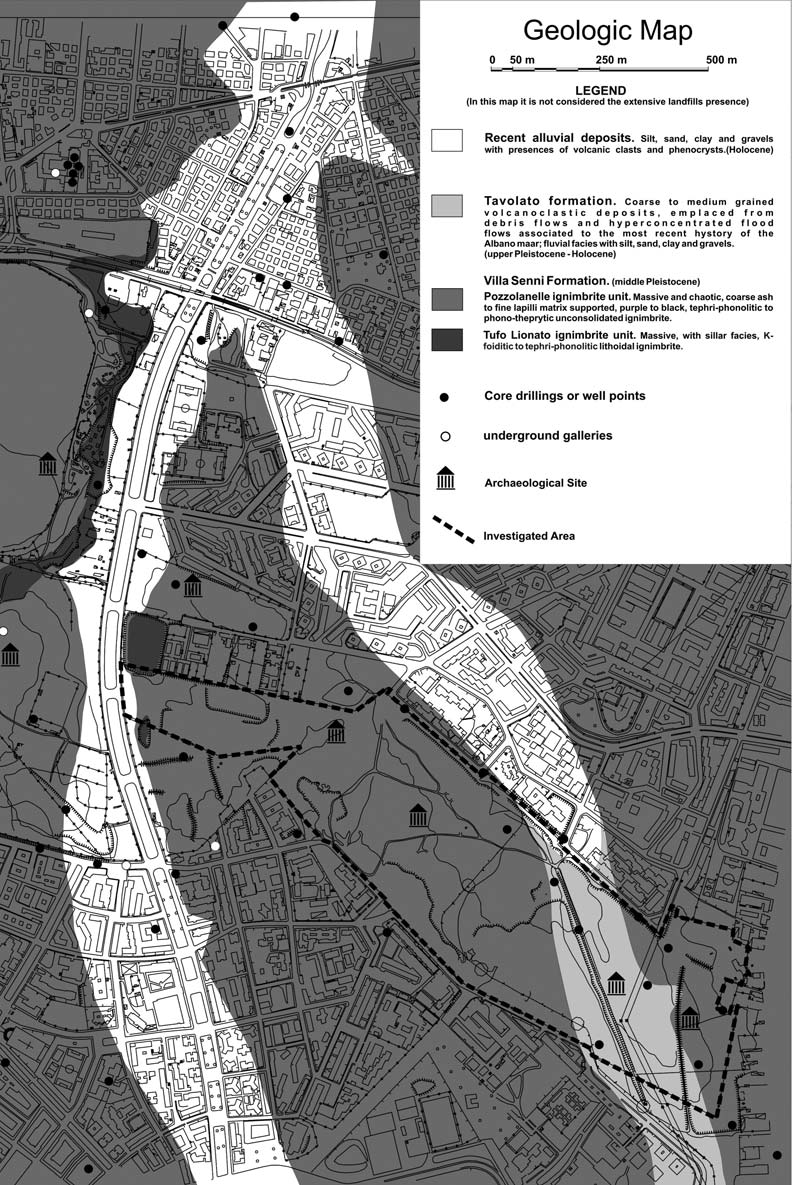} \\
{\scriptsize \textit{Geological Map of Torrespaccata. \cite{Gioia:ref1}.}} 
\subsection{Morphological Lines}
The morphological characteristics and structural evolution of the survey area have been fundamentally shaped by Late Quaternary volcanic deposition originating from the nearby Colli Albani Volcanic District (specifically the Vulcano Laziale eruptive phases) and subsequent fluvial incision within the Aniene River basin. The landscape is dominated by a sub-horizontal or gently sloping ignimbrite plateau, ranging in elevation between 50 and 70 meters above sea level, composed of coherent and semi-coherent pyroclastic successions. Exogenous morphogenetic processes have carved relatively shallow, flat-bottomed valleys within the less consolidated tuffs, with the historical minor drainage network primarily represented by the Tor Tre Teste stream quadrant. Widespread post-war anthropogenic urbanization has heavily modified and leveled the original surface topography through extensive soil sealing and the artificial culverting of seasonal creeks. A major geomorphological exception within this framework is the unbuilt ``Pratone di Torre Spaccata'' sector, which preserves the pristine morphology of the gently incised paleovolcanic plateau, functioning as a key localized natural retention basin during extreme run-off events.

\subsection{Geological Characteristics}
As already mentioned, volcanic and pyroclastic rocks are the most widespread lithotypes within the survey area, whereas formations belonging to the deep sedimentary substrate emerge only marginally. The following geological description is therefore proportionate to the outcropping extension and to the structural importance that both complexes represent for the engineering and environmental purposes of this study.

The subsurface of Torrespaccata (eastern Rome quadrant) is dominated by volcanic deposits associated with the ``Tufi Terrosi'' complex and the ``Tufo Rosso a Scorie Nere'' formation (approximately 0.45 Ma), unconformably overlying lower Pleistocene marine and continental sediments. The area exhibits a flat to gently undulating topography, locally incised by minor drainage networks. The key stratigraphic and geotechnical characteristics of the district are summarized below:
\begin{itemize}
	\item \(\textbf{Volcanic Cover:}\) Thick banks of lithified and semi-coherent tuff (eruptive products of the Colli Albani Volcanic District), frequently weathered at the surface into reworked soils, clayey silts, and pozzolanic eluvium;
	\item \(\textbf{Plio-Pleistocene Substrate:}\) The deep geological basement is composed of highly consolidated marly clays, silty pelites, and fine sands belonging to the Monte Mario and Argille Azzurre Formations;
	\item \(\textbf{Hydrogeological Setting:}\) A shallow unconfined (phreatic) aquifer is hosted within the permeable pyroclastic cover, subject to seasonal water table fluctuations. Conversely, a deeper confined/artesian aquifer flows within the underlying sandy-gravelly layers of the marine substrate;
	\item \(\textbf{Geomorphology and Geohazards:}\) The district is situated within a peripheral suburban sector characterized by moderate localized hydraulic vulnerability and localized flash-flood risks, primarily concentrated near historical, now-culverted drainage channels (e.g., the historical Centocelle and Togliatti drainage fosses) due to intensive anthropogenic soil sealing.
\end{itemize}

\subsection{Volcanic Formations}
The geological-deformation history of the study area of this article is closely connected with that of the Lazio volcanic region, which must be framed within the context of the tectonic-dynamic evolution of central Italy. The geological evolution of this area is the result of the geodynamic processes and the extensional tectonics of the belt between the Apennine chain and the Tyrrhenian coast, following the Apennine orogeny and active in the period between the upper Pliocene up to the Late Quaternary. Starting from the Pliocene, an intense magmatic activity begins that involves, along the Apennine route, a vast range from Monte Amiata to the Campania region. The volcanic successions present in the specific area of this study belong exclusively to the Colli Albani Volcanic District (Vulcano Laziale complex). This volcanic system exhibits a distinctly alkaline-potassium character and initiated its paroxysmal active eruptive state approximately 0.6 Ma ago, depositing widespread sheets of pyroclastic products and localized lava flows directly over the sedimentary substrates of the Roman basin. The following figure identifies this regional classification scheme.\\ \\
\includegraphics[width=0.49\textwidth{}]{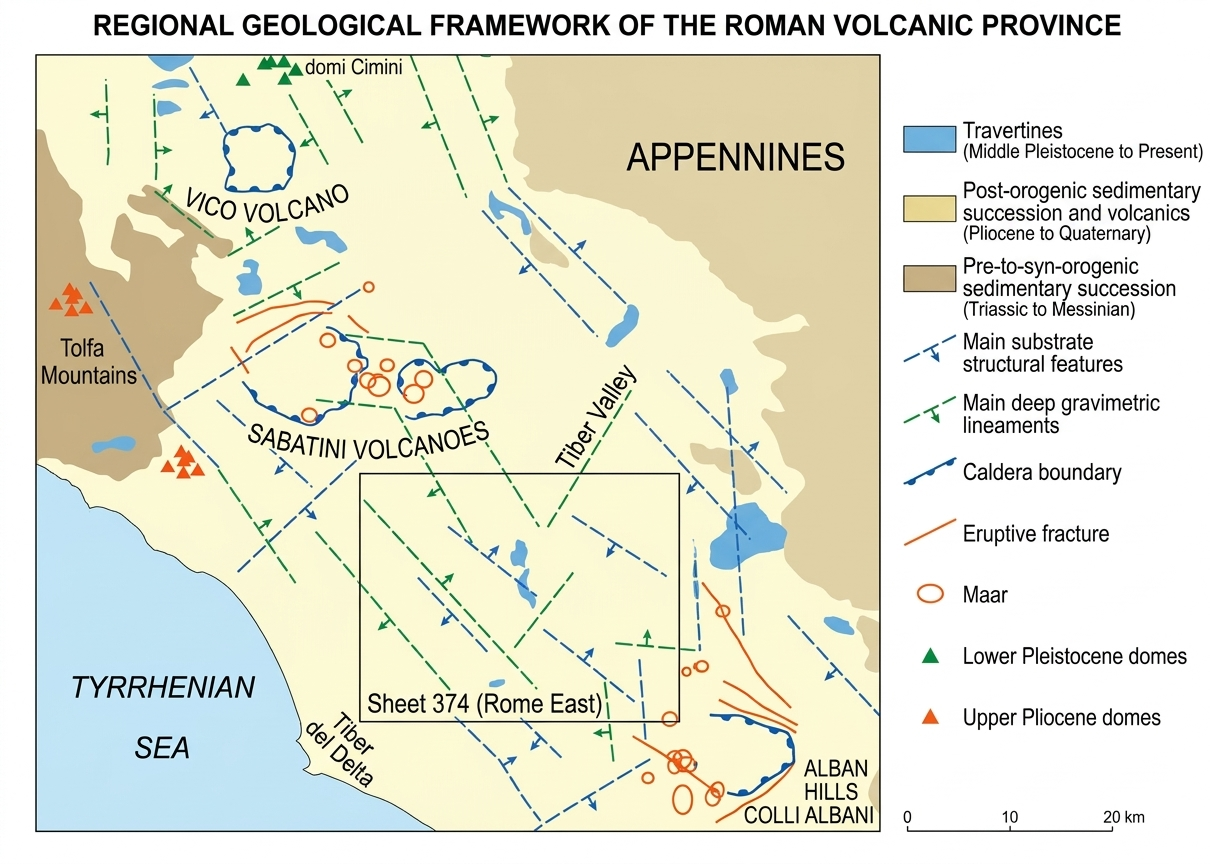} \\
{\scriptsize \textit{Regional classification scheme. Legend (from top to bottom): Travertines - Post-orogenic sedimentary succession and vulcanites - Pre-syn-orogenic sedimentary succession - Main substrate features - Main deep gravimetric features - Caldera - Eruptive fracture - Maar - Lower Pleistocene domes - Pleistocene domes superior}}.
\subsubsection{Alban Hills District} 
The Alban Hills (or Vulcano Laziale) is a large stratovolcano featuring a central caldera, located approximately 20~km southeast of Rome. The Alban Hills complex developed within a zone characterized by extensional fault systems oriented NW-SE, NE-SW, and N-S. 

The volcanic products consist essentially of pyroclastic deposits (peperino) and minor lava flows, erupted over a timespan ranging from 0.6~Ma down to the Roman historical era. These volcanic rocks cover an area of approximately 1000~km$^2$ and overlie Plio-Pleistocene marine deposits that fill extensional basins, which originally developed within flysch sequences and carbonate platforms (Sabina facies and Umbro-Latium Facies). The Alban Hills are geologically situated on the western side of the Ancona-Anzio tectonic line.

The Alban Hills volcano has attracted the attention of geologists for a long time; numerous volcanological, petrological, and morphological studies have allowed the reconstruction of its main eruptive phases:
\begin{itemize}
	\item \textbf{Tuscolano-Artemisio Phase (from approx. 0.6 to 0.3~Ma):} This phase involved the construction of the primary volcanic edifice and concluded with a calderic collapse. It produced approximately 280~km$^3$ of volcanic materials, consisting mainly of ignimbrites and fallout deposits. The largest ignimbrite eruption (approx. 350~ka) generated a massive pyroclastic flow (Villa Senni Tuff) and triggered the calderic collapse. Symmetrical syn- and post-collapse eruptions occurred subsequently.
	\item \textbf{Faete Phase (or Campi di Annibale phase, from approx. 0.3 to 0.2~Ma):} This phase was characterized by the formation of the intra-calderic Faete cone, followed by an intra-calderic collapse. It generated approximately 6~km$^3$ of volcanic materials, which constitute the Faete volcanic cone, massive intra-calderic pyroclastic deposits (Campi di Annibale Tuff), and minor lava flows along the caldera flanks. The Faete activity concluded with the collapse of the nested Campi di Annibale caldera and circum-calderic Strombolian eruptions.
	\item \textbf{Hydromagmatic Phase (from approx. 0.2 to 0.02 Ma):} This phase was dominated by violent phreatomagmatic eruptions. It developed along numerous volcanic centers situated in the western sector of the main volcanic edifice, producing approximately 1~km$^3$ of volcanic materials (surges, lahars, and rare lava flows). The most prominent eruptive centers are the maars of Nemi and Albano Lake. The youngest ages for the Alban Hills volcanic deposits, measured in the Albano Lake area, indicate an age younger than 19~ka.
\end{itemize} 

The volcanic rocks of the Alban Hills display a strongly ultrapotassic and highly undersaturated composition; the products range from tephrites to foidites (leucitites) and tephri-phonolites.

\vspace{1em}
\noindent\includegraphics[width=\columnwidth]{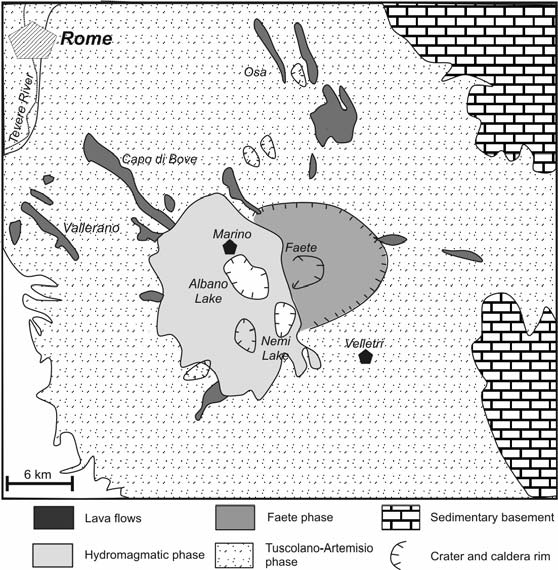}

\noindent{\small\textbf{Figure 3:} \textit{Geological structure of the Colli Albani (Alban Hills) complex \cite{DeRita:ref1}.}}
\vspace{1em}
The lava flows appear more mafic than the pyroclastic deposits originating from the same eruptive centers, while maintaining similar alkalinity values. The majority of the lavas exhibit a poorly porphyritic texture, featuring phenocrysts of leucite and clinopyroxene, along with minor olivine, embedded in a groundmass composed of the same phases with the addition of melilite, nepheline, phlogopite, and calcite. Plagioclase is practically absent in the Alban Hills products, with the exception of the groundmass in certain specific lava flows.

The pyroclastic deposits of the Alban Hills consist of scoria and pumice, showing a porphyritic structure with leucite and clinopyroxene crystals. Sizable lithic xenoliths of varying nature are frequently hosted within these pyroclastic sequences, including fragments of lava flows, microgranular plutonic rocks representing the intrusive equivalents of the erupted lavas, cumulates, and skarns.

The recent history of the Alban Hills has highlighted volcanic activity even during the Roman era (certain Roman authors reported anomalous phenomena such as stone showers, sudden fires, and explosions), and numerous archaeological data exist, including pottery and other pre-Roman human artifacts discovered beneath pyroclastic layers. These data indicate that the volcanic activity is considerably younger than the age derived from Albano Lake (19~ka). Currently, based on seismic and historical data, the Alban Hills complex is considered to be in a phase of long-term quiescence \cite{Peccerillo A:2005}.
\subsubsection{Volcanic area of Rome}
The initial activity of the Rome complex is mainly identified with the Ospedaletto eruption, characterized by a Plinian eruptive column about 20 km high which gave rise to deposits of pumice covering a very large area in the eastern and southern sectors.\\ The eruptive scenarios of the Rome volcanic area were influenced by the structure of the carbonate base. NNO-SSE fractures have in fact determined an intense pre and post-calderic effusive activity in the central band; in the southern one, it was an O-E fault that represented the supply route for scoria cones and vast lava flows.\\ The explosive activity in the Roma area was very intense and mainly of the hydromagmatic type.\\
The Roma Ignimbrite and the formation of the relative caldera are connected to the eruption richer in energy. The basal ignimbrite, which crops out in the southern and eastern sector, appears as a massive deposit from light gray to dark gray.\\
A second central explosive eruption determines the formation of Lava Drop Ignimbrite. A further collapse of a part of the northern sector of the caldera itself, while a resumption of activity with hydromagmatic phases, gave rise to the chaotic collapse of the eastern sector.\\
Among the lava products of the Rome complex, there is a prevalence of leucitites, tephra, and basanites.
\subsection{Hydrological and Hydrogeological Framework}
\label{sec:idrologia_torrespaccata}
The hydrological and hydrogeological setting of the Torre Spaccata district (Eastern Rome) results from the complex interaction between the original volcaniclastic morphology, Quaternary continental sedimentation, and intense anthropogenic urbanization. The latter has radically modified natural surface drainage patterns through the rectification and culverting of the minor hydrographic network.
\subsubsection{Surface Hydrography and Catchment Basin}
The surveyed territory lies within the hydrographic sub-basin of the Aniene River, the primary left-bank tributary of the Tiber River. Originally, the area was drained by a low-to-medium hierarchy drainage network, characterized by modest drainage furrows and seasonal torrential creeks. The primary collector of this quadrant is identified as the \textit{Tor Tre Teste stream} (historically known also as the Marrana di Tor Sapienza), which flows in a North/North-West direction toward the Aniene valley.
Due to intensive urban development (public housing programs and directional districts), almost the entirety of this minor hydrographic network has been integrated, culverted, or diverted into the mixed sewerage system of Roma Capitale. A localized partial exception is represented by the non-urbanized area known as the \textit{Pratone di Torre Spaccata}, which preserves the pristine morphology of a gently incised plateau, acting as a local natural retention basin and partially mitigating the flood peak induced by the surrounding surface \textit{run-off}.
\subsubsection{Hydrogeology and Aquifer Systems}
From a hydrogeological perspective, the subsurface architecture of Torre Spaccata can be schematized as a multi-aquifer system governed by the hydraulic conductivity of the different lithological complexes, as synthesized in Table 10.
\begin{enumerate}
	\item \textbf{Surficial Unconfined (Phreatic) Aquifer:} Hosted within the pyroclastic cover of the Alban Hills Volcanic District (primarily the \textit{Pozzolane Rosse} and \textit{Tufo Lionato} formations). The primary intergranular porosity (within the pozzolane) and secondary fracturing (within lithified tuffs) yield a medium-to-high permeability ($K \approx 10^{-4} \div 10^{-5}$~m/s). Subsurface circulation is directly recharged by effective meteoric precipitation where the soil canvas is not sealed.
	\item \textbf{Semi-confined and Artesian Aquifers:} Positioned at greater depths within the polygenic gravel and sandy-quartzose lenses of the Middle Pleistocene fluvio-lacustrine deposits (e.g., Valle Giulia Synthem). These permeable horizons are vertically bounded above and below by low-conductivity silty-clayey lenses.
	\item \textbf{Basal Aquiclude:} Formed by the thick marly clay and pelitic successions of the Plio-Pleistocene marine substrate (\textit{Monte Mario Formation} / \textit{Monte Vaticano Unit}). This complex is hydraulically impermeable ($K < 10^{-9}$~m/s), blocking vertical deep drainage and sustaining the overall subsurface hydrodynamics of the urban quadrant.
\end{enumerate}
\begin{table}[htbp]
	\centering
	\caption{Macroscopic hydrogeological properties of the outcropping and sub-outcropping units in the Torre Spaccata area (reprocessed from ISPRA Sheet 374 data).}
	\label{tab:idrogeologia}
	\begin{tabular}{llll}
		\hline
		\textbf{Geological Unit} & \textbf{Prevalent Lithology} & \textbf{Permeability} & \textbf{Hydrogeological Function} \\ \hline
		Holocene Alluvium & Clayey silts, sands & Moderate/Low & Surface drainage \\
		Pyroclastic Products & Pozzolane, fractured tuffs & Medium/High & Main unconfined aquifer \\
		Continental Deposits & Gravels, sands, silts & Discontinuous & Semi-confined aquifers \\
		Monte Mario Fm. & Blue marly clays & Practically nil & Regional aquiclude (basement) \\ \hline
	\end{tabular}
\end{table}
\subsubsection{Hydraulic Vulnerability and Hydrological Balance}
The local hydrological balance is critically impacted by the high soil sealing index ($I_{imp} > 70\%$), which is typical of consolidated peripheral urban fabrics. The surface run-off coefficient ($\mathcal{C}$) exhibits severe spatial fluctuations: 
\begin{itemize}
	\item $\mathcal{C} \approx 0.15 \div 0.25$ within park sectors and the \textit{Pratone} area, characterized by andosols with high water-retention and infiltration capacities;
	\item $\mathcal{C} \approx 0.85 \div 0.95$ within densely built-up areas and paved transport infrastructures.
\end{itemize}
Consequently, the district is not prone to large-scale river basin flooding, but is highly vulnerable to urban \textit{flash floods} (localized flooding events). During short-duration, high-intensity convective rainfall events ("water bombs"), the artificial drainage network rapidly saturates, generating a hyper-concentrated \textit{run-off} along major roadways and within topographic depressions of the paleovolcanic morphology.
\subsection{The Sedimentary Units}
\noindent\includegraphics[width=\columnwidth]{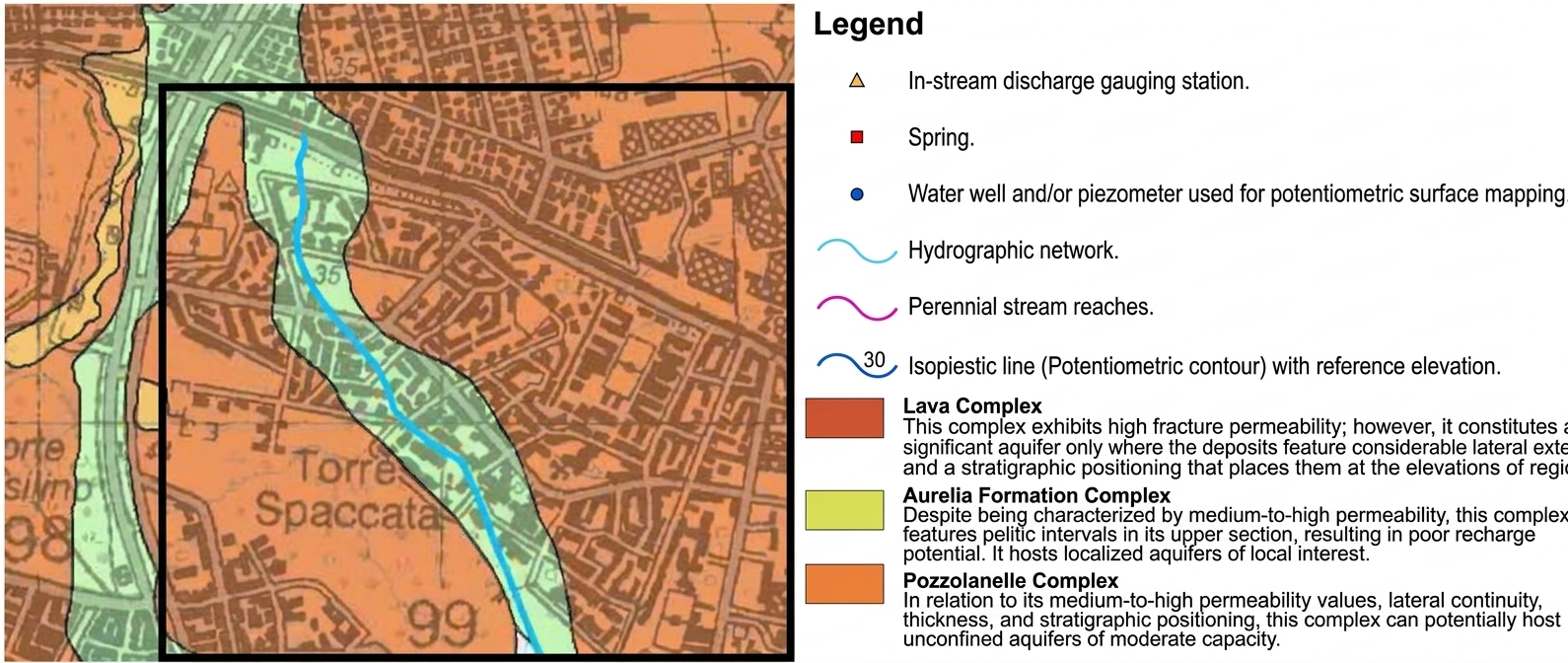}
\noindent{\small\textbf{} \textit{Lithostratigraphic and surficial geological map of the Torrespaccata district. The spatial distribution of the main lithological complexes—specifically the high-permeability fractured Lava Complex, the pelite-rich Aurelia Formation Complex, and the unconfined aquifer-hosting Pozzolanelle Complex—is mapped alongside local hydrographic features, stream reaches, and active gauging and monitoring stations.}}
\vspace{1em}
\subsubsection{Characterization of Sedimentary Units}
The Torrespaccata district, although morphologically dominated by the pyroclastic cover of the Alban Hills Volcanic District, features an articulate sedimentary component. This component develops across three main domains: the deep marine basement (pre-volcanic), the transitional fluvio-lacustrine deposits (interstratified), and the Holocene-to-present alluvial sediments that fill the valley incisions.
\subsubsection{The Plio-Pleistocene Marine Substrate}
At the base of the entire stratigraphic succession in the East Rome quadrant lies the lower Pleistocene marine unit, known in the literature as the Monte Mario Formation (or Monte Vaticano Unit).
\begin{itemize}
	\item \textbf{Lithology and Facies:} The unit consists of a thick and homogeneous succession of compacted gray-blue marly clays, silty pelites, and fine sands. At the top of the formation, local transitions to fossiliferous yellowish fine sands are observed, rich in macrofauna (benthic mollusks) and microfauna (foraminifera).
	\item \textbf{Structure and Geometry:} In Torrespaccata, this unit does not outcrop at the surface, being buried beneath a volcanic and continental cover of variable thickness (with the roof estimated between 20 and 50 meters below ground level, depending on paleomorphologies).
	\item \textbf{Hydrogeological Role:} Having an extremely low hydraulic conductivity ($K < 10^{-9}$ m/s), the formation acts as a regional aquiclude. It blocks vertical groundwater flow and sustains the overlying hydrogeological complexes.
\end{itemize}

\subsubsection{Continental and Fluvio-Lacustrine Deposits of the Middle Pleistocene}
During the Middle Pleistocene, prior to and concurrent with the paroxysmal phases of the Vulcano Laziale, the fluvial dynamics of the paleo-Tiber and paleo-Anien deposited continental sedimentary bodies. These are associated with the Valle Giulia Synthem and the San Paolo Unit.

\begin{itemize}
	\item \textbf{Lithology:} The succession is markedly heterogeneous and consists of alternating well-rounded, polygenic gravels (within a sandy matrix), medium-to-coarse quartz-feldspathic sands, and discontinuous lenses of grayish clayey silts, occasionally rich in organic matter and plant debris.
	\item \textbf{Stratigraphic Relationships:} These deposits are frequently found interbedded or in lateral facies transitions (heteropia) with the earliest pyroclastic products (e.g., Pozzolane Rosse and Tufo Lionato), showing a continuous interaction between volcanic/pyroclastic damming and the reorganization of the Pleistocene drainage network.
	\item \textbf{Geotechnical and Hydrological Properties:} The gravelly-sandy layers exhibit medium-to-high primary porosity permeability, acting as confined or semi-confined aquifers. Conversely, the silty-clayey lenses show high compressibility, potentially leading to differential settlement under anthropic loads.
\end{itemize}

\subsubsection{Holocene and Present-Day Alluvial Deposits}
The surface morphology of the remaining non-urbanized areas (such as the protected area of the Pratone di Torre Spaccata) preserves the products of recent sedimentation linked to the minor hydrographic network tributary of the Aniene River.

\begin{itemize}
	\item \textbf{Sediment Characteristics:} These consist of alluvial deposits filling the valley floors of the Tor Tre Teste stream and its minor tributaries. They are predominantly composed of clayey silts, fine sands, and volcaniclastic fractions reworked and redeposited by surface runoff.
	\item \textbf{Hydro-Geomorphological Dynamics:} These surficial sedimentary soils exhibit moderate-to-low hydraulic conductivity due to the abundant fine matrix. During intense weather events, the low vertical infiltration capacity of these valley-floor sedimentary belts promotes surficial water ponding and localized anoxia, sharply contrasting with the adjacent pyroclastic plateaus (characterized by highly well-drained andosols).
\end{itemize}

\subsubsection{Hydrogeological Area}
\includegraphics[width=0.49\textwidth{}]{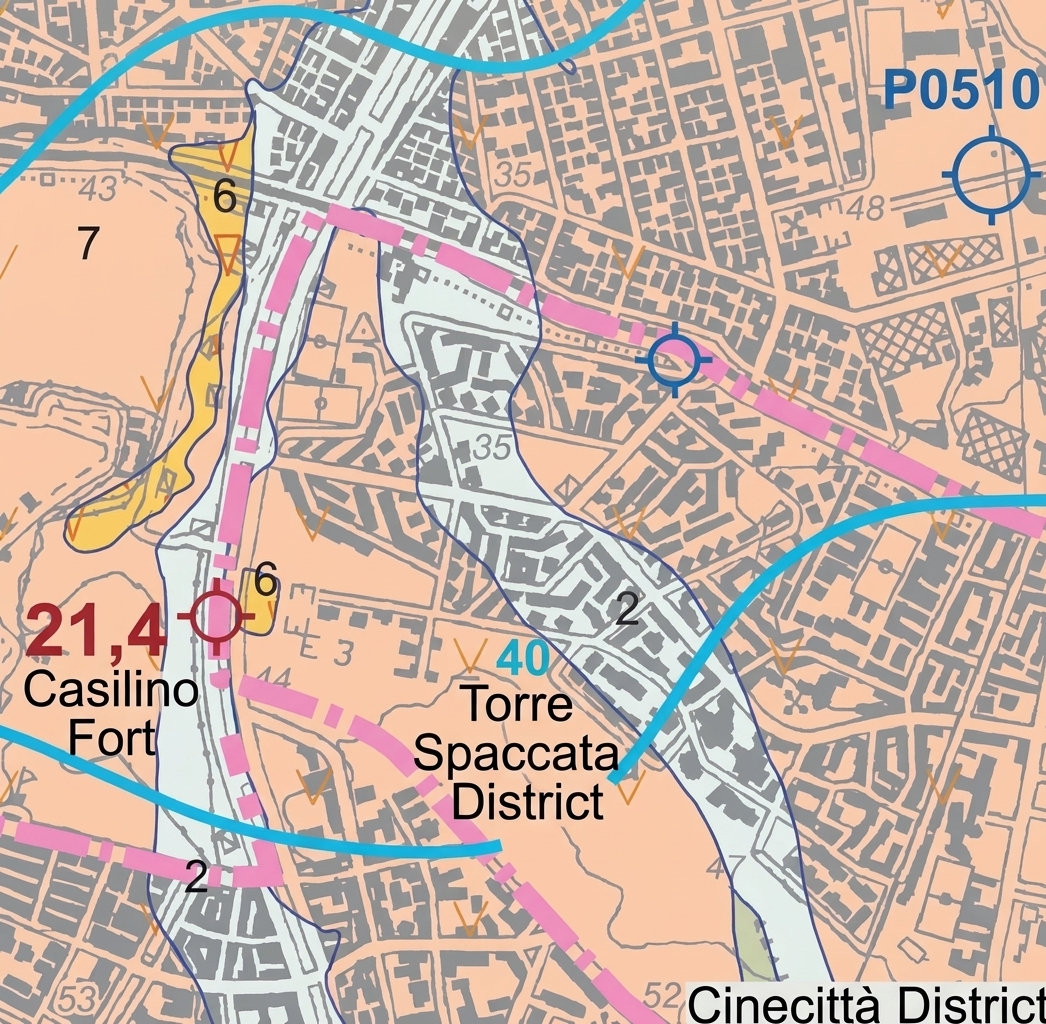} \\\includegraphics[width=0.49\textwidth{}]{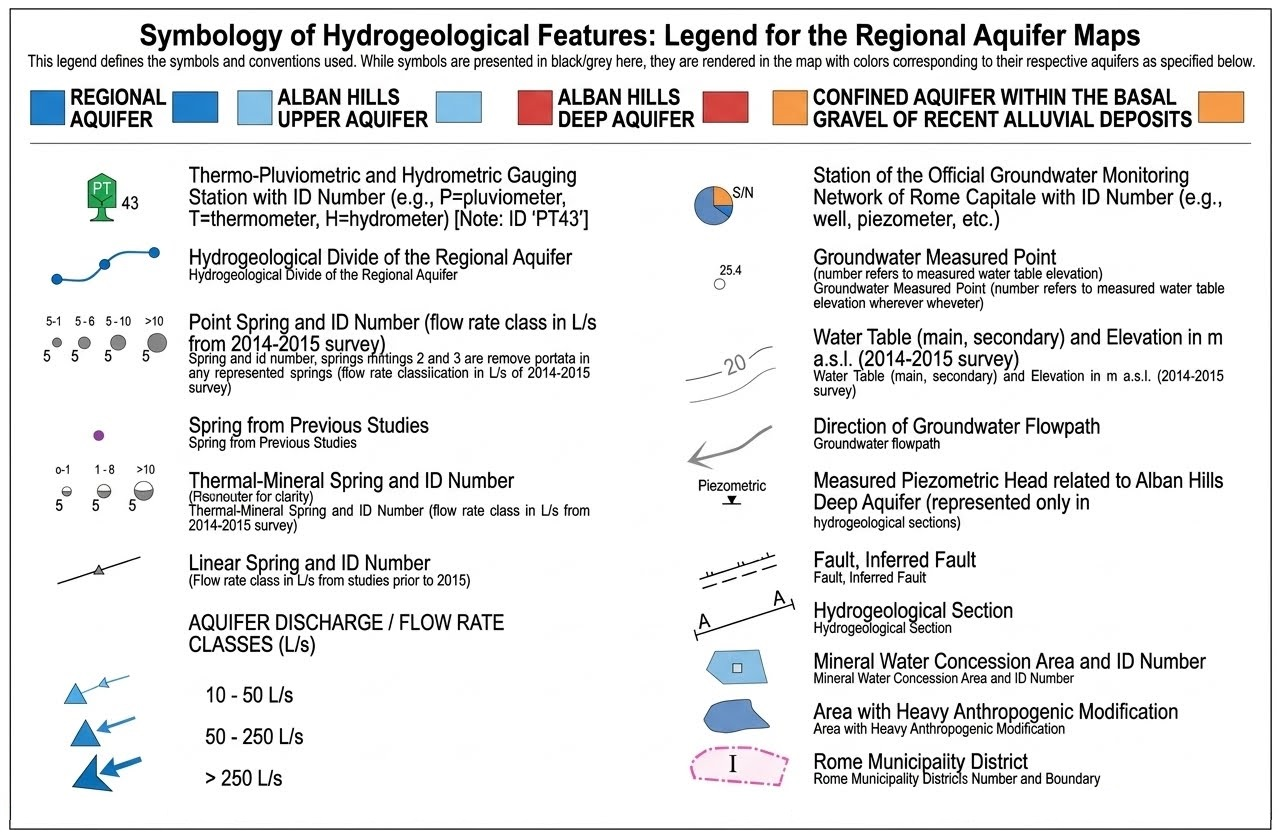} 
{\scriptsize \textit{Hydrogeological map of the Torrespaccata district.}} \\ \\
The Torrespaccata district (Eastern Rome) is characterized by a highly anthropized hydrological setting. Surface drainage networks are managed through regulated channels. The territory is part of the Aniene River catchment basin (the primary left-bank tributary of the Tiber). Surface waters flow through modest valleys draining northwestward, such as the Marrana di Tor Sapienza (or Tor Tre Teste stream). Due to intensive urbanization, most of the minor drainage network has been culverted or integrated into the city's mixed sewerage system.\\
\begin{itemize}
	\item \textbf{Permeability}: Unconsolidated volcanic soils and lava flows show medium-to-high permeability, promoting rainwater infiltration; 
	\item \textbf{Aquifers}: The substrate hosts shallow phreatic aquifers (often located at the stratigraphic contact between volcanic products and Pliocene sediments) and deeper aquifers confined within sedimentary layers, although no significant spring basins are present in Torrespaccata. 
	\item \textbf{Morphology}:The area features gentle slopes with elevations ranging approximately between 50 and 70 m a.s.l. Being a densely urbanized plain with minor reliefs, the primary hazard is related to urban flash floods ("water bombs" causing drainage system failures) rather than river flooding, given the absence of torrential watercourses in the immediate vicinity of the built-up area.
\end{itemize} 

\subsection{Earthquake and Seismic Response Spectrum}
\normalsize The Municipality of Rome is characterized by a territory with medium-high seismicity, \\ \includegraphics[width=0.49\textwidth{}]{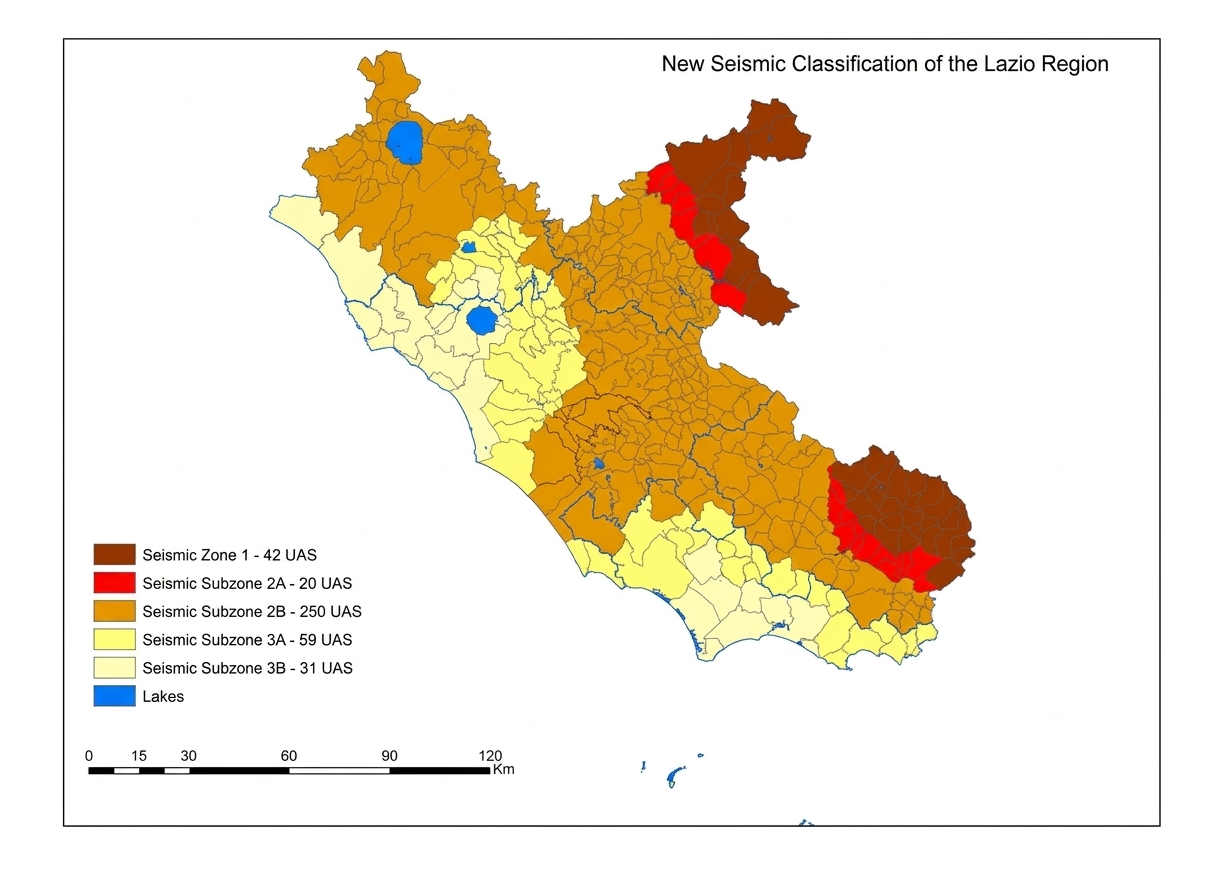} \\
{\scriptsize \textit{Lazio Seismic Hazard Map (MPS). The acceleration values refer to a return of 475 years (INGV2004). Legend:
Seismic zone 1 - 42 UAS - Seismic sub-area 2A - 20 UAS - Seismic sub-area 2B - 250 UAS -
Seismic sub-area 3A - 59 UAS - Seismic sub-area 3B - 31 UAS - Lakes}} $^{\scriptsize \cite{INGV:2004}}$ \\ \\
Indeed, the seismic hazard estimates are highlighted in Table A attached to the Decree of 01/14/2008, drawn up by the Ministry of Infrastructure. From these estimates, the elastic response spectrum (both horizontal and vertical) of the seismic actions was determined, as illustrated in the following graph.

\includegraphics[width=0.49\textwidth{}]{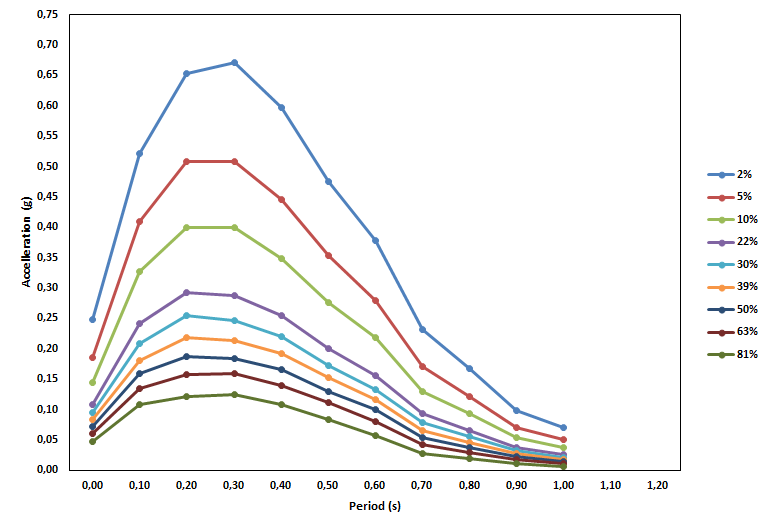} \\
{\scriptsize \textit{The spectra indicate the calculated shaking values for 11 spectral periods, ranging from 0 to 2 seconds. The PGA corresponds to the period of 0 seconds. The graph relates to the median estimates (50th percentile) proposed by hazard model. The different spectra in the graph relate to different probabilities of overflow (PoE) over 50 years.}} $^{\scriptsize \cite{INGV:2004:ref2} \cite{Stucchi:ref1}}$ 

In the following figures, the chronological distribution and characteristics of local seismic activity are presented, while \textbf{Table 19} details the number of earthquake events recorded by year within the investigated area. \\ \\
\includegraphics[width=0.5\textwidth{}]{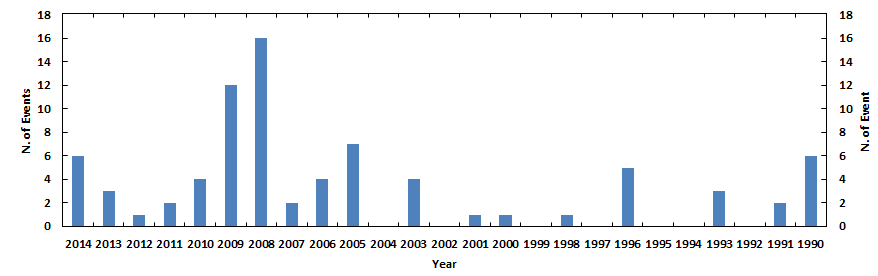}
{\scriptsize \textit{Number of earthquakes from 1990 to 2014}}
\begin{center}
	\scriptsize
	\textit{Table 19: Chronological breakdown of recorded local seismic events in Rome by year (1990--2014).} \\ \vspace{2mm}
	\begin{tabular}{|r|c|r|c|}
		\hline
		\textbf{Year} & \textbf{N. of Events} & \textbf{Year} & \textbf{N. of Events} \\ \hline
		2014 & 6 & 2013 & 2 \\
		2012 & 1 & 2011 & 4 \\
		2010 & 0 & 2009 & 1 \\
		2008 & 0 & 2007 & 0 \\
		2006 & 2 & 2005 & 1 \\
		2004 & 0 & 2003 & 0 \\
		2002 & 1 & 2001 & 0 \\
		2000 & 3 & 1999 & 0 \\
		1998 & 0 & 1997 & 0 \\
		1996 & 0 & 1995 & 0 \\
		1994 & 1 & 1993 & 2 \\
		1992 & 1 & 1991 & 0 \\ 
		1990 & 0 &      &   \\ \hline
	\end{tabular}
\end{center}
The study evidence that the major number of Earthquake are production on deep from 5 to 10 Km. Next graphics evidence this. \\ \\
\includegraphics[width=0.5\textwidth{}]{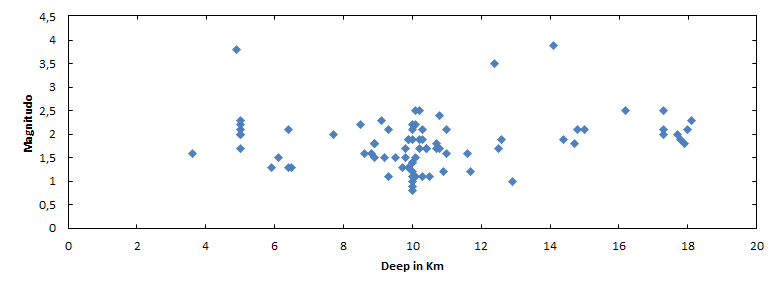} \\
{\scriptsize \bf \textit{Magnitudo/Deep for 1990 to 2014}}
\normalsize
\section{Conclusion} 
We presented for the first time a comprehensive analysis of geological and long-term temperature data obtained directly from a local meteorological site in Rome. This station is located within a dense urban area and well above the atmospheric inversion layer.

The data analysis, particularly regarding annual rainfall regimes and their temporal evolution, indicates a recorded increase in cumulative precipitation, alongside a higher concentration of events during the wettest periods. If confirmed over the long term, the extremes of this trend would lead to a radical shift in the local climate. The area would no longer be characterized by the complete absence of a dry season. Instead, it would exhibit a well-defined rainy season with heavy, concentrated precipitation, contrasting with a distinct dry season mirroring the typical features of a Mediterranean climate.

Moreover, this study establishes a critical, direct link between multi-decadal microclimatic transitions and local hydrogeological dynamics within the framework of the NGICS project. The dual pressure exerted by the sharp contraction in winter Heating Degree Days (HDD)—driven by the Urban Heat Island (UHI) effect—and the documented increase in short-term, convective rainfall events significantly alters the local hydrological balance, which can be mathematically formulated through the conservation of mass equation:

\begin{equation}
	\Delta S = P - (ET_o \cdot k_c) - Q_{run-off} - I_{deep}
\end{equation}

where $\Delta S$ represents the net change in subsurface moisture storage within the phreatic aquifer hosted in the pyroclastic cover, $P$ is the total cumulative precipitation, $ET_o$ is the reference evapotranspiration modulated by the crop coefficient $k_c$, $Q_{run-off}$ denotes the surface discharge volume, and $I_{deep}$ is the deep vertical drainage limited by the underlying low-conductivity basal aquiclude of the Monte Mario Formation ($K < 10^{-9}$ m/s). 

In a highly anthropized canvas where intensive postwar urbanization has led to a critical soil sealing index ($I_{imp} > 70\%$), the effective surface run-off coefficient is heavily altered. The composite run-off coefficient ($C_{comp}$) for the district can be structured as a weighted average proportional to the spatial partitioning of the territory:

\begin{equation}
	C_{comp} = \frac{\sum_{i=1}^{n} C_i \cdot A_i}{\sum_{i=1}^{n} A_i}
\end{equation}

where $C_i$ and $A_i$ represent the specific run-off coefficient and surface area of the $i$-th land-use sector, respectively. Given that built-up infrastructures exhibit extreme values ($C \approx 0.85 \div 0.95$), the macroclimatic trend toward short-duration high-intensity convective rainfall events triggers a rapid, hyper-concentrated surface saturation.

Consequently, non-urbanized sectors such as the ``Pratone di Torre Spaccata'' experience a two-fold environmental stress. Symmetrically acting as a localized natural retention basin, this unbuilt paleovolcanic flatland is forced to absorb the excess $Q_{run-off}$ generated by the surrounding paved canopy layer, while serving as the primary remaining recharge node sustaining the hydraulic head of the unconfined phreatic aquifer.

The thermal inversion in Rome elicits contrasting responses among the population. It is highly appreciated during the peak of summer, when it provides significantly lower and more refreshing temperatures in the early morning hours. Conversely, during winter, it is perceived as a disadvantage due to the persistently low temperatures and the frequent occurrence of concurrent fog and frost events.

Given the specific geological structure of the substrate and the local territory, combined with the significant amount of rainfall recorded over the analyzed period (as detailed in Section 3.4 and Table 4) and the mathematical constraints of urban run-off partitioning, closer attention to the continuous monitoring of these hydro-climatic phenomena is strongly recommended. Shifting from stationary engineering baselines to dynamic, climate-adaptive models is essential to mitigate future urban flash-flood hazards and safeguard localized groundwater reserves.

\section{Acknowledgments}
\normalsize We would like to thank Dr.ssa Fiorella Cerrone for her helpful suggestions and support to make the paper more complete. The constructive comments are highly appreciated. 
\markboth{Data Appendix}{Data Appendix}
\end{multicols}
\newpage
\begin{figure}
\begin{center}
\includegraphics[width=0.8\textwidth]{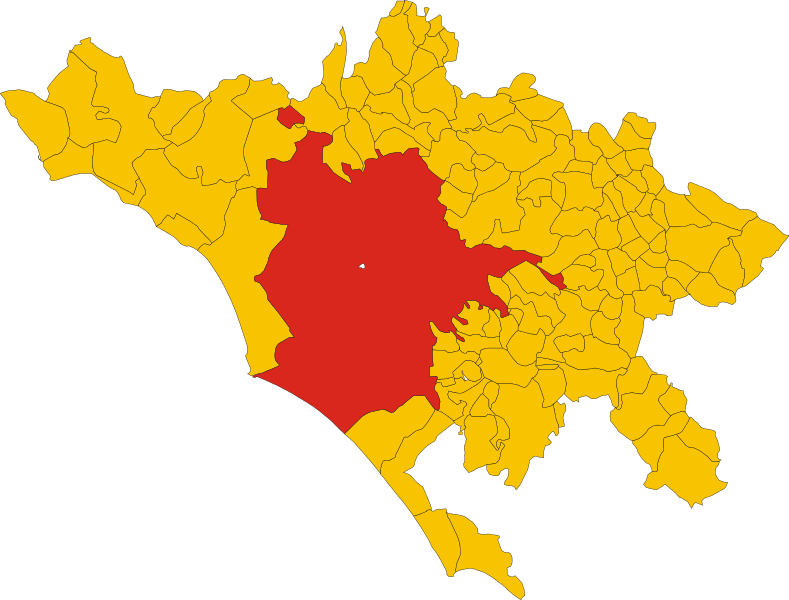}
\caption{Identification of Rome in Lazio Region}
\end{center}
\end{figure}
\clearpage
\centering
\begin{table}[htbp]
	\centering
	\caption{Diurnal Thermal Range Range ($\Delta T = T_{max} - T_{min}$) and Historical Monthly Averages (1990--2014).}
	\label{tab:diurnal_range_clean}
	\tiny
	\begin{tabular}{lccccccccccccc}
		\toprule
		\textbf{Year} & \textbf{Jan} & \textbf{Feb} & \textbf{Mar} & \textbf{Apr} & \textbf{May} & \textbf{Jun} & \textbf{Jul} & \textbf{Aug} & \textbf{Sep} & \textbf{Oct} & \textbf{Nov} & \textbf{Dec} & \textbf{Annual} \\
		\midrule
		2014  & 7 & 8  & 11 & 9  & 10 & 11 & 9  & 9  & 9  & 9  & 7 & 7 & \textbf{8.96} \\
		2013  & 8 & 9  & 8  & 10 & 8  & 11 & 12 & 12 & 9  & 9  & 8 & 8 & \textbf{9.30} \\
		2012  & 10 & 8 & 12 & 9  & 11 & 12 & 12 & 12 & 10 & 9  & 9 & 8 & \textbf{10.10} \\
		2011  & 8 & 11 & 9  & 11 & 12 & 10 & 10 & 12 & 13 & 10 & 10 & 9 & \textbf{10.33} \\
		2010  & 7 & 7  & 9  & 10 & 10 & 11 & 11 & 11 & 10 & 9  & 8 & 7 & \textbf{9.28} \\
		2009  & 7 & 9  & 9  & 10 & 12 & 11 & 11 & 12 & 10 & 10 & 9 & 7 & \textbf{9.76} \\
		2008  & 9 & 11 & 10 & 10 & 12 & 10 & 12 & 13 & 11 & 10 & 8 & 7 & \textbf{10.11} \\
		2007  & 9 & 9  & 10 & 12 & 11 & 12 & 14 & 12 & 11 & 11 & 9 & 9 & \textbf{10.65} \\
		2006  & 8 & 8  & 8  & 10 & 11 & 12 & 12 & 11 & 11 & 11 & 10 & 8 & \textbf{10.09} \\
		2005  & 8 & 8  & 9  & 9  & 10 & 11 & 11 & 10 & 9  & 8  & 8 & 7 & \textbf{9.10} \\
		2004  & 8 & 8  & 8  & 7  & 9  & 10 & 12 & 10 & 10 & 8  & 7 & 7 & \textbf{8.62} \\
		2003  & 8 & 11 & 12 & 11 & 12 & 12 & 11 & 13 & 11 & 9  & 8 & 8 & \textbf{10.34} \\
		2002  & 9 & 9  & 11 & 10 & 10 & 11 & 10 & 9  & 10 & 9  & 7 & 7 & \textbf{9.34} \\
		2001  & 7 & 9  & 9  & 10 & 11 & 12 & 12 & 12 & 11 & 11 & 9 & 8 & \textbf{10.10} \\
		2000  & 9 & 10 & 8  & 9  & 10 & 12 & 12 & 13 & 11 & 9  & 7 & 8 & \textbf{9.96} \\
		1999  & 10 & 11 & 10 & 9  & 10 & 11 & 11 & 11 & 10 & 9  & 8 & 8 & \textbf{9.83} \\
		1998  & 8 & 10 & 10 & 8  & 10 & 11 & 11 & 11 & 10 & 8  & 7 & 8 & \textbf{9.48} \\
		1997  & 8 & 9  & 11 & 10 & 11 & 10 & 11 & 11 & 11 & 8  & 7 & 7 & \textbf{9.40} \\
		1996  & 7 & 8  & 8  & 9  & 9  & 10 & 11 & 10 & 10 & 9  & 7 & 7 & \textbf{8.84} \\
		1995  & 8 & 9  & 8  & 9  & 9  & 10 & 11 & 10 & 8  & 12 & 9 & 6 & \textbf{9.01} \\
		1994  & 7 & 9  & 10 & 8  & 9  & 8  & 10 & 11 & 9  & 8  & 8 & 6 & \textbf{8.71} \\
		1993  & 6 & 6  & 9  & 13 & 18 & 23 & 24 & 26 & 21 & 17 & 10 & 9 & \textbf{15.23} \\
		1992  & 9 & 11 & 9  & 9  & 12 & 10 & 10 & 11 & 11 & 7  & 9 & 8 & \textbf{9.60} \\
		1991  & 9 & 9  & 9  & 10 & 10 & 10 & 12 & 12 & 10 & 10 & 8 & 10 & \textbf{9.86} \\
		1990  & 10 & 11 & 10 & 9  & 11 & 11 & 11 & 11 & 10 & 8  & 8 & 7 & \textbf{9.73} \\
		\midrule
		\textbf{Average} & \textbf{8.13} & \textbf{9.03} & \textbf{9.42} & \textbf{9.58} & \textbf{10.70} & \textbf{11.30} & \textbf{11.77} & \textbf{11.84} & \textbf{10.73} & \textbf{9.50} & \textbf{8.25} & \textbf{7.67} & \textbf{9.83} \\
		\bottomrule
	\end{tabular}
\end{table}

\begin{table}[htbp]
	\centering
	\caption{Monthly and Annual Cumulative Precipitation Regimes in Torrespaccata District (1990--2014).}
	\label{tab:rain_regimes_clean}
	\tiny
	\begin{tabular}{lcccccccccccccc}
		\toprule
		\textbf{Year} & \textbf{Jan} & \textbf{Feb} & \textbf{Mar} & \textbf{Apr} & \textbf{May} & \textbf{Jun} & \textbf{Jul} & \textbf{Aug} & \textbf{Sep} & \textbf{Oct} & \textbf{Nov} & \textbf{Dec} & \textbf{Total (mm)} & \textbf{\% Year} \\
		\midrule
		2014  & 195.00 & 232.70 & 71.00  & 210.00 & 65.00  & 200.00 & 140.00 & 40.00  & 130.00 & 80.00  & 235.00 & 50.00  & \textbf{1648.70} & 5.06\% \\
		2013  & 120.00 & 130.00 & 150.00 & 75.00  & 135.00 & 50.00  & 155.00 & 70.00  & 110.00 & 100.00 & 190.00 & 70.00  & \textbf{1355.00} & 4.15\% \\
		2012  & 50.00  & 105.00 & 40.00  & 145.00 & 70.00  & 15.00  & 20.00  & 15.00  & 105.00 & 115.00 & 130.00 & 145.00 & \textbf{955.00}  & 2.93\% \\
		2011  & 110.00 & 65.00  & 175.00 & 95.00  & 150.00 & 50.00  & 80.00  & 15.00  & 50.00  & 60.00  & 60.00  & 100.00 & \textbf{1010.00} & 3.10\% \\
		2010  & 135.00 & 205.00 & 70.00  & 100.00 & 195.00 & 80.00  & 35.00  & 35.00  & 70.00  & 95.00  & 515.00 & 220.00 & \textbf{1755.00} & 5.38\% \\
		2009  & 113.00 & 88.80  & 73.00  & 130.00 & 55.00  & 155.00 & 5.00   & 65.00  & 110.00 & 60.00  & 55.00  & 110.00 & \textbf{1019.80} & 3.13\% \\
		2008  & 154.80 & 46.50  & 208.50 & 59.20  & 162.30 & 13.60  & 20.50  & 38.10  & 50.50  & 149.00 & 335.70 & 21.20  & \textbf{1259.90} & 3.86\% \\
		2007  & 62.30  & 151.60 & 112.00 & 64.70  & 67.10  & 20.00  & 0.50   & 27.00  & 68.30  & 89.70  & 37.60  & 8.70   & \textbf{709.50}  & 2.18\% \\
		2006  & 38.90  & 85.60  & 75.10  & 91.40  & 36.50  & 56.00  & 107.20 & 68.70  & 114.20 & 72.80  & 100.10 & 0.00   & \textbf{846.50}  & 2.60\% \\
		2005  & 73.90  & 84.00  & 60.20  & 77.70  & 62.00  & 35.90  & 38.00  & 105.30 & 181.30 & 77.30  & 51.40  & 42.40  & \textbf{889.40}  & 2.73\% \\
		2004  & 80.90  & 140.90 & 98.50  & 229.00 & 69.60  & 18.40  & 71.90  & 10.00  & 125.80 & 170.60 & 152.70 & 99.10  & \textbf{1267.40} & 3.89\% \\
		2003  & 142.60 & 46.40  & 62.70  & 32.90  & 53.90  & 0.80   & 45.00  & 72.60  & 92.30  & 125.20 & 110.50 & 48.60  & \textbf{833.50}  & 2.56\% \\
		2002  & 29.50  & 84.60  & 26.40  & 68.40  & 83.80  & 74.70  & 83.70  & 179.80 & 95.50  & 106.10 & 145.50 & 142.50 & \textbf{1120.50} & 3.44\% \\
		2001  & 187.50 & 51.50  & 55.60  & 115.20 & 91.30  & 20.10  & 6.60   & 10.00  & 61.10  & 12.80  & 134.10 & 35.50  & \textbf{781.30}  & 2.40\% \\
		2000  & 85.00  & 45.00  & 70.00  & 105.00 & 30.00  & 26.60  & 70.80  & 51.60  & 52.90  & 134.40 & 194.80 & 70.40  & \textbf{936.50}  & 2.87\% \\
		1999  & 55.00  & 85.00  & 125.000 & 105.00 & 60.00  & 60.00  & 95.00  & 35.00  & 150.00 & 175.00 & 205.00 & 180.00 & \textbf{1330.00} & 4.08\% \\
		1998  & 60.00  & 25.00  & 25.00  & 105.00 & 125.00 & 15.00  & 5.00   & 45.00  & 95.00  & 130.00 & 45.00  & 65.00  & \textbf{740.00}  & 2.27\% \\
		1997  & 60.00  & 40.00  & 55.00  & 90.00  & 75.00  & 80.00  & 96.90  & 100.00 & 15.00  & 135.00 & 200.00 & 120.00 & \textbf{1066.90} & 3.27\% \\
		1996  & 105.00 & 120.00 & 50.00  & 85.00  & 85.00  & 95.00  & 65.00  & 75.00  & 185.00 & 165.00 & 360.00 & 160.00 & \textbf{1550.00} & 4.75\% \\
		1995  & 105.00 & 175.00 & 315.00 & 205.00 & 145.00 & 65.00  & 30.00  & 255.00 & 155.00 & 10.00  & 60.00  & 405.00 & \textbf{1925.00} & 5.90\% \\
		1994  & 110.00 & 110.00 & 0.00   & 140.10 & 30.00  & 90.00  & 70.00  & 5.00   & 70.00  & 155.00 & 65.00  & 120.00 & \textbf{965.10}  & 2.96\% \\
		1993  & 40.00  & 70.00  & 90.00  & 225.00 & 240.00 & 80.00  & 12.20  & 17.40  & 95.00  & 115.00 & 300.00 & 180.00 & \textbf{1464.60} & 4.49\% \\
		1992  & 65.00  & 45.00  & 150.00 & 205.00 & 95.00  & 450.00 & 70.00  & 65.00  & 55.00  & 970.00 & 140.00 & 270.00 & \textbf{2580.00} & 7.91\% \\
		1991  & 20.00  & 330.00 & 105.00 & 280.00 & 245.00 & 60.00  & 35.00  & 60.00  & 170.00 & 185.00 & 450.00 & 50.00  & \textbf{1990.00} & 6.10\% \\
		1990  & 30.00  & 65.00  & 285.00 & 355.00 & 300.00 & 105.00 & 75.00  & 75.00  & 235.00 & 480.00 & 195.00 & 415.00 & \textbf{2615.00} & 8.02\% \\
		\midrule
		\textbf{Average} & 107.25 & 89.14 & 105.10 & 101.92 & 135.74 & 109.06 & 76.64 & 57.33 & 61.42 & 105.68 & 158.72 & 178.70 & \textbf{32614.60} & \\
		\midrule
		\multicolumn{12}{r}{\textbf{Annual Average Precipitation (1990--2014):}} & \multicolumn{2}{l}{\textbf{1304.58 mm}} \\
		\bottomrule
	\end{tabular}
\end{table}

\begin{table}[htbp]
	\centering
	\caption{Wind in Torrespaccata District (1990-2014).}
	\label{tab:rain_regimes_clean}
	\tiny
\begin{tabular}{|c|c|c|c|c|c|c|c|c|c|c|c|c|c|c|c|} 
\hline
\multicolumn{ 14}{|c}{} &            &            \\
\hline
\multicolumn{ 1}{|c|}{Day} &        Jan &        Feb &        Mar &        Apr &        May &        Jun &        Jul &        Ago &        Sep &        Oct &        Nov &        Dec &\bf Average &  \bf Min &  \bf Max  \\
\hline
\multicolumn{ 1}{|c|}{} &            &            &            &            &            &            &            &            &            &            &            &            &            &            &            \\
1     & 23.68 & 21.54 & 19.95 & 21.33 & 15.12 & 18.66 & 16.93 & 18.78 & 21.02 & 14.69 & 18.85 & 22.68 & 19.44 & 14.69 & 23.68 \\
2     & 18.82 & 19.62 & 21.64 & 18.73 & 16.57 & 16.90 & 17.82 & 18.75 & 19.21 & 15.70 & 18.89 & 19.30 & 18.50 & 15.70 & 21.64 \\
3     & 22.05 & 17.88 & 21.00 & 18.08 & 19.80 & 17.23 & 18.13 & 18.78 & 18.15 & 15.77 & 17.57 & 20.89 & 18.78 & 15.77 & 22.05 \\
4     & 22.18 & 19.70 & 22.20 & 20.48 & 19.90 & 17.39 & 17.98 & 19.15 & 19.54 & 18.13 & 19.72 & 17.33 & 19.47 & 17.33 & 22.20 \\
5     & 19.09 & 20.06 & 23.24 & 21.88 & 15.89 & 17.70 & 17.91 & 20.00 & 19.33 & 18.13 & 18.59 & 17.13 & 19.08 & 15.89 & 23.24 \\
6     & 17.73 & 20.44 & 20.05 & 18.38 & 18.02 & 17.25 & 17.76 & 20.21 & 17.19 & 16.42 & 17.30 & 20.83 & 18.46 & 16.42 & 20.83 \\
7     & 15.66 & 18.64 & 21.73 & 20.38 & 17.14 & 17.39 & 19.67 & 17.55 & 18.85 & 19.17 & 21.04 & 21.76 & 19.08 & 15.66 & 21.76 \\
8     & 16.16 & 21.18 & 23.14 & 16.87 & 15.38 & 18.00 & 20.15 & 19.18 & 18.73 & 17.27 & 19.35 & 20.74 & 18.85 & 15.38 & 23.14 \\
9     & 14.48 & 20.68 & 20.81 & 19.89 & 16.36 & 18.43 & 17.54 & 17.73 & 19.77 & 16.98 & 17.72 & 22.89 & 18.61 & 14.48 & 22.89 \\
10    & 14.36 & 20.10 & 21.40 & 21.09 & 15.20 & 17.60 & 17.61 & 20.29 & 17.93 & 16.81 & 19.22 & 22.41 & 18.67 & 14.36 & 22.41 \\
11    & 17.61 & 20.38 & 20.69 & 20.93 & 13.98 & 17.71 & 19.09 & 19.57 & 18.00 & 16.52 & 20.79 & 17.89 & 18.60 & 13.98 & 20.93 \\
12    & 19.24 & 18.70 & 17.52 & 20.80 & 16.68 & 17.25 & 17.91 & 20.42 & 21.09 & 17.17 & 18.48 & 17.70 & 18.58 & 16.68 & 21.09 \\
13    & 17.73 & 18.26 & 18.91 & 19.91 & 16.18 & 16.13 & 17.63 & 17.71 & 18.52 & 17.75 & 18.52 & 19.52 & 18.06 & 16.13 & 19.91 \\
14    & 18.83 & 17.78 & 16.25 & 18.98 & 17.28 & 16.45 & 17.57 & 19.11 & 20.48 & 18.65 & 20.77 & 21.34 & 18.62 & 16.25 & 21.34 \\
15    & 20.85 & 21.73 & 17.20 & 18.46 & 17.36 & 17.70 & 18.91 & 18.17 & 17.89 & 19.08 & 21.45 & 20.98 & 19.15 & 17.20 & 21.73 \\
16    & 18.87 & 23.35 & 19.11 & 20.52 & 16.90 & 16.98 & 18.77 & 17.96 & 18.17 & 18.13 & 25.20 & 19.15 & 19.43 & 16.90 & 25.20 \\
17    & 18.89 & 21.08 & 19.43 & 19.07 & 17.64 & 17.26 & 18.98 & 18.00 & 17.15 & 18.83 & 21.32 & 19.52 & 18.93 & 17.15 & 21.32 \\
18    & 17.59 & 21.40 & 19.75 & 19.35 & 19.30 & 16.82 & 17.87 & 16.71 & 16.95 & 17.26 & 19.48 & 17.57 & 18.34 & 16.71 & 21.40 \\
19    & 18.70 & 20.34 & 20.78 & 18.37 & 20.40 & 17.64 & 18.43 & 17.40 & 16.11 & 18.57 & 18.84 & 17.59 & 18.60 & 16.11 & 20.78 \\
20    & 17.91 & 18.54 & 18.74 & 16.72 & 18.83 & 18.61 & 18.33 & 17.37 & 18.77 & 18.15 & 20.77 & 19.39 & 18.51 & 16.72 & 20.77 \\
21    & 20.16 & 17.06 & 19.57 & 19.57 & 17.52 & 16.70 & 19.13 & 18.72 & 19.13 & 21.54 & 20.09 & 19.26 & 19.04 & 16.70 & 21.54 \\
22    & 20.93 & 18.60 & 19.89 & 17.02 & 16.90 & 19.57 & 18.72 & 17.52 & 15.85 & 17.46 & 17.95 & 19.89 & 18.36 & 15.85 & 20.93 \\
23    & 21.46 & 21.94 & 20.09 & 16.84 & 18.05 & 17.67 & 18.74 & 18.04 & 18.50 & 19.09 & 17.84 & 23.20 & 19.29 & 16.84 & 23.20 \\
24    & 24.02 & 19.58 & 19.46 & 16.43 & 18.23 & 16.83 & 21.80 & 16.80 & 18.08 & 18.13 & 17.29 & 22.50 & 19.10 & 16.43 & 24.02 \\
25    & 24.26 & 18.88 & 21.11 & 17.83 & 17.88 & 18.09 & 18.38 & 17.61 & 19.04 & 17.57 & 16.55 & 22.14 & 19.11 & 16.55 & 24.26 \\
26    & 22.17 & 19.02 & 20.94 & 17.65 & 16.90 & 17.00 & 17.93 & 16.41 & 21.19 & 17.83 & 18.07 & 23.04 & 19.01 & 16.41 & 23.04 \\
27    & 19.04 & 18.75 & 22.42 & 16.98 & 17.08 & 17.42 & 17.87 & 18.50 & 19.08 & 16.67 & 18.05 & 19.78 & 18.47 & 16.67 & 22.42 \\
28    & 17.52 & 22.00 & 20.27 & 14.48 & 16.83 & 18.43 & 18.13 & 19.55 & 17.69 & 17.46 & 18.57 & 21.00 & 18.49 & 14.48 & 22.00 \\
29    & 21.02 & 16.40 & 19.69 & 16.21 & 19.88 & 18.09 & 18.09 & 19.76 & 17.65 & 18.23 & 18.18 & 19.83 & 18.58 & 16.21 & 21.02 \\
30    & 21.48 &       & 20.15 & 15.88 & 17.18 & 17.65 & 18.39 & 20.24 & 17.67 & 18.09 & 19.07 & 18.70 & 18.59 & 15.88 & 21.48 \\
31    & 21.17 &       & 22.44 &       & 16.43 &       & 19.64 & 20.59 &       & 20.50 &       & 21.82 & 20.37 & 16.43 & 22.44 \\
\hline
\bf{Average} & \bf{19.47} & \bf{19.78} & \bf{20.31} & \bf{18.64} & \bf{17.32} & \bf{17.55} & \bf{18.45} & \bf{18.60} & \bf{18.56} & \bf{17.80} & \bf{19.18} & \bf{20.25} & \bf{Average} & \bf{16.06} & \bf{22.09} \\
\bf{Min} & \bf{14.36} & \bf{16.40} & \bf{16.25} & \bf{14.48} & \bf{13.98} & \bf{16.13} & \bf{16.93} & \bf{16.41} & \bf{15.85} & \bf{14.69} & \bf{16.55} & \bf{17.13} & \bf{15.76} &       &  \\
\bf{Max} & \bf{24.26} & \bf{23.35} & \bf{23.24} & \bf{21.88} & \bf{20.40} & \bf{19.57} & \bf{21.80} & \bf{20.59} & \bf{21.19} & \bf{21.54} & \bf{25.20} & \bf{23.20} & \bf{22.19} &       &  \\

\hline
\end{tabular}  
\end{table}
\centering
\begin{table}[htbp]
	\centering
	\caption{Seasonal Trend of Mean Temperatures ($^{\circ}\text{C}$) during Core Winter (Jan--Feb--Mar) and Summer (Jun--Jul--Aug) Quarters (1990--2014).}
	\label{tab:seasonal_temperature_trends}
	\small
	\begin{tabular}{lcccccc}
		\toprule
		\textbf{Year} & \textbf{Jan} & \textbf{Feb} & \textbf{Mar} & \textbf{Jun} & \textbf{Jul} & \textbf{Aug} \\
		\midrule
		2014 & 9.27 & 11.00 & 10.84 & 21.83 & 23.29 & 24.13 \\
		2013 & 8.13 & 6.75  & 10.55 & 21.30 & 25.35 & 25.81 \\
		2012 & 7.53 & 2.98  & 13.15 & 23.78 & 26.39 & 27.47 \\
		2011 & 8.19 & 8.45  & 10.45 & 22.60 & 24.06 & 26.23 \\
		2010 & 6.47 & 8.13  & 9.92  & 21.57 & 26.71 & 25.05 \\
		2009 & 8.08 & 7.75  & 11.27 & 22.45 & 25.84 & 27.23 \\
		2008 & 8.59 & 8.49  & 10.72 & 22.48 & 25.29 & 25.66 \\
		2007 & 9.62 & 9.94  & 10.90 & 22.75 & 25.17 & 25.14 \\
		2006 & 5.85 & 7.93  & 9.87  & 22.10 & 26.93 & 24.40 \\
		2005 & 5.47 & 5.76  & 9.87  & 23.06 & 25.38 & 23.51 \\
		2004 & 5.45 & 8.15  & 9.41  & 21.94 & 24.83 & 24.63 \\
		2003 & 8.59 & 5.23  & 10.68 & 26.10 & 26.75 & 27.94 \\
		2002 & 6.00 & 9.80  & 12.90 & 23.40 & 25.30 & 23.60 \\
		2001 & 9.71 & 8.99  & 13.32 & 21.67 & 24.89 & 25.98 \\
		2000 & 5.20 & 7.91  & 10.48 & 22.29 & 23.32 & 25.19 \\
		1999 & 7.58 & 6.51  & 10.56 & 21.83 & 23.60 & 24.90 \\
		1998 & 8.13 & 9.45  & 9.67  & 22.70 & 24.79 & 26.06 \\
		1997 & 8.76 & 8.18  & 10.61 & 22.50 & 23.77 & 24.24 \\
		1996 & 9.81 & 7.37  & 9.41  & 21.93 & 23.58 & 23.66 \\
		1995 & 5.97 & 9.30  & 9.53  & 19.75 & 25.58 & 23.73 \\
		1994 & 8.76 & 8.24  & 11.85 & 21.17 & 26.32 & 27.20 \\
		1993 & 6.60 & 6.89  & 9.35  & 22.33 & 23.58 & 26.15 \\
		1992 & 7.06 & 7.90  & 10.58 & 20.25 & 23.55 & 26.10 \\
		1991 & 7.10 & 7.81  & 12.60 & 20.45 & 25.08 & 25.66 \\
		1990 & 6.98 & 10.20 & 11.15 & 21.78 & 24.84 & 24.27 \\
		\midrule
		\textbf{Average 1990--2014} & \textbf{7.56} & \textbf{7.96} & \textbf{10.79} & \textbf{22.16} & \textbf{24.97} & \textbf{25.36} \\
		\midrule
		\multicolumn{7}{l}{Absolute Minimum Historical Record: $2.98 \ \text{}^{\circ}\text{C}$ (February 2012)} \\
		\multicolumn{7}{l}{Absolute Maximum Historical Record: $27.94 \ \text{}^{\circ}\text{C}$ (August 2003)} \\
		\bottomrule
	\end{tabular}
\end{table}
\newpage \clearpage
\begin{table}[htbp]
	\centering
	\caption{Chronological Distribution of Snow Days in the Torrespaccata District (1990--2014).}
	\label{tab:snow_days_clean}
	\small
	\begin{tabular}{lccccccccccccr}
		\toprule
		\textbf{Year} & \textbf{Jan} & \textbf{Feb} & \textbf{Mar} & \textbf{Apr} & \textbf{May} & \textbf{Jun} & \textbf{Jul} & \textbf{Aug} & \textbf{Sep} & \textbf{Oct} & \textbf{Nov} & \textbf{Dec} & \textbf{Total} \\
		\midrule
		2014 & 0 & 0 & 0 & 0 & 0 & 0 & 0 & 0 & 0 & 0 & 0 & 0 & \textbf{0} \\
		2013 & 0 & 0 & 0 & 0 & 0 & 0 & 0 & 0 & 0 & 0 & 0 & 0 & \textbf{0} \\
		2012 & 0 & 1 & 0 & 0 & 0 & 0 & 0 & 0 & 0 & 0 & 0 & 0 & \textbf{1} \\
		2011 & 0 & 0 & 0 & 0 & 0 & 0 & 0 & 0 & 0 & 0 & 0 & 0 & \textbf{0} \\
		2010 & 0 & 0 & 0 & 0 & 0 & 0 & 0 & 0 & 0 & 0 & 0 & 0 & \textbf{0} \\
		2009 & 0 & 0 & 0 & 0 & 0 & 0 & 0 & 0 & 0 & 0 & 0 & 0 & \textbf{0} \\
		2008 & 0 & 0 & 0 & 0 & 0 & 0 & 0 & 0 & 0 & 0 & 0 & 0 & \textbf{0} \\
		2007 & 0 & 0 & 0 & 0 & 0 & 0 & 0 & 0 & 0 & 0 & 0 & 0 & \textbf{0} \\
		2006 & 0 & 0 & 0 & 0 & 0 & 0 & 0 & 0 & 0 & 0 & 0 & 0 & \textbf{0} \\
		2005 & 0 & 0 & 0 & 0 & 0 & 0 & 0 & 0 & 0 & 0 & 0 & 0 & \textbf{0} \\
		2004 & 0 & 0 & 0 & 0 & 0 & 0 & 0 & 0 & 0 & 0 & 0 & 0 & \textbf{0} \\
		2003 & 0 & 0 & 0 & 0 & 0 & 0 & 0 & 0 & 0 & 0 & 0 & 0 & \textbf{0} \\
		2002 & 0 & 0 & 0 & 0 & 0 & 0 & 0 & 0 & 0 & 0 & 0 & 0 & \textbf{0} \\
		2001 & 0 & 0 & 0 & 0 & 0 & 0 & 0 & 0 & 0 & 0 & 0 & 0 & \textbf{0} \\
		2000 & 0 & 0 & 0 & 0 & 0 & 0 & 0 & 0 & 0 & 0 & 0 & 0 & \textbf{0} \\
		1999 & 0 & 0 & 0 & 0 & 0 & 0 & 0 & 0 & 0 & 0 & 0 & 0 & \textbf{0} \\
		1998 & 0 & 0 & 0 & 0 & 0 & 0 & 0 & 0 & 0 & 0 & 0 & 0 & \textbf{0} \\
		1997 & 0 & 0 & 0 & 0 & 0 & 0 & 0 & 0 & 0 & 0 & 0 & 0 & \textbf{0} \\
		1996 & 0 & 0 & 0 & 0 & 0 & 0 & 0 & 0 & 0 & 0 & 0 & 1 & \textbf{1} \\
		1995 & 0 & 0 & 0 & 0 & 0 & 0 & 0 & 0 & 0 & 0 & 0 & 0 & \textbf{0} \\
		1994 & 1 & 2 & 0 & 0 & 0 & 0 & 0 & 0 & 0 & 0 & 0 & 0 & \textbf{3} \\
		1993 & 1 & 0 & 1 & 0 & 0 & 0 & 0 & 0 & 0 & 0 & 0 & 0 & \textbf{2} \\
		1992 & 0 & 0 & 0 & 0 & 0 & 0 & 0 & 0 & 0 & 0 & 0 & 0 & \textbf{0} \\
		1991 & 0 & 0 & 0 & 0 & 0 & 0 & 0 & 0 & 0 & 0 & 0 & 2 & \textbf{2} \\
		1990 & 0 & 1 & 1 & 0 & 0 & 0 & 0 & 0 & 0 & 0 & 0 & 0 & \textbf{2} \\
		\midrule
		\textbf{Total} & \textbf{2} & \textbf{4} & \textbf{2} & \textbf{0} & \textbf{0} & \textbf{0} & \textbf{0} & \textbf{0} & \textbf{0} & \textbf{0} & \textbf{0} & \textbf{3} & \textbf{11} \\
		\bottomrule
	\end{tabular}
\end{table}
\begin{figure}
	\begin{center}
		\includegraphics[width=1.3\textwidth{}]{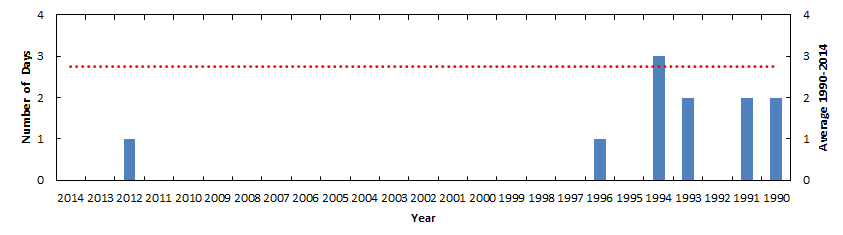}
		\caption{Snow (solid line) and Average Snow Variations 1990-2014 (dotted line) }
	\end{center}
\end{figure}
\begin{table}[] 
	\centering
	\caption{Average wind speed in the period 1990-2014 for Rome - RM.}
	\tiny \rotatebox{90}{
	\begin{tabular}{|lllllllllllllllllllllllllll|} \hline
		\multicolumn{1}{|c|}{Year} & \multicolumn{2}{|c|}{Jan} & \multicolumn{2}{|c|}{Feb} & \multicolumn{2}{|c|}{Mar} & \multicolumn{2}{c}{Apr} & \multicolumn{2}{|c|}{May} & \multicolumn{2}{|c|}{Jun} & \multicolumn{2}{|c|}{Jul} & \multicolumn{2}{|c|}{Aug} & \multicolumn{2}{|c|}{Sep} & \multicolumn{2}{|c|}{Oct} & \multicolumn{2}{|c|}{Nov} & \multicolumn{2}{|c|}{Dec} &        &        \\ \hline
		\multicolumn{1}{|c|}{}                               & Min        & Max        & Min        & Max        & Min        & Max        & Min        & Max        & Min        & Max        & Min        & Max        & Min        & Max        & Min        & Max        & Min        & Max        & Min        & Max        & Min        & Max        & Min        & Max        &        &        \\ \hline
2014                                               & 7.00       & 46.00      & 8.00       & 50.00      & 8.00       & 43.00      & 6.00       & 37.00      & 8.00       & 43.00      & 11.00      & 37.00      & 8.00       & 35.00      & 8.00       & 37.00      & 8.00       & 43.00      & 7.00       & 41.00      & 7.00       & 43.00      & 5.00       & 48.00      &        &        \\
2013                                               & 5.00       & 41.00      & 8.00       & 44.00      & 9.00       & 52.00      & 8.00       & 32.00      & 7.00       & 41.00      & 9.00       & 30.00      & 10.00      & 65.00      & 10.00      & 70.00      & 8.00       & 41.00      & 8.00       & 30.00      & 6.00       & 57.00      & 8.00       & 59.00      &        &        \\
2012                                               & -          & -          & -          & -          & -          & -          & -          & -          & -          & -          & 4.00       & 22.00      & -          & -          & 11.00      & 32.00      & 10.00      & 43.00      & 8.00       & 41.00      & 7.00       & 44.00      & 8.00       & 37.00      &        &        \\
2011                                               & 3.00       & 31.00      & 5.00       & 36.00      & 5.00       & 33.00      & 4.00       & 35.00      & 4.00       & 25.00      & 7.00       & 22.00      & 7.00       & 37.00      & 10.00      & 41.00      & 8.00       & 39.00      & 8.00       & 46.00      & 9.00       & 9.00       & 9.00       & 9.00       &        &        \\
2010                                               & 5.00       & 24.00      & 3.00       & 37.00      & 4.00       & 24.00      & 4.00       & 24.00      & 3.00       & 22.00      & 10.00      & 30.00      & 5.00       & 30.00      & 7.00       & 47.00      & 7.00       & 29.00      & 5.00       & 30.00      & 4.00       & 38.00      & 6.00       & 31.00      &        &        \\
2009                                               & -          & -          & -          & -          & 8.00       & 33.00      & 5.00       & 33.00      & 2.00       & 21.00      & 5.00       & 21.00      & 5.00       & 18.00      & 6.00       & 20.00      & 4.00       & 21.00      & 4.00       & 33.00      & 4.00       & 31.00      & 5.00       & 31.00      &        &        \\
2008                                               & 5.00       & 55.00      & 6.00       & 55.00      & 8.00       & 42.00      & 15.00      & 47.00      & -          & -          & 5.00       & 19.00      & 14.00      & 14.00      & 10.00      & 23.00      & 6.00       & 6.00       & -          & -          & -          & -          & -          & -          &        &        \\
2007                                               & 7.00       & 63.00      & 8.00       & 39.00      & 6.00       & 52.00      & 8.00       & 34.00      & 11.00      & 47.00      & 9.00       & 50.00      & 14.00      & 48.00      & 10.00      & 47.00      & 12.00      & 48.00      & 9.00       & 55.00      & 8.00       & 48.00      & 8.00       & 50.00      &        &        \\
2006                                               & 10.00      & 63.00      & 8.00       & 68.00      & 10.00      & 65.00      & 6.00       & 47.00      & 9.00       & 39.00      & 11.00      & 44.00      & 13.00      & 50.00      & 9.00       & 41.00      & 8.00       & 35.00      & 6.00       & 41.00      & 6.00       & 55.00      & 6.00       & 39.00      &        &        \\
2005                                               & 1.00       & 59.00      & 9.00       & 61.00      & 7.00       & 41.00      & 11.00      & 47.00      & 8.00       & 41.00      & 10.00      & 48.00      & 15.00      & 44.00      & 11.00      & 42.00      & 8.00       & 44.00      & 7.00       & 35.00      & 7.00       & 42.00      & 4.00       & 54.00      &        &        \\
2004                                               & 5.00       & 55.00      & 4.00       & 47.00      & 5.00       & 39.00      & 7.00       & 42.00      & 5.00       & 41.00      & 5.00       & 34.00      & 7.00       & 35.00      & 7.00       & 28.00      & 7.00       & 48.00      & 2.00       & 34.00      & 4.00       & 67.00      & 2.00       & 47.00      &        &        \\
2003                                               & 5.00       & 59.00      & 6.00       & 47.00      & 4.00       & 54.00      & 5.00       & 39.00      & 10.00      & 72.00      & 12.00      & 41.00      & 13.00      & 50.00      & 11.00      & 55.00      & 10.00      & 44.00      & 8.00       & 55.00      & 6.00       & 47.00      & 7.00       & 68.00      &        &        \\
2002                                               & 4.00       & 72.00      & 8.00       & 76.00      & 8.00       & 79.00      & 6.00       & 54.00      & 8.00       & 54.00      & 8.00       & 43.00      & 6.00       & 42.00      & 7.00       & 35.00      & 6.00       & 34.00      & 5.00       & 37.00      & 5.00       & 57.00      & 3.00       & 39.00      &        &        \\
2001                                               & 6.00       & 57.00      & 7.00       & 68.00      & 7.00       & 50.73      & 10.00      & 68.00      & 8.00       & 58.00      & 9.00       & 58.00      & 11.00      & 61.00      & 10.00      & 50.00      & 10.00      & 79.00      & 4.00       & 47.00      & 6.00       & 79.00      & 9.00       & 86.00      &        &        \\
2000                                               & 4.00       & 34.00      & 3.00       & 52.00      & 6.00       & 52.00      & 5.00       & 31.00      & 3.00       & 28.00      & 8.00       & 29.00      & 9.00       & 43.00      & 10.00      & 42.00      & 8.00       & 47.00      & 3.00       & 31.00      & 5.00       & 41.00      & 4.00       & 54.00      &        &        \\
1999                                               & 5.00       & 55.00      & 6.00       & 54.00      & 7.00       & 55.00      & 8.00       & 37.00      & 8.00       & 26.00      & 10.00      & 28.00      & 5.00       & 41.00      & 9.00       & 34.00      & 7.00       & 37.00      & 4.00       & 35.00      & 7.00       & 61.00      & 6.00       & 59.00      &        &        \\
1998                                               & 6.00       & 57.00      & 8.00       & 42.00      & 7.00       & 57.00      & 8.00       & 42.00      & 5.00       & 34.00      & 13.00      & 25.00      & 9.00       & 34.00      & 8.00       & 39.00      & 7.00       & 41.00      & 5.00       & 37.00      & 6.00       & 50.00      & 6.00       & 48.00      &        &        \\
1997                                               & 7.00       & 50.00      & 7.00       & 57.00      & 8.00       & 54.00      & 7.00       & 47.00      & 8.00       & 34.00      & 3.00       & 48.00      & 8.00       & 35.00      & 7.00       & 34.00      & 8.00       & 39.00      & 7.00       & 57.00      & 6.00       & 35.00      & 5.00       & 41.00      &        &        \\
1996                                               & 6.00       & 41.00      & 7.00       & 52.00      & 4.00       & 54.00      & 4.00       & 42.00      & 6.00       & 42.00      & 7.00       & 52.00      & 7.00       & 52.00      & 8.00       & 34.00      & 7.00       & 42.00      & 7.00       & 37.00      & 9.00       & 57.00      & 8.00       & 65.00      &        &        \\
1995                                               & 8.00       & 61.00      & 4.00       & 41.00      & 8.00       & 59.00      & 6.00       & 37.00      & 4.00       & 47.00      & 7.00       & 31.00      & 7.00       & 31.00      & 5.00       & 42.00      & 5.00       & 61.00      & 4.00       & 24.00      & 6.00       & 47.00      & 4.00       & 44.00      &        &        \\
1994                                               & 7.00       & 44.00      & 6.00       & 39.00      & 6.00       & 41.00      & 6.00       & 42.00      & 5.00       & 42.00      & 4.00       & 29.00      & 7.00       & 37.00      & 3.00       & 39.00      & 2.00       & 37.00      & 5.00       & 41.00      & 4.00       & 29.00      & 3.00       & 41.00      &        &        \\
1993                                               & 5.00       & 55.00      & 7.00       & 50.00      & 5.00       & 55.00      & 6.00       & 46.00      & 5.00       & 18.00      & 7.00       & 18.00      & 7.00       & 31.00      & 7.00       & 39.00      & 6.00       & 63.00      & 8.00       & 39.00      & 4.00       & 42.00      & 5.00       & 47.00      &        &        \\
1992                                               & 7.00       & 50.00      & 5.00       & 47.00      & 5.00       & 52.00      & 7.00       & 63.00      & 7.00       & 61.00      & 6.00       & 29.00      & 6.00       & 28.00      & 9.00       & 35.00      & 7.00       & 33.00      & 8.00       & 42.00      & 7.00       & 39.00      & 8.00       & 67.00      &        &        \\
1991                                               & 7.00       & 58.00      & 6.00       & 48.00      & 8.00       & 50.00      & 8.00       & 61.00      & 7.00       & 47.00      & 4.00       & 29.00      & 5.00       & 47.00      & 10.00      & 37.00      & 10.00      & 41.00      & 9.00       & 47.00      & 8.00       & 61.00      & 10.00      & 63.00      &        &        \\
1990                                               & 1.00       & 42.00      & 1.00       & 61.00      & 1.00       & 42.00      & 4.00       & 41.00      & 2.00       & 41.00      & 3.00       & 29.00      & 3.00       & 41.00      & 4.00       & 42.00      & 2.00       & 41.00      & 4.00       & 42.00      & 10.00      & 72.00      & 7.00       & 61.00      &        &        \\ \hline
Average 1990-2014                                  & 5.48       & 50.96      & 6.09       & 50.91      & 6.42       & 49.20      & 6.83       & 42.83      & 6.22       & 40.17      & 7.48       & 33.84      & 8.38       & 39.54      & 8.28       & 39.40      & 7.24       & 41.44      & 6.04       & 39.88      & 6.29       & 47.96      & 6.08       & 49.50      &        &        \\
		\hline
		Average Min                                        & 5.48       & 50.96      & 6.09       & 50.91      & 6.42       & 49.20      & 6.83       & 42.83      & 6.22       & 40.17      & 7.48       & 33.84      & 8.38       & 39.54      & 8.28       & 39.40      & 7.24       & 41.44      & 6.04       & 39.88      & 6.29       & 47.96      & 6.08       & 49.50      & 6.08   &        \\
		Average Max                                        & 5.48       & 50.96      & 6.09       & 50.91      & 6.42       & 49.20      & 6.83       & 42.83      & 6.22       & 40.17      & 7.48       & 33.84      & 8.38       & 39.54      & 8.28       & 39.40      & 7.24       & 41.44      & 6.04       & 39.88      & 6.29       & 47.96      & 6.08       & 49.50      & 49.50  &        \\
		\hline
		Average Min                                                & 1.00       &            & 1.00       &            & 1.00       &            & 4.00       &            & 2.00       &            & 3.00       &            & 3.00       &            & 3.00       &            & 2.00       &            & 2.00       &            & 4.00       &            & 2.00       &            & 1.00   & mar-90 \\
		Average Max                                                &            & 72.00      &            & 76.00      &            & 50.73      &            & 68.00      &            & 72.00      &            & 58.00      &            & 65.00      &            & 70.00      &            & 48.00      &            & 57.00      &            & 79.00      &            & 86.00      & 50.73  & mar-01 \\
		\hline
	\end{tabular}
}\smallskip
\begin{flushleft}
\footnotesize{\textit{Note: Missing entries (--) represent unrecorded data before statistical imputation.}}
\end{flushleft}
\end{table}

\begin{table}[] 
	\centering
	\caption{$\Delta$ Temperature Average 1990-2014 for Rome - RM}
	\tiny \rotatebox{90}{
		\begin{tabular}{|l|l|l|l|l|l|l|l|l|l|l|l|l|l|l|l|l|l|l|l|l|l|l|l|l|} \hline
			\multicolumn{1}{|c|}{Year} & \multicolumn{2}{|c|}{Jan} & \multicolumn{2}{|c|}{Feb} & \multicolumn{2}{|c|}{Mar} & \multicolumn{2}{c}{Apr} & \multicolumn{2}{|c|}{May} & \multicolumn{2}{|c|}{Jun} & \multicolumn{2}{|c|}{Jul} & \multicolumn{2}{|c|}{Aug} & \multicolumn{2}{|c|}{Sep} & \multicolumn{2}{|c|}{Oct} & \multicolumn{2}{|c|}{Nov} & \multicolumn{2}{|c|}{Dec} \\ \hline
			\multicolumn{1}{|c|}{}                               & Min        & Max        & Min        & Max        & Min        & Max        & Min        & Max        & Min        & Max        & Min        & Max        & Min        & Max        & Min        & Max        & Min        & Max        & Min        & Max        & Min        & Max        & Min        & Max              \\ \hline
\bf{Average 1990-2003} & -3.63 & 15.04 & -3.30 & 16.16 & -1.07 & 20.54 & 0.09  & 22.64 & 6.53  & 27.29 & 9.47  & 32.06 & 12.60 & 34.59 & 12.81 & 35.17 & 8.93  & 28.99 & 4.73  & 24.75 & -0.24 & 19.94 & -2.64 & 15.96 \\

\bf{Average 2003-2014} & -3.41 & 15.74 & -3.16 & 15.94 & -1.66 & 20.19 & 3.13  & 24.18 & 6.64  & 28.36 & 10.56 & 33.62 & 12.56 & 35.96 & 14.27 & 35.09 & 9.18  & 31.65 & 5.15  & 26.46 & 0.70  & 21.12 & -2.96 & 16.53 \\
\bf{Average 1990-2003} & -3.63 & 15.04 & -3.30 & 16.16 & -1.07 & 20.54 & 0.09  & 22.64 & 6.53  & 27.29 & 9.47  & 32.06 & 12.60 & 34.59 & 12.81 & 35.17 & 8.93  & 28.99 & 4.73  & 24.75 & -0.24 & 19.94 & -2.64 & 15.96 \\
\bf{Average 2003-2014} & -3.41 & 15.74 & -3.16 & 15.94 & -1.66 & 20.19 & 3.13  & 24.18 & 6.64  & 28.36 & 10.56 & 33.62 & 12.56 & 35.96 & 14.27 & 35.09 & 9.18  & 31.65 & 5.15  & 26.46 & 0.70  & 21.12 & -2.96 & 16.53 \\
\hline
\bf{$\Delta$ Average 1990-2014} & \bf 0.22  & \bf 0.69  & \bf 0.14  & \bf -0.23 & \bf -0.59 & \bf -0.34 & \bf 3.04  & \bf 1.54  & \bf 0.11  & \bf 1.07  & \bf 1.09  & \bf 1.56  & \bf -0.04 & \bf 1.38  & \bf 1.46  & \bf -0.08 & \bf 0.25  & \bf 2.66  & \bf 0.42  & \bf 1.71  & \bf 0.94  & \bf 1.18  & \bf -0.32 & \bf 0.56 \\
\hline
\end{tabular}
}
\end{table}

\begin{table}[htbp]
	\centering
	\footnotesize
	\setlength{\tabcolsep}{3pt}
	\caption{Monthly climatological comparison and Heating Degree Days (HDD) variations for the city of Rome: Regulatory Data (D.P.R. 412/1993) vs. Real Measured Data (1990–2014 period).}
	\label{tab:hdd_comparison}
	\vspace{0.5em}
	\begin{tabular}{lrrr}
		\hline
		\textbf{Month} & \textbf{Real $T_{\text{mean}}$ ($^\circ\text{C}$)} & \textbf{Regulatory HDD} & \textbf{Real HDD} \\ \hline
		January & 7.6 & 350 & 322 \\
		February & 8.2 & 290 & 265 \\
		March & 10.6 & 210 & 185 \\
		April & 13.4 & 55 & 78 \\
		May & 18.1 & 0 & 12 \\
		June & 22.2 & 0 & 0 \\
		July & 25.1 & 0 & 0 \\
		August & 25.1 & 0 & 0 \\
		September & 21.3 & 0 & 3 \\
		October & 16.9 & 0 & 38 \\
		November & 12.0 & 200 & 168 \\
		December & 8.6 & 310 & 291 \\ \hline
		\textbf{TOTAL} & \textbf{15.8} & \textbf{1,415} & \textbf{1,362} \\ \hline
	\end{tabular}
\end{table}
\begin{table}[htbp]
	\centering
	\footnotesize
	\setlength{\tabcolsep}{3pt}
	\caption{Chronostratigraphic framework and physical properties of the sedimentary units in the Torre Spaccata subsurface.}
	\label{tab:sedimentary_properties}
	\vspace{0.5em}
	\begin{tabular}{llll}
		\hline
		\textbf{Group / Unit} & \textbf{Chronostrat.} & \textbf{Lithological Facies} & \textbf{Perm. ($K$, m/s)} \\ \hline
		Holocene Alluvium & Holocene - Present & Clayey silts and fine sands & Low-to-Med \\
		Valle Giulia Synthem & Mid Pleistocene & Tuffaceous gravels, sands & Med-to-High \\
		Monte Mario Fm. & Low Pleist. - Pliocene & Compact marly clays & $< 10^{-9}$ \\ \hline
	\end{tabular}
\end{table}

\begin{figure}
\begin{center}
\end{center}
\end{figure}


\begin{figure}
\begin{center}
\includegraphics[width=1.3\textwidth{}]{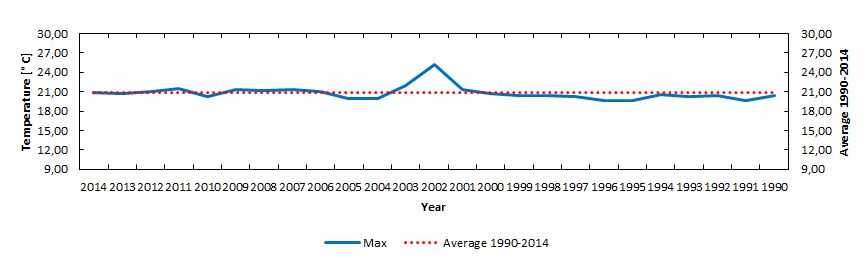}
\caption{Maximum Temperature (dotted line) and Average Temperature Variations 1990-2014 (solid line)}
\end{center}
\end{figure}
\begin{figure}
\begin{center}
\includegraphics[width=1.3\textwidth{}]{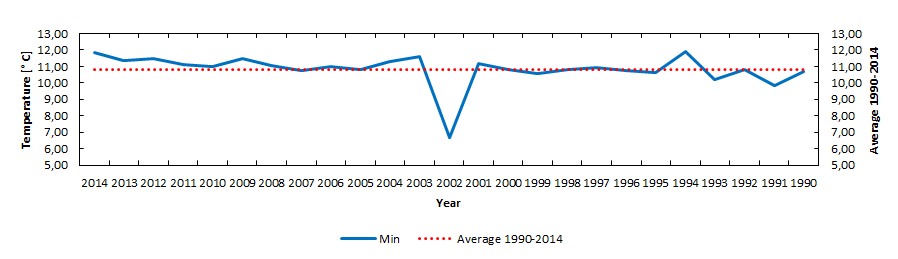}
\caption{Minimum Temperature (dotted line) and Average Temperature Variations 1990-2014 (solid line)}
\end{center}
\end{figure}
\begin{figure}
\begin{center}
\includegraphics[width=1.3\textwidth{}]{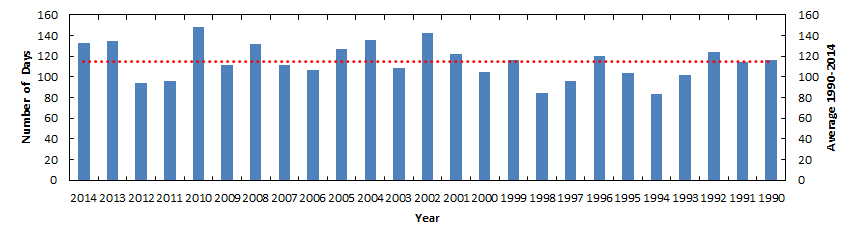}
\caption{Total Days of Rain by Year (solid line) and Average Variations 1990-2014 (dotted line)}
\end{center}
\end{figure}
\begin{figure}
\begin{center}
\includegraphics[width=1.3\textwidth{}]{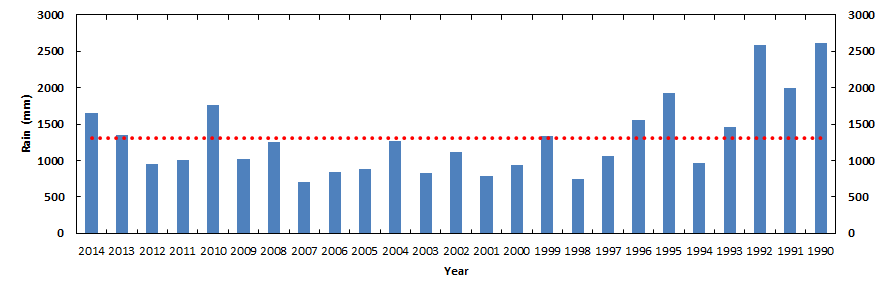}
\caption{Total rain [mm] by year (solid line) and Average Variations 1990-2014 (dotted line)}
\end{center}
\end{figure}
\begin{table}[htbp]
	\centering
	\caption{Monthly distribution of Fog Days in Torrespaccata quadrant (1990-2014).}
	\label{tab:fog_days_real}
	\tiny
	\begin{tabular}{lccccccccccccc}
		\toprule
		\textbf{Year} & \textbf{Jan} & \textbf{Feb} & \textbf{Mar} & \textbf{Apr} & \textbf{May} & \textbf{Jun} & \textbf{Jul} & \textbf{Aug} & \textbf{Sep} & \textbf{Oct} & \textbf{Nov} & \textbf{Dec} & \textbf{Tot} \\
		\midrule
		2014  & 1 & 0 & 0 & 2 & 1 & 0 & 0 & 0 & 1 & 1 & 1 & 1 & \textbf{8} \\
		2013  & 3 & 0 & 0 & 0 & 0 & 0 & 0 & 0 & 0 & 2 & 0 & 4 & \textbf{9} \\
		2012  & 3 & 1 & 2 & 3 & 3 & 0 & 1 & 1 & 2 & 1 & 0 & 2 & \textbf{19} \\
		2011  & 6 & 3 & 2 & 1 & 0 & 0 & 1 & 0 & 0 & 1 & 1 & 0 & \textbf{15} \\
		2010  & 4 & 0 & 0 & 2 & 0 & 0 & 0 & 2 & 0 & 0 & 1 & 0 & \textbf{9} \\
		2009  & 6 & 0 & 0 & 2 & 0 & 0 & 0 & 0 & 0 & 0 & 5 & 0 & \textbf{13} \\
		2008  & 5 & 1 & 2 & 0 & 0 & 0 & 0 & 0 & 0 & 0 & 2 & 4 & \textbf{16} \\
		2007  & 6 & 2 & 2 & 0 & 0 & 2 & 0 & 0 & 0 & 0 & 0 & 0 & \textbf{12} \\
		2006  & 4 & 2 & 3 & 2 & 1 & 1 & 0 & 1 & 1 & 2 & 5 & 1 & \textbf{23} \\
		2005  & 4 & 0 & 1 & 0 & 0 & 1 & 0 & 0 & 0 & 6 & 1 & 2 & \textbf{15} \\
		2004  & 5 & 1 & 5 & 4 & 0 & 0 & 0 & 0 & 1 & 2 & 1 & 0 & \textbf{19} \\
		2003  & 6 & 0 & 2 & 1 & 1 & 2 & 0 & 0 & 0 & 1 & 3 & 0 & \textbf{16} \\
		2002  & 1 & 2 & 6 & 0 & 1 & 0 & 0 & 0 & 0 & 0 & 0 & 4 & \textbf{14} \\
		2001  & 5 & 0 & 1 & 0 & 2 & 1 & 0 & 0 & 0 & 0 & 0 & 0 & \textbf{9} \\
		2000  & 3 & 3 & 0 & 3 & 1 & 0 & 0 & 2 & 0 & 0 & 1 & 0 & \textbf{13} \\
		1999  & 4 & 2 & 1 & 3 & 2 & 0 & 0 & 0 & 0 & 1 & 1 & 1 & \textbf{15} \\
		1998  & 5 & 1 & 0 & 1 & 2 & 0 & 0 & 0 & 0 & 0 & 0 & 0 & \textbf{9} \\
		1997  & 6 & 1 & 2 & 1 & 2 & 0 & 1 & 1 & 0 & 2 & 0 & 2 & \textbf{18} \\
		1996  & 2 & 0 & 0 & 2 & 3 & 0 & 0 & 1 & 0 & 0 & 0 & 4 & \textbf{12} \\
		1995  & 1 & 2 & 0 & 1 & 3 & 0 & 0 & 0 & 0 & 0 & 2 & 2 & \textbf{11} \\
		1994  & 2 & 1 & 4 & 0 & 2 & 0 & 0 & 0 & 0 & 0 & 4 & 0 & \textbf{15} \\
		1993  & 8 & 0 & 2 & 1 & 2 & 1 & 0 & 0 & 0 & 1 & 1 & 0 & \textbf{16} \\
		1992  & 6 & 0 & 3 & 4 & 1 & 0 & 1 & 2 & 0 & 2 & 4 & 1 & \textbf{24} \\
		1991  & 2 & 5 & 2 & 1 & 0 & 0 & 0 & 0 & 2 & 2 & 4 & 2 & \textbf{20} \\
		1990  & 2 & 5 & 5 & 0 & 1 & 0 & 0 & 0 & 1 & 3 & 0 & 0 & \textbf{14} \\
		\midrule
		\textbf{Mean} & \textbf{4} & \textbf{2} & \textbf{2} & \textbf{1} & \textbf{1} & \textbf{0} & \textbf{0} & \textbf{0} & \textbf{1} & \textbf{2} & \textbf{2} & \textbf{2} & \textbf{254} \\
		\bottomrule
	\end{tabular}
\end{table}

\begin{figure}
\begin{center}
\includegraphics[width=1.3\textwidth{}]{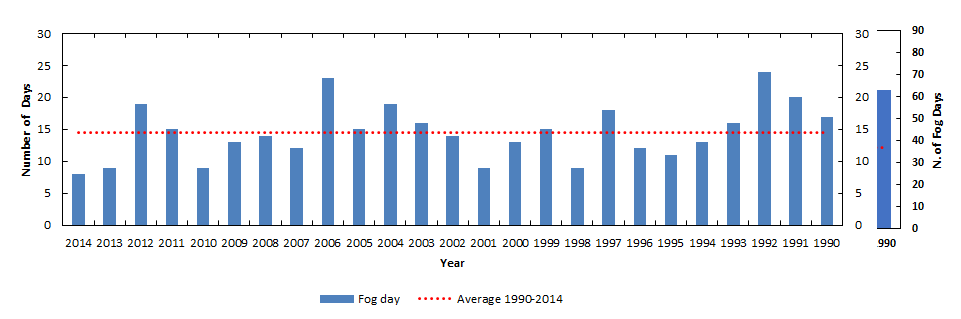}
\caption{Total day of Fog by year (solid line) and Average Variations 1990-2014 (dotted line)}
\end{center}
\end{figure}
\begin{table*}[htbp]
	\centering
	\caption{Comprehensive Chronological Matrix of Mean Monthly Relative Humidity ($RH$, \%) in the Torrespaccata District (1990--2014).}
	\label{tab:humidity_matrix_clean}
	\scriptsize
	\begin{tabular}{lccccccccccccc}
		\toprule
		\textbf{Year} & \textbf{Jan} & \textbf{Feb} & \textbf{Mar} & \textbf{Apr} & \textbf{May} & \textbf{Jun} & \textbf{Jul} & \textbf{Aug} & \textbf{Sep} & \textbf{Oct} & \textbf{Nov} & \textbf{Dec} & \textbf{Annual} \\
		\midrule
		2014 & 86 & 80 & 74 & 66 & 69 & 67 & 70 & 67 & 74 & 73 & 79 & 76 & \textbf{73} \\
		2013 & 84 & 74 & 80 & 73 & 76 & 67 & 67 & 64 & 76 & 84 & 87 & 87 & \textbf{77} \\
		2012 & 76 & 79 & 69 & 79 & 73 & 63 & 62 & 61 & 76 & 82 & 85 & 80 & \textbf{74} \\
		2011 & 84 & 73 & 79 & 74 & 69 & 68 & 68 & 64 & 69 & 71 & 83 & 82 & \textbf{74} \\
		2010 & 79 & 77 & 75 & 74 & 76 & 68 & 64 & 66 & 69 & 76 & 83 & 76 & \textbf{74} \\
		2009 & 83 & 71 & 71 & 72 & 62 & 65 & 61 & 60 & 69 & 75 & 82 & 80 & \textbf{71} \\
		2008 & 77 & 72 & 81 & 68 & 64 & 65 & 61 & 59 & 63 & 75 & 77 & 80 & \textbf{70} \\
		2007 & 81 & 79 & 74 & 70 & 70 & 65 & 55 & 59 & 65 & 70 & 75 & 75 & \textbf{70} \\
		2006 & 76 & 75 & 76 & 75 & 72 & 63 & 65 & 65 & 71 & 76 & 76 & 79 & \textbf{72} \\
		2005 & 77 & 69 & 73 & 77 & 67 & 65 & 60 & 72 & 82 & 89 & 93 & 80 & \textbf{75} \\
		2004 & 80 & 80 & 79 & 86 & 74 & 68 & 59 & 68 & 74 & 85 & 83 & 82 & \textbf{76} \\
		2003 & 78 & 63 & 63 & 62 & 57 & 56 & 59 & 55 & 69 & 79 & 89 & 79 & \textbf{67} \\
		2002 & 79 & 84 & 71 & 78 & 77 & 70 & 72 & 81 & 83 & 88 & 84 & 90 & \textbf{80} \\
		2001 & 88 & 81 & 84 & 78 & 79 & 60 & 61 & 64 & 72 & 80 & 80 & 75 & \textbf{75} \\
		2000 & 81 & 78 & 79 & 78 & 75 & 63 & 60 & 64 & 73 & 86 & 87 & 89 & \textbf{76} \\
		1999 & 82 & 74 & 76 & 80 & 81 & 68 & 72 & 74 & 81 & 87 & 87 & 86 & \textbf{79} \\
		1998 & 84 & 78 & 68 & 78 & 78 & 70 & 66 & 69 & 79 & 86 & 87 & 85 & \textbf{77} \\
		1997 & 81 & 72 & 66 & 70 & 66 & 68 & 65 & 67 & 72 & 79 & 87 & 79 & \textbf{73} \\
		1996 & 84 & 78 & 74 & 74 & 76 & 68 & 66 & 72 & 79 & 79 & 80 & 78 & \textbf{76} \\
		1995 & 67 & 80 & 72 & 75 & 73 & 70 & 70 & 72 & 78 & 76 & 75 & 89 & \textbf{75} \\
		1994 & 81 & 78 & 77 & 74 & 65 & 62 & 59 & 59 & 64 & 76 & 83 & 86 & \textbf{72} \\
		1993 & 79 & 66 & 71 & 76 & 71 & 64 & 58 & 59 & 73 & 81 & 84 & 84 & \textbf{72} \\
		1992 & 80 & 73 & 76 & 74 & 71 & 71 & 69 & 67 & 70 & 83 & 86 & 84 & \textbf{75} \\
		1991 & 75 & 73 & 69 & 75 & 71 & 67 & 60 & 64 & 74 & 80 & 85 & 71 & \textbf{72} \\
		1990 & 79 & 79 & 75 & 74 & 71 & 63 & 63 & 67 & 73 & 79 & 79 & 78 & \textbf{73} \\
		\midrule
		\textbf{Clim. Mean} & \textbf{80.05} & \textbf{75.40} & \textbf{74.03} & \textbf{74.41} & \textbf{71.40} & \textbf{65.83} & \textbf{63.59} & \textbf{65.56} & \textbf{73.12} & \textbf{79.70} & \textbf{83.06} & \textbf{81.16} & \textbf{73.94} \\
		\bottomrule
	\end{tabular}
\end{table*}

\begin{figure}
\begin{center}
\includegraphics[width=1.3\textwidth{}]{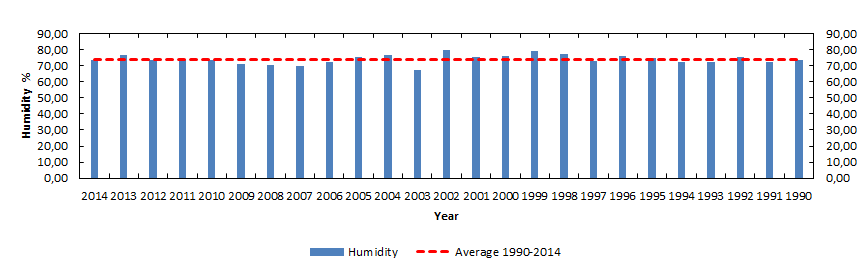}
\caption{Humidity (solid line) and Average Humidity Variations 1990-2014 (dotted line) }
\end{center}
\end{figure}
\begin{table*}[htbp]
	\centering
	\caption{Comprehensive Chronological Matrix of Mean Monthly Surface Atmospheric Pressure ($P_{\text{surf}}$, hPa) in the Torrespaccata District (1990--2014).}
	\label{tab:pressure_matrix_clean}
	\scriptsize
	\begin{tabular}{lccccccccccccc}
		\toprule
		\textbf{Year} & \textbf{Jan} & \textbf{Feb} & \textbf{Mar} & \textbf{Apr} & \textbf{May} & \textbf{Jun} & \textbf{Jul} & \textbf{Aug} & \textbf{Sep} & \textbf{Oct} & \textbf{Nov} & \textbf{Dec} & \textbf{Annual} \\
		\midrule
		2014 & 1012 & 1012 & 1014 & 1011 & 1014 & 1014 & 1011 & 1013 & 1016 & 1017 & 1013 & 1016 & \textbf{1014} \\
		2013 & 1011 & 1009 & 1006 & 1013 & 1012 & 1015 & 1016 & 1014 & 1015 & 1018 & 1011 & 1022 & \textbf{1013} \\
		2012 & 1019 & 1015 & 1022 & 1008 & 1014 & 1014 & 1013 & 1014 & 1014 & 1012 & 980  & 1014 & \textbf{1012} \\
		2011 & 1017 & 1015 & 1019 & 1014 & 1016 & 1014 & 1012 & 1014 & 1015 & 1018 & 1020 & 1018 & \textbf{1016} \\
		2010 & 1011 & 1005 & 1016 & 1016 & 1012 & 1012 & 1014 & 1013 & 1013 & 1012 & 1008 & 1011 & \textbf{1012} \\
		2009 & 1011 & 1012 & 1010 & 976  & 1017 & 1013 & 1015 & 1015 & 1015 & 1014 & 1014 & 1007 & \textbf{1010} \\
		2008 & 1021 & 1026 & 1008 & 1011 & 1013 & 1014 & 1014 & 1013 & 1014 & 1018 & 1014 & 1015 & \textbf{1015} \\
		2007 & 1022 & 1013 & 1012 & 1017 & 1012 & 1013 & 1014 & 1013 & 1016 & 1016 & 1015 & 1019 & \textbf{1015} \\
		2006 & 1014 & 1020 & 1010 & 1014 & 1013 & 1014 & 1014 & 1013 & 1015 & 1017 & 1014 & 1017 & \textbf{1015} \\
		2005 & 1016 & 1018 & 1017 & 1014 & 1013 & 1014 & 1014 & 1013 & 1015 & 1017 & 1014 & 1000 & \textbf{1014} \\
		2004 & 1015 & 1019 & 1014 & 1014 & 1019 & 1014 & 1015 & 1019 & 1014 & 1014 & 1019 & 1020 & \textbf{1016} \\
		2003 & 1016 & 1024 & 1020 & 1008 & 1024 & 1014 & 1016 & 1012 & 1016 & 1012 & 1016 & 1012 & \textbf{1016} \\
		2002 & 1015 & 1020 & 1015 & 1012 & 1017 & 1014 & 1015 & 1014 & 1015 & 1015 & 1016 & 1012 & \textbf{1015} \\
		2001 & 1016 & 1019 & 1013 & 1008 & 1016 & 1014 & 1009 & 1015 & 1015 & 1015 & 1016 & 1011 & \textbf{1014} \\
		2000 & 1017 & 1020 & 1014 & 1012 & 1016 & 1014 & 1014 & 1014 & 1015 & 1016 & 1015 & 1014 & \textbf{1015} \\
		1999 & 1016 & 1020 & 1016 & 1011 & 1018 & 1014 & 1014 & 1015 & 1015 & 1015 & 1016 & 1011 & \textbf{1015} \\
		1998 & 1016 & 1020 & 1015 & 1012 & 1017 & 1014 & 1014 & 1014 & 1015 & 1015 & 1016 & 1012 & \textbf{1015} \\
		1997 & 1016 & 1019 & 1014 & 1009 & 1016 & 1014 & 1014 & 1014 & 1015 & 1015 & 1015 & 1013 & \textbf{1015} \\
		1996 & 1016 & 1020 & 1015 & 1011 & 1017 & 1014 & 1014 & 1014 & 1015 & 1015 & 1016 & 1012 & \textbf{1015} \\
		1995 & 1013 & 1013 & 1011 & 1015 & 1002 & 1015 & 1013 & 1013 & 1014 & 1015 & 1015 & 1014 & \textbf{1013} \\
		1994 & 1014 & 1015 & 1015 & 1007 & 1016 & 1013 & 1014 & 1014 & 1015 & 1015 & 1015 & 1012 & \textbf{1014} \\
		1993 & 1016 & 1020 & 1014 & 1011 & 1016 & 1014 & 1013 & 1014 & 1015 & 1015 & 1015 & 1012 & \textbf{1015} \\
		1992 & 1015 & 1017 & 1014 & 1009 & 1016 & 1014 & 1013 & 1014 & 1015 & 1015 & 1015 & 1012 & \textbf{1014} \\
		1991 & 1013 & 1011 & 1015 & 1002 & 1015 & 1013 & 1013 & 1014 & 1015 & 1015 & 1014 & 1012 & \textbf{1013} \\
		1990 & 1014 & 1014 & 1015 & 1005 & 1015 & 1013 & 1013 & 1014 & 1015 & 1015 & 1015 & 1012 & \textbf{1013} \\
		\midrule
		\textbf{Clim. Mean} & \textbf{1015} & \textbf{1015} & \textbf{1017} & \textbf{1014} & \textbf{1009} & \textbf{1015} & \textbf{1014} & \textbf{1014} & \textbf{1014} & \textbf{1015} & \textbf{1015} & \textbf{1014} & \textbf{1013} \\
		\bottomrule
	\end{tabular}
\end{table*}

\begin{figure}
\begin{center}
\includegraphics[width=1.3\textwidth{}]{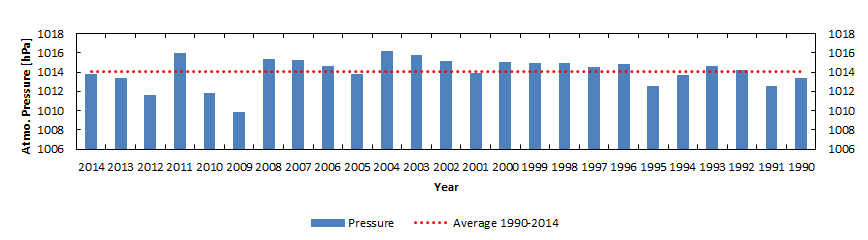}
\caption{Pressure (solid line) and Average Pressure Variations 1990-2014 (dotted line) }
\end{center}
\end{figure}
\begin{table*}[htbp]
	\centering
	\caption{Comprehensive Chronological Matrix of Mean Monthly Dew Point Temperatures ($T_{\text{dew}}$, $^{\circ}\text{C}$) in the Torrespaccata District (1990--2014).}
	\label{tab:dew_point_matrix_clean}
	\scriptsize
	\begin{tabular}{lccccccccccccc}
		\toprule
		\textbf{Year} & \textbf{Jan} & \textbf{Feb} & \textbf{Mar} & \textbf{Apr} & \textbf{May} & \textbf{Jun} & \textbf{Jul} & \textbf{Aug} & \textbf{Sep} & \textbf{Oct} & \textbf{Nov} & \textbf{Dec} & \textbf{Annual} \\
		\midrule
		2014 & 7.0 & 8.0 & 6.0 & 8.0 & 11.0 & 15.0 & 17.0 & 17.0 & 16.0 & 13.0 & 11.0 & 6.0 & \textbf{11.0} \\
		2013 & 6.0 & 3.0 & 7.0 & 10.0 & 12.0 & 14.0 & 18.0 & 18.0 & 16.0 & 16.0 & 10.0 & 6.0 & \textbf{11.0} \\
		2012 & 4.0 & 3.0 & 6.0 & 10.0 & 12.0 & 16.0 & 18.0 & 18.0 & 17.0 & 15.0 & 11.0 & 5.0 & \textbf{11.0} \\
		2011 & 5.0 & 4.0 & 7.0 & 9.0 & 12.0 & 16.0 & 17.0 & 16.0 & 16.0 & 11.0 & 9.0 & 7.0 & \textbf{11.0} \\
		2010 & 3.0 & 5.0 & 6.0 & 9.0 & 12.0 & 15.0 & 18.0 & 18.0 & 14.0 & 12.0 & 10.0 & 5.0 & \textbf{11.0} \\
		2009 & 5.0 & 4.0 & 6.0 & 9.0 & 12.0 & 15.0 & 17.0 & 18.0 & 16.0 & 10.0 & 9.0 & 6.0 & \textbf{11.0} \\
		2008 & 5.0 & 3.0 & 7.0 & 8.0 & 11.0 & 15.0 & 17.0 & 17.0 & 12.0 & 13.0 & 9.0 & -5.0 & \textbf{9.0} \\
		2007 & 6.0 & 6.0 & 6.0 & 10.0 & 13.0 & 15.0 & 15.0 & 16.0 & 13.0 & 10.0 & 6.0 & 3.0 & \textbf{10.0} \\
		2006 & 2.0 & 3.0 & 6.0 & 10.0 & 13.0 & 15.0 & 19.0 & 17.0 & 16.0 & 13.0 & 8.0 & 6.0 & \textbf{11.0} \\
		2005 & 1.0 & 0.0 & 5.0 & 8.0 & 12.0 & 16.0 & 17.0 & 18.0 & 18.0 & 14.0 & 13.0 & 4.0 & \textbf{11.0} \\
		2004 & 2.0 & 4.0 & 6.0 & 11.0 & 11.0 & 16.0 & 16.0 & 18.0 & 16.0 & 16.0 & 9.0 & 7.0 & \textbf{11.0} \\
		2003 & 5.0 & -1.0 & 3.0 & 7.0 & 11.0 & 17.0 & 16.0 & 18.0 & 15.0 & 13.0 & 12.0 & 5.0 & \textbf{10.0} \\
		2002 & 2.0 & 8.0 & 7.0 & 10.0 & 14.0 & 18.0 & 19.0 & 20.0 & 16.0 & 15.0 & 12.0 & 8.0 & \textbf{12.0} \\
		2001 & 8.0 & 5.0 & 11.0 & 9.0 & 15.0 & 13.0 & 17.0 & 18.0 & 14.0 & 16.0 & 9.0 & 2.0 & \textbf{11.0} \\
		2000 & 2.0 & 4.0 & 7.0 & 10.0 & 14.0 & 15.0 & 15.0 & 17.0 & 16.0 & 14.0 & 11.0 & 8.0 & \textbf{11.0} \\
		1999 & 4.0 & 2.0 & 6.0 & 10.0 & 15.0 & 16.0 & 18.0 & 20.0 & 17.0 & 15.0 & 10.0 & 6.0 & \textbf{12.0} \\
		1998 & 5.0 & 5.0 & 4.0 & 10.0 & 13.0 & 17.0 & 18.0 & 19.0 & 16.0 & 14.0 & 8.0 & 4.0 & \textbf{11.0} \\
		1997 & 5.0 & 4.0 & 4.0 & 5.0 & 11.0 & 16.0 & 16.0 & 17.0 & 16.0 & 13.0 & 11.0 & 6.0 & \textbf{10.0} \\
		1996 & 7.0 & 3.0 & 5.0 & 9.0 & 14.0 & 16.0 & 17.0 & 18.0 & 14.0 & 11.0 & 9.0 & 4.0 & \textbf{11.0} \\
		1995 & 0.0 & 6.0 & 4.0 & 8.0 & 11.0 & 14.0 & 19.0 & 18.0 & 16.0 & 13.0 & 7.0 & 9.0 & \textbf{10.0} \\
		1994 & 5.0 & 4.0 & 8.0 & 8.0 & 11.0 & 15.0 & 18.0 & 19.0 & 15.0 & 12.0 & 10.0 & 7.0 & \textbf{11.0} \\
		1993 & 3.0 & 4.0 & 4.0 & 9.0 & 13.0 & 15.0 & 15.0 & 17.0 & 16.0 & 14.0 & 8.0 & 6.0 & \textbf{10.0} \\
		1992 & 4.0 & 3.0 & 6.0 & 8.0 & 13.0 & 14.0 & 17.0 & 19.0 & 15.0 & 14.0 & 11.0 & 6.0 & \textbf{11.0} \\
		1991 & 3.0 & 3.0 & 7.0 & 7.0 & 8.0 & 14.0 & 17.0 & 18.0 & 17.0 & 12.0 & 8.0 & -1.0 & \textbf{9.0} \\
		1990 & 3.0 & 6.0 & 6.0 & 8.0 & 12.0 & 14.0 & 17.0 & 17.0 & 16.0 & 14.0 & 8.0 & 3.0 & \textbf{10.0} \\
		\midrule
		\textbf{Clim. Mean} & \textbf{4.12} & \textbf{3.98} & \textbf{5.98} & \textbf{8.77} & \textbf{12.30} & \textbf{15.27} & \textbf{17.11} & \textbf{17.90} & \textbf{15.64} & \textbf{13.24} & \textbf{9.50} & \textbf{4.98} & \textbf{10.73} \\
		\bottomrule
	\end{tabular}
\end{table*}

\begin{figure}
\begin{center}
\includegraphics[width=1.3\textwidth{}]{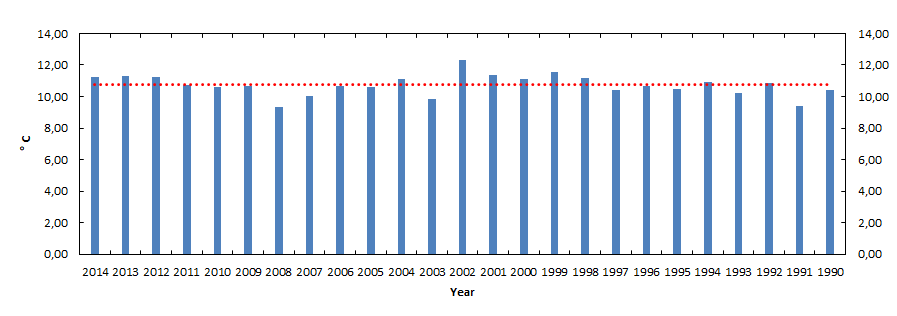}
\caption{Dew Point (solid line) and Average Dew Point Variations 1990-2014 (dotted line) }
\end{center}
\end{figure}
\begin{table*}[p]
	\centering
	\caption{Comprehensive Climatological Matrix of Mean Daily Wind Speeds ($v_{\text{wind}}$, \text{km/h}) in the Torrespaccata District broken down by Day of the Month (1990--2014).}
	\label{tab:daily_wind_matrix}
	\tiny
	\begin{tabular}{lcccccccccccccclc}
		\toprule
		\textbf{Day} & \textbf{Jan} & \textbf{Feb} & \textbf{Mar} & \textbf{Apr} & \textbf{May} & \textbf{Jun} & \textbf{Jul} & \textbf{Aug} & \textbf{Sep} & \textbf{Oct} & \textbf{Nov} & \textbf{Dec} & \textbf{Avg 90--14} & \textbf{Min} & \textbf{Max} & \textbf{Midrange} \\
		\midrule
		1  & 23.12 & 19.64 & 23.92 & 23.60 & 22.33 & 23.58 & 22.20 & 19.79 & 25.96 & 19.28 & 22.64 & 19.64 & \textbf{22.14} & 19.28 & 25.96 & 22.62 \\
		2  & 23.36 & 23.24 & 26.92 & 25.12 & 24.92 & 23.13 & 24.12 & 20.28 & 21.56 & 21.44 & 20.64 & 20.44 & \textbf{22.93} & 20.28 & 26.92 & 23.60 \\
		3  & 21.00 & 19.28 & 26.00 & 27.80 & 26.67 & 23.80 & 25.44 & 21.36 & 21.08 & 18.76 & 19.29 & 20.68 & \textbf{22.60} & 18.76 & 27.80 & 23.28 \\
		4  & 21.32 & 19.16 & 25.08 & 25.32 & 24.57 & 22.36 & 22.96 & 22.00 & 23.48 & 21.52 & 21.44 & 20.00 & \textbf{22.43} & 19.16 & 25.32 & 22.24 \\
		5  & 19.72 & 21.28 & 26.36 & 28.20 & 20.26 & 24.68 & 22.44 & 24.12 & 23.24 & 20.52 & 23.16 & 20.64 & \textbf{22.89} & 19.72 & 28.20 & 23.96 \\
		6  & 21.96 & 24.00 & 22.79 & 22.60 & 22.87 & 25.28 & 24.44 & 21.60 & 20.28 & 22.76 & 25.68 & 19.12 & \textbf{22.78} & 19.12 & 25.68 & 22.40 \\
		7  & 24.48 & 20.00 & 23.40 & 25.84 & 21.38 & 23.68 & 24.00 & 20.24 & 22.92 & 21.00 & 22.68 & 21.84 & \textbf{22.62} & 20.00 & 25.84 & 22.92 \\
		8  & 17.96 & 22.72 & 21.71 & 26.29 & 22.33 & 22.76 & 22.96 & 21.48 & 26.29 & 19.96 & 21.88 & 20.32 & \textbf{22.22} & 17.96 & 26.29 & 22.13 \\
		9  & 16.46 & 22.12 & 23.44 & 25.42 & 21.04 & 23.64 & 22.00 & 22.04 & 21.24 & 16.52 & 18.50 & 24.08 & \textbf{21.37} & 16.46 & 25.42 & 20.94 \\
		10 & 17.41 & 20.12 & 21.96 & 27.52 & 22.25 & 24.16 & 23.40 & 24.56 & 20.04 & 18.25 & 21.44 & 21.64 & \textbf{21.90} & 17.41 & 27.52 & 22.46 \\
		11 & 19.21 & 22.29 & 21.56 & 27.72 & 19.92 & 24.60 & 24.84 & 21.76 & 19.48 & 18.67 & 23.24 & 17.00 & \textbf{21.69} & 17.00 & 27.72 & 22.36 \\
		12 & 18.17 & 21.08 & 19.84 & 25.44 & 22.83 & 22.63 & 23.48 & 22.88 & 23.36 & 22.48 & 24.04 & 17.48 & \textbf{21.98} & 17.48 & 25.44 & 21.46 \\
		13 & 17.88 & 17.92 & 23.16 & 24.28 & 21.08 & 20.76 & 22.92 & 20.56 & 24.20 & 22.52 & 23.17 & 20.88 & \textbf{21.61} & 17.88 & 24.28 & 21.08 \\
		14 & 18.91 & 18.25 & 18.36 & 23.52 & 24.24 & 23.08 & 25.92 & 21.04 & 24.88 & 21.60 & 23.96 & 23.00 & \textbf{22.23} & 18.25 & 25.92 & 22.09 \\
		15 & 16.17 & 24.92 & 18.52 & 25.13 & 21.12 & 20.17 & 25.92 & 21.42 & 23.68 & 19.96 & 24.25 & 22.76 & \textbf{22.00} & 16.17 & 25.92 & 21.04 \\
		16 & 15.50 & 22.38 & 23.80 & 28.42 & 23.48 & 22.38 & 26.84 & 21.92 & 22.40 & 18.96 & 28.33 & 21.08 & \textbf{22.96} & 15.50 & 28.42 & 21.96 \\
		17 & 18.96 & 22.25 & 23.20 & 28.63 & 23.96 & 24.71 & 21.40 & 19.64 & 20.32 & 21.80 & 24.83 & 18.20 & \textbf{22.32} & 18.20 & 28.63 & 23.41 \\
		18 & 17.80 & 23.38 & 19.60 & 25.92 & 23.64 & 22.48 & 22.16 & 19.88 & 19.52 & 19.44 & 22.17 & 20.96 & \textbf{21.41} & 17.80 & 25.92 & 21.86 \\
		19 & 19.16 & 22.76 & 21.84 & 24.13 & 23.24 & 22.50 & 24.56 & 21.08 & 21.60 & 19.88 & 21.08 & 19.48 & \textbf{21.78} & 19.16 & 24.56 & 21.86 \\
		20 & 18.00 & 22.67 & 21.32 & 24.00 & 25.28 & 22.38 & 22.72 & 20.84 & 21.20 & 21.20 & 21.52 & 22.12 & \textbf{21.94} & 18.00 & 25.28 & 21.64 \\
		21 & 17.92 & 22.68 & 21.80 & 27.22 & 21.40 & 22.58 & 21.08 & 21.00 & 19.76 & 24.56 & 21.05 & 20.54 & \textbf{21.80} & 17.92 & 27.22 & 22.57 \\
		22 & 18.36 & 21.24 & 21.84 & 21.09 & 20.84 & 27.16 & 22.16 & 21.44 & 20.16 & 18.72 & 24.57 & 22.91 & \textbf{21.71} & 18.36 & 27.16 & 22.76 \\
		23 & 22.44 & 25.46 & 26.29 & 21.79 & 23.64 & 22.08 & 23.68 & 20.32 & 21.36 & 17.92 & 22.83 & 22.09 & \textbf{22.49} & 17.92 & 26.29 & 22.11 \\
		24 & 23.16 & 23.00 & 25.40 & 25.88 & 22.39 & 21.04 & 24.40 & 20.44 & 23.42 & 18.48 & 21.48 & 21.41 & \textbf{22.54} & 18.48 & 25.88 & 22.18 \\
		25 & 20.36 & 20.16 & 26.56 & 23.50 & 19.20 & 22.68 & 24.16 & 23.12 & 20.48 & 19.84 & 18.13 & 22.43 & \textbf{21.72} & 18.13 & 26.56 & 22.34 \\
		26 & 23.76 & 23.08 & 23.88 & 21.75 & 19.79 & 20.32 & 21.92 & 21.16 & 23.12 & 22.56 & 20.42 & 27.91 & \textbf{22.47} & 19.79 & 27.91 & \textbf{23.85} \\
		27 & 20.64 & 23.72 & 23.80 & 26.04 & 21.58 & 21.92 & 19.48 & 21.56 & 17.36 & 20.20 & 19.21 & 23.43 & \textbf{21.58} & 17.36 & 26.04 & 21.70 \\
		28 & 20.52 & 26.24 & 25.36 & 20.83 & 21.63 & 24.67 & 21.40 & 26.42 & 19.40 & 19.80 & 17.79 & 23.57 & \textbf{22.30} & 17.79 & 26.42 & 22.10 \\
		29 & 21.44 & 19.29 & 22.84 & 22.42 & 23.13 & 22.60 & 22.16 & 22.72 & 21.12 & 20.68 & 19.96 & 20.65 & \textbf{21.58} & 19.29 & 23.13 & 21.21 \\
		30 & 20.92 & ---   & 24.60 & 21.79 & 18.50 & 22.08 & 23.04 & 20.76 & 19.40 & 23.12 & 17.13 & 21.14 & \textbf{21.13} & 17.13 & 24.60 & 20.86 \\
		31 & 20.80 & ---   & 25.24 & ---   & 19.26 & ---   & 21.96 & 22.44 & ---   & 23.72 & ---   & 22.64 & \textbf{22.29} & 19.26 & 25.24 & 22.25 \\
		\midrule
		\textbf{Mean} & \textbf{19.90} & \textbf{21.87} & \textbf{23.24} & \textbf{24.91} & \textbf{22.22} & \textbf{23.00} & \textbf{23.23} & \textbf{21.61} & \textbf{21.74} & \textbf{20.52} & \textbf{21.88} & \textbf{21.29} & \textbf{22.06} & \textbf{15.50} & \textbf{28.63} & \textbf{22,06} \\
		\bottomrule
	\end{tabular}
\end{table*}
\begin{figure}
\begin{center}
\includegraphics[width=1.3\textwidth{}]{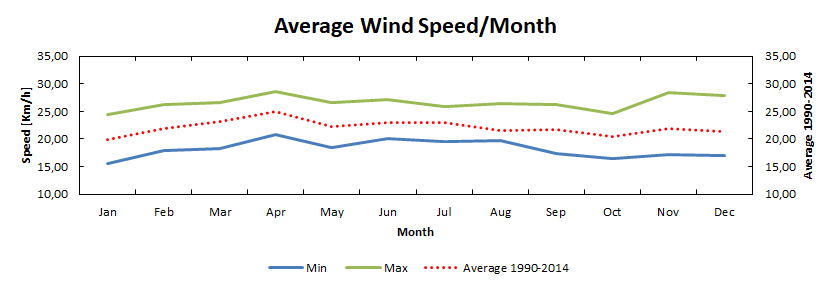}
\caption{Monthly distribution of minimum, maximum, and multi-year average surface wind speeds (km/h) recorded in the Torrespaccata district over the 1990–2014 climatological baseline.}
\end{center}
\end{figure}
\begin{figure}
	\begin{center}
		\includegraphics[width=1.3\textwidth{}]{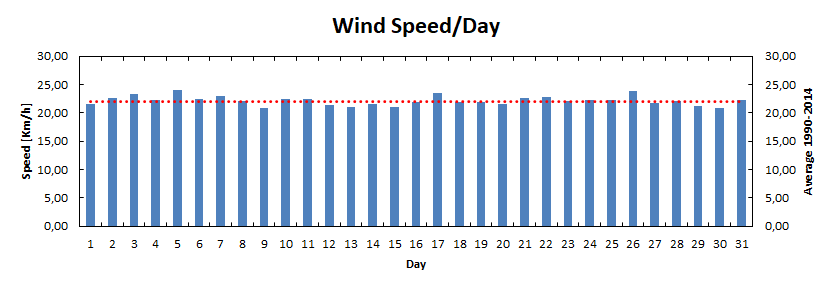}
		\caption{Distribution of interannual multi-year average daily wind speeds (km/h) across the 1-31 day of the month baseline profile.}
	\end{center}
\end{figure}
\clearpage
\begin{figure}
\includegraphics[width=1.3\textwidth{}]{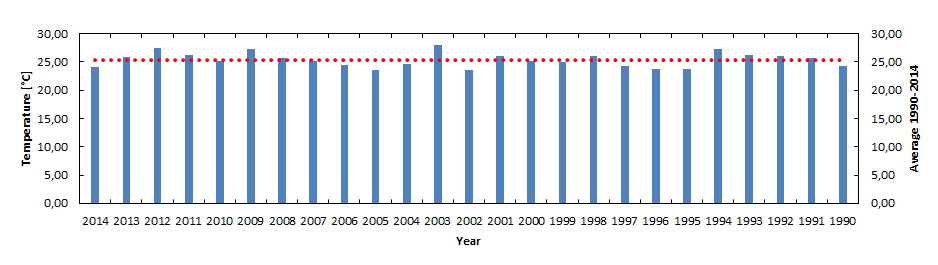}
\caption{August Temperature (solid line) and Average August Temperature 1990-2014 (dotted line) }
\end{figure}
\begin{figure}
\includegraphics[width=1.3\textwidth{}]{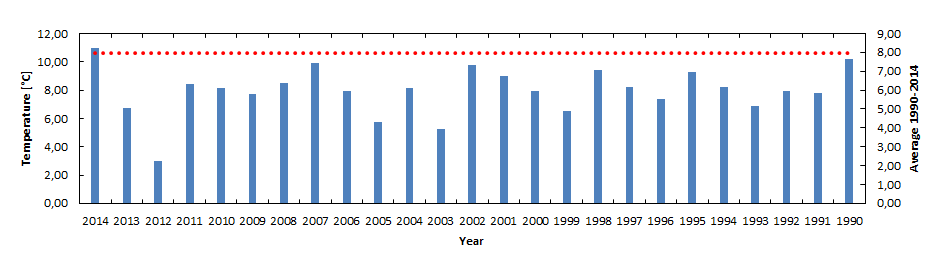}
\caption{February Temperature (solid line) and Average February Temperature 1990-2014 (dotted line)}
\end{figure}

\centering
\begin{figure}
\begin{center}
\includegraphics[width=1.3\textwidth{}]{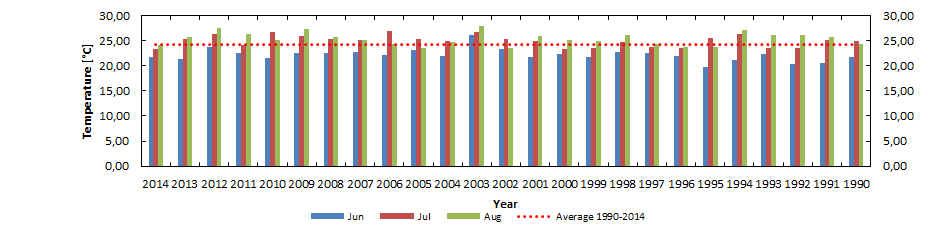}
\caption{Summer Temperature (solid line) and Average Temperature Variations 1990-2014 (dotted line). Jun: blu, Jul: red Aug: green}
\end{center}
\end{figure}
\begin{figure}
\begin{center}
\includegraphics[width=1.3\textwidth{}]{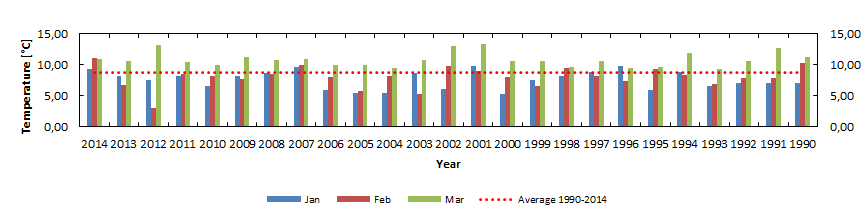}
\caption{Winter Temperature (solid line) and Average Temperature Variations 1990-2014 (dotted line) Jan: blu, Feb: red Mar: green}
\end{center}
\end{figure}
\begin{table*}[t]
	\centering
	\caption{Interannual Thermal Deviations ($\Delta T_{\text{blackbird}}$, $^{\circ}\text{C}$) for the Traditional Blackbird Days in Rome (Baseline Period: 1990-2014)}
	\label{tab:blackbird_days_clean}
	\small
	\begin{tabular}{lrcr}
		\toprule
		\textbf{Year} & \textbf{$\Delta T_{\text{blackbird}}$} & \textbf{Year} & \textbf{$\Delta T_{\text{blackbird}}$} \\
		\midrule
		2014 & 6.67  & 2013 & 3.67 \\
		2012 & 3.33  & 2011 & 6.00 \\
		2010 & 2.67  & 2009 & 1.67 \\
		2008 & 4.80  & 2007 & 2.00 \\
		2006 & 6.53  & 2005 & -0.27 \\
		2004 & -1.00 & 2003 & 3.00 \\
		2002 & 8.53  & 2001 & 6.33 \\
		2000 & 3.67  & 1999 & 0.33 \\
		1998 & 3.48  & 1997 & 2.33 \\
		1996 & 6.67  & 1995 & 6.67 \\
		1994 & 1.87  & 1993 & 5.33 \\
		1992 & 2.00  & 1991 & -2.67 \\
		1990 & 7.33  & ---  & --- \\
		\bottomrule
	\end{tabular}
\end{table*}

\begin{figure}
	\begin{center}
		\includegraphics[width=1.3\textwidth{}]{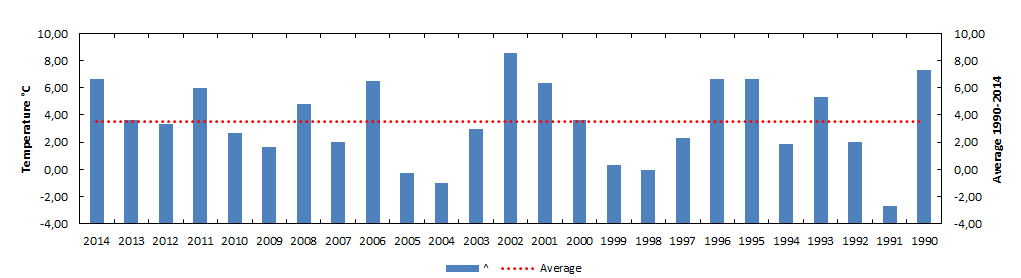}
		\caption{$\Delta$ Blackbird days (solid line) and Average Variations 1990-2014 (dotted line)}
	\end{center}
\end{figure}

\begin{figure}
\includegraphics[width=0.55\textwidth{}]{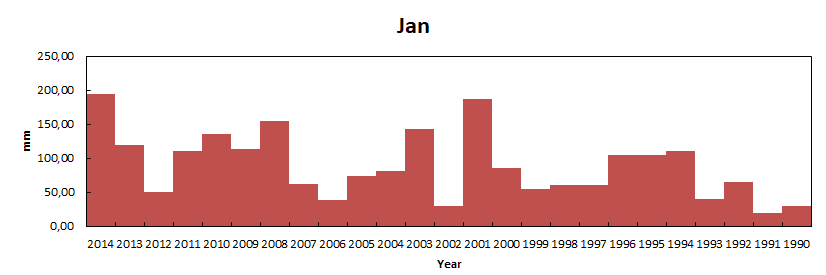} \quad 
\includegraphics[width=0.55\textwidth{}]{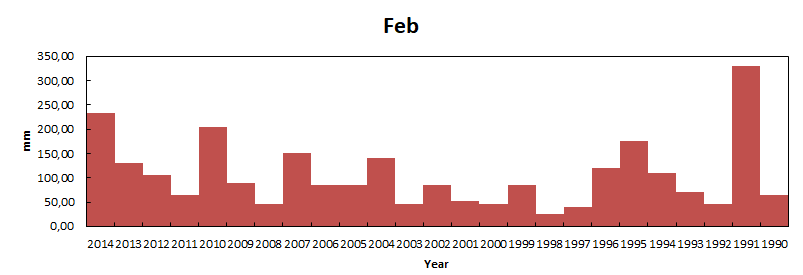} \quad
\includegraphics[width=0.55\textwidth{}]{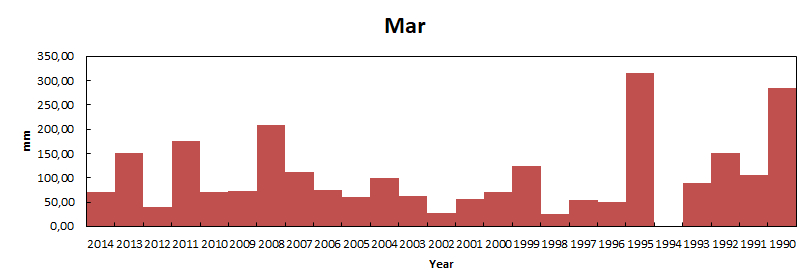} \quad
\includegraphics[width=0.55\textwidth{}]{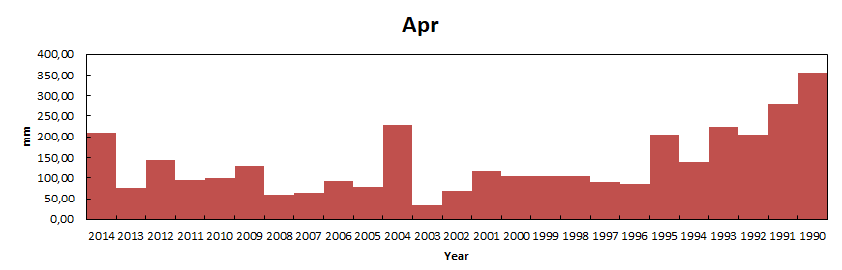} \quad
\includegraphics[width=0.55\textwidth{}]{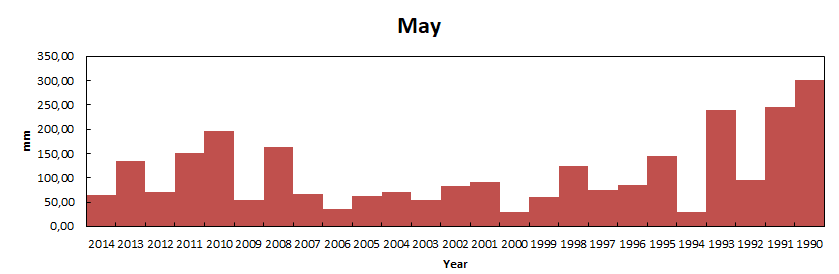} \quad 
\includegraphics[width=0.55\textwidth{}]{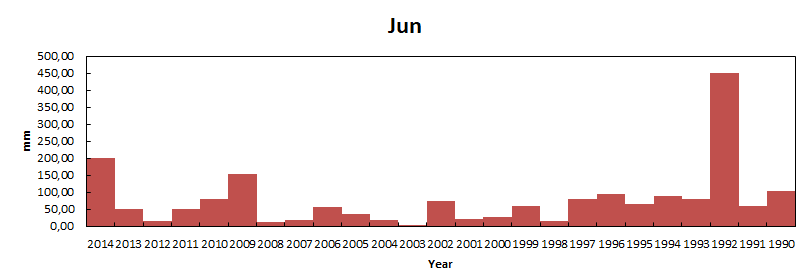} \quad
\includegraphics[width=0.55\textwidth{}]{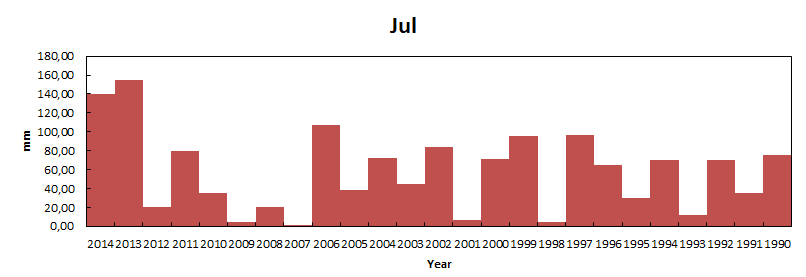} \quad
\includegraphics[width=0.55\textwidth{}]{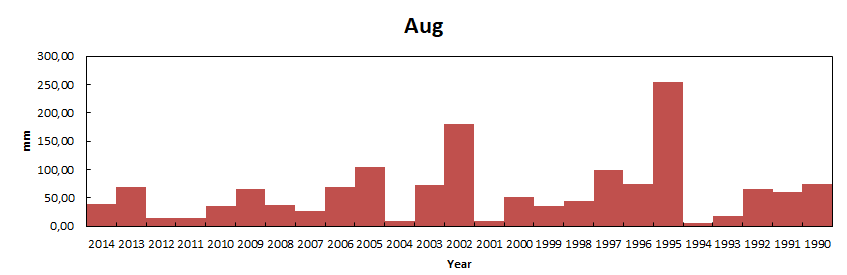} \quad
\includegraphics[width=0.55\textwidth{}]{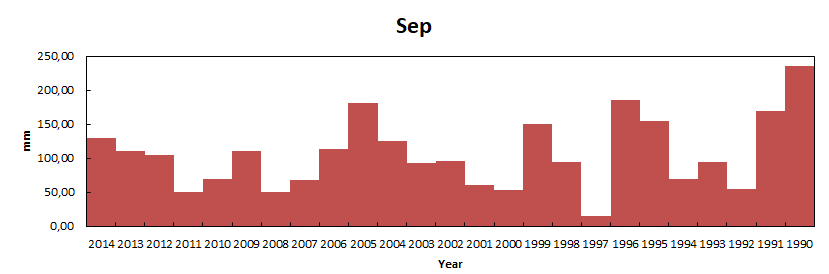} \quad 
\includegraphics[width=0.55\textwidth{}]{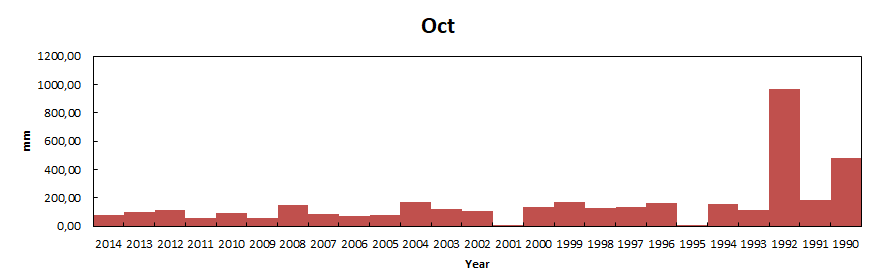} \quad
\includegraphics[width=0.55\textwidth{}]{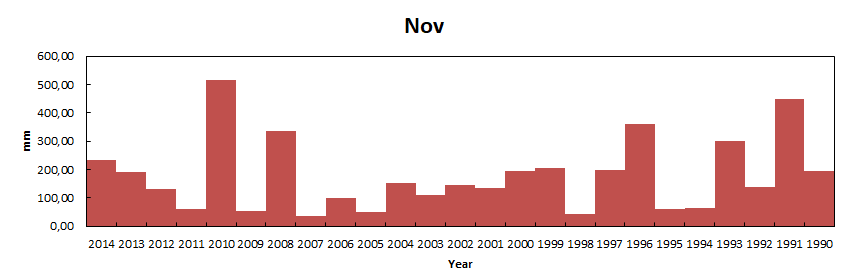} \quad
\includegraphics[width=0.55\textwidth{}]{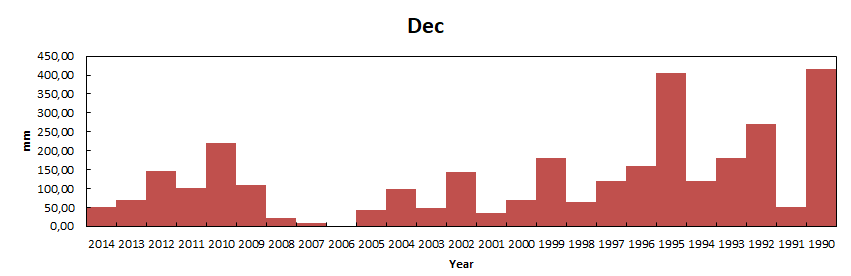} \\
\caption{Total Rain (mm) by Month}
\end{figure}
\begin{figure}
\includegraphics[width=1\textwidth{}]{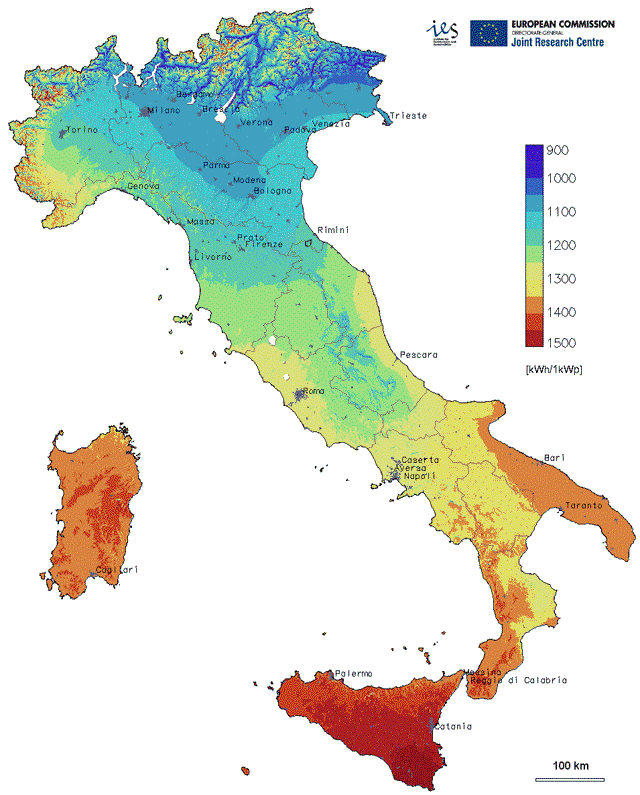}
\end{figure}
\begin{figure}
\includegraphics[width=0.9\textwidth{}]{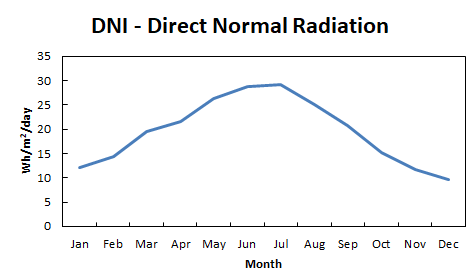}
\caption{DNI Direct Normal Radiation (Wh/m$^2$/day)} 
\end{figure}
\begin{figure}
	\includegraphics[width=0.9\textwidth{}]{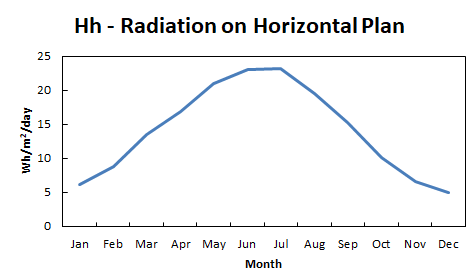}
	\caption{Hh Radiation on Horizontal Plan (Wh/m$^2$/day)} 
\end{figure}
\begin{figure}
	\includegraphics[width=0.9\textwidth{}]{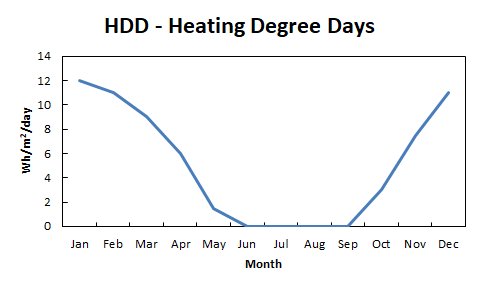}
	\caption{Number of heating degree days (HDD) expressed in $\circ C$} 
\end{figure}
\begin{figure}
	\includegraphics[width=1.3\textwidth{}]{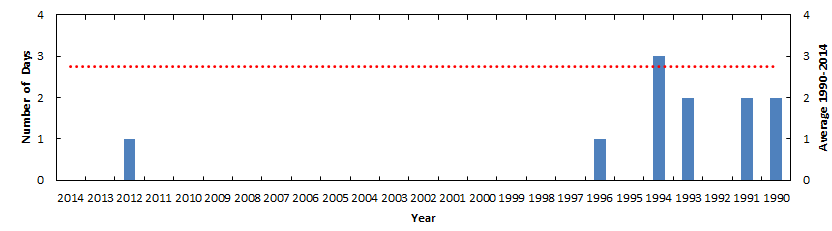}
	\caption{Days of Snow for years} 
\end{figure}
\begin{figure}
	\includegraphics[width=0.55\textwidth{}]{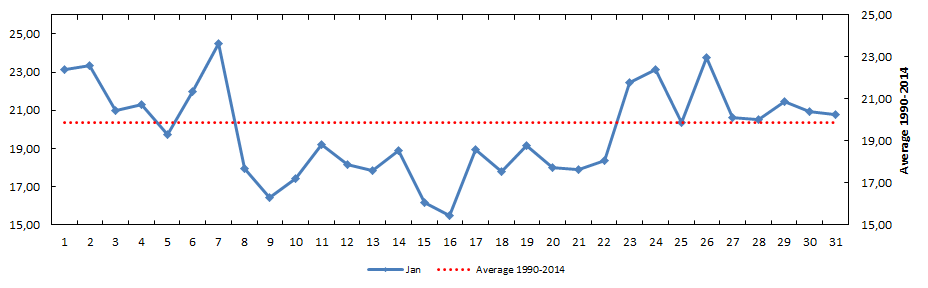} \quad 
	\includegraphics[width=0.55\textwidth{}]{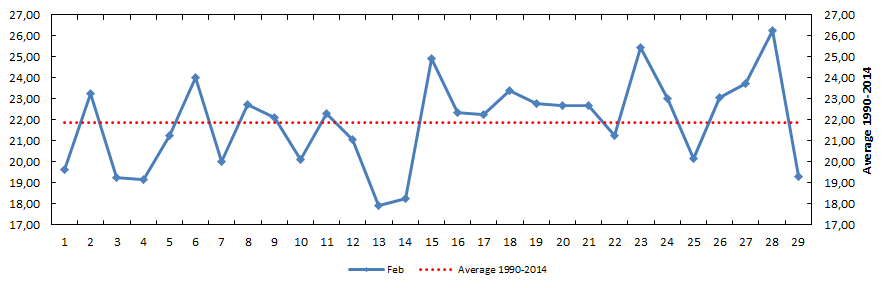} \quad
	\includegraphics[width=0.55\textwidth{}]{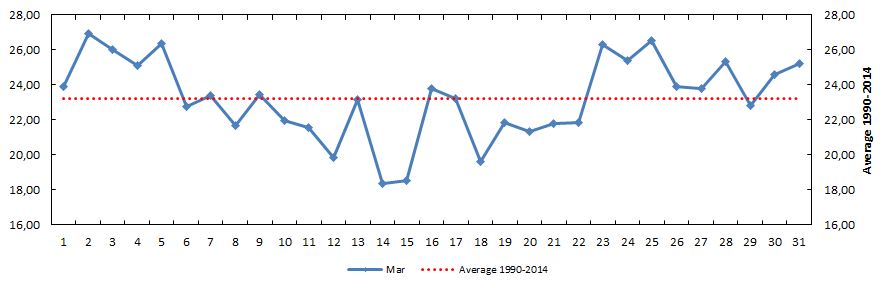} \quad
	\includegraphics[width=0.55\textwidth{}]{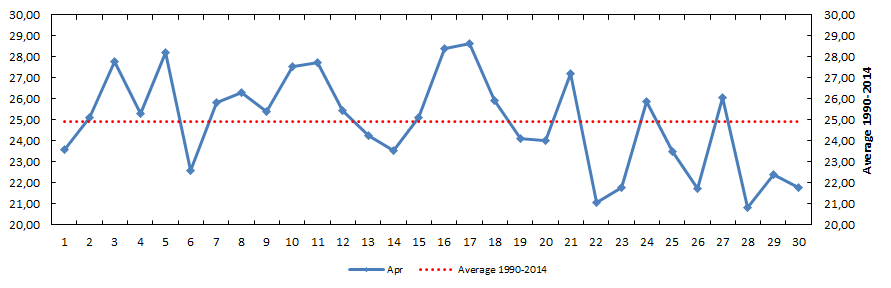} \quad
	\includegraphics[width=0.55\textwidth{}]{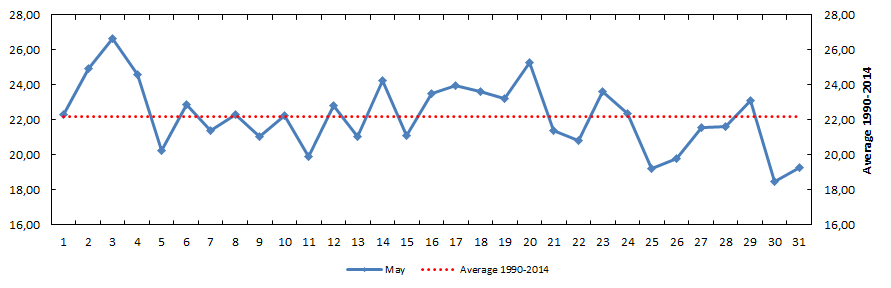} \quad 
	\includegraphics[width=0.55\textwidth{}]{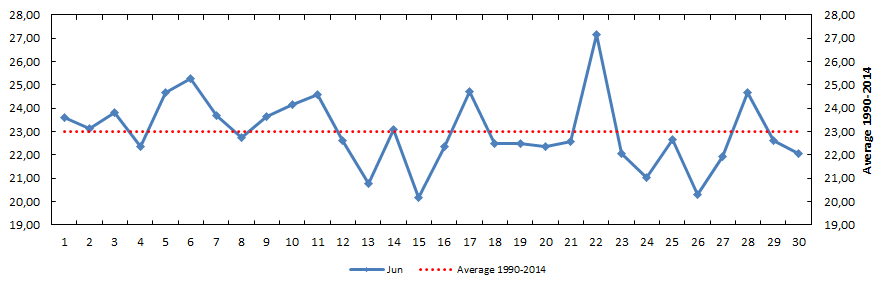} \quad
	\includegraphics[width=0.55\textwidth{}]{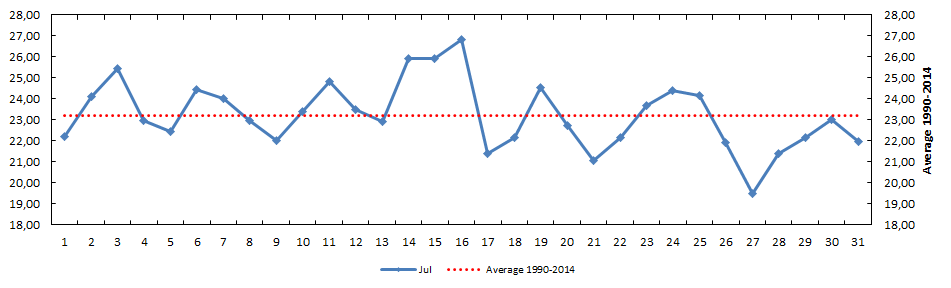} \quad
	\includegraphics[width=0.55\textwidth{}]{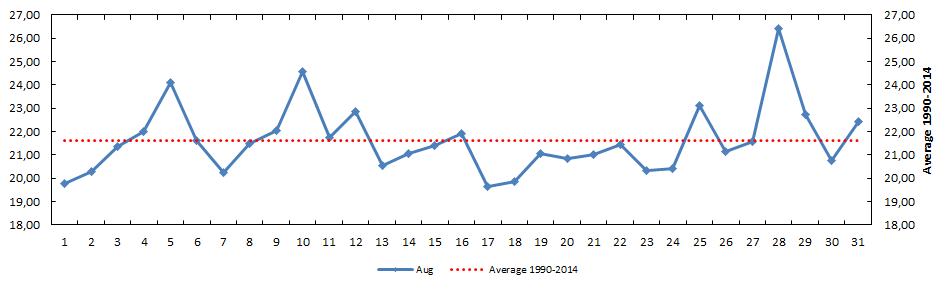} \quad
	\includegraphics[width=0.55\textwidth{}]{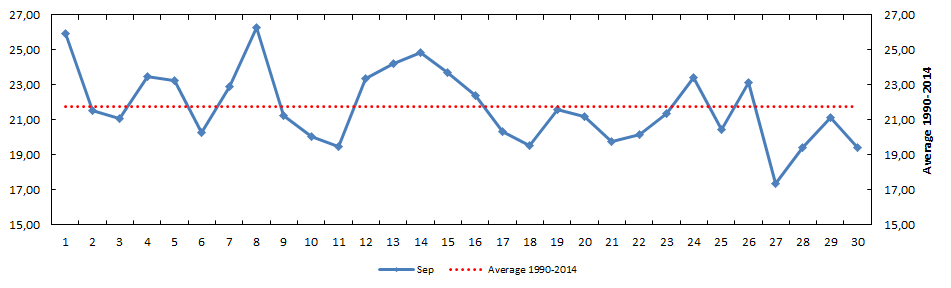} \quad 
	\includegraphics[width=0.55\textwidth{}]{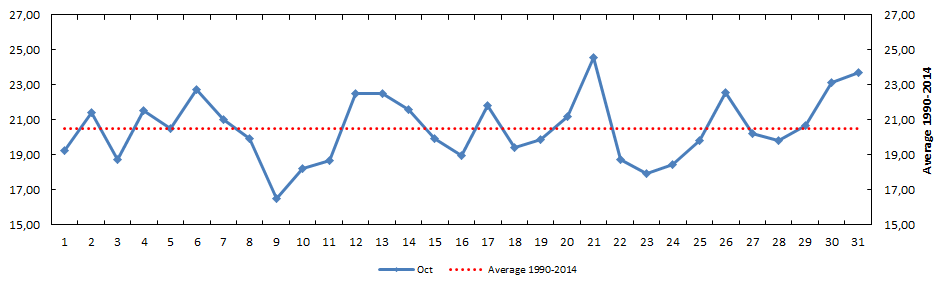} \quad
	\includegraphics[width=0.55\textwidth{}]{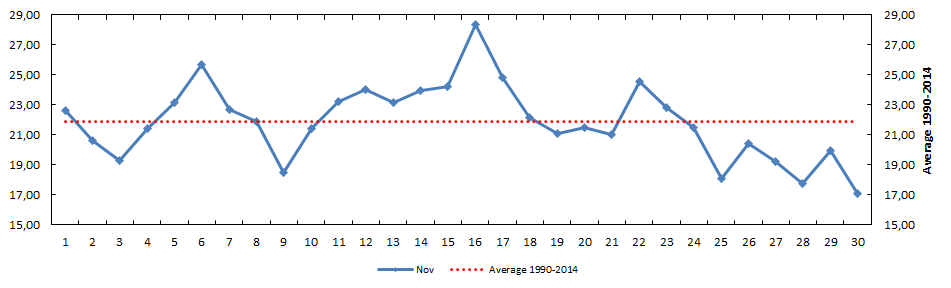} \quad
	\includegraphics[width=0.55\textwidth{}]{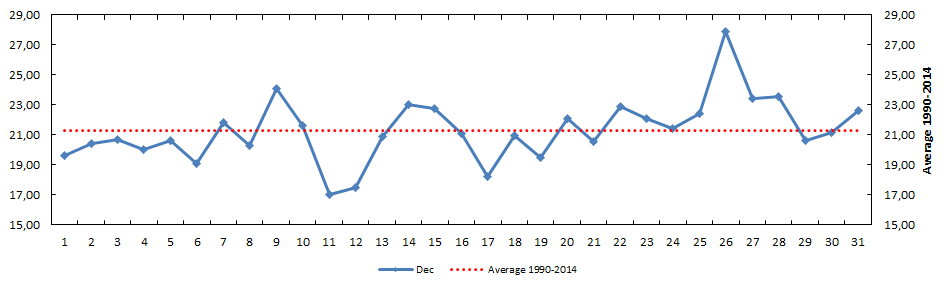} \\
		\caption{Chronological multi-panel matrix of mean monthly surface wind speeds (solid line) compared against the historical interannual 1990--2014 regional average baseline (red line) for Rome--RM.}
\end{figure}
\clearpage 
\begin{table*}[htbp]
	\centering
	\caption{Comprehensive Chronological Matrix of Mean Monthly Surface Air Temperatures ($T_{\text{mean}}$, $^{\circ}\text{C}$) in the Torrespaccata District (1990-2014).}
	\label{tab:complete_monthly_temperatures}
	\scriptsize
	\begin{tabular}{lccccccccccccc}
		\toprule
		\textbf{Year} & \textbf{Jan} & \textbf{Feb} & \textbf{Mar} & \textbf{Apr} & \textbf{May} & \textbf{Jun} & \textbf{Jul} & \textbf{Aug} & \textbf{Sep} & \textbf{Oct} & \textbf{Nov} & \textbf{Dec} & \textbf{Annual} \\
		\midrule
		2014  & 9.27 & 11.00 & 10.84 & 14.32 & 16.78 & 21.83 & 23.29 & 24.13 & 21.13 & 18.61 & 15.21 & 9.73  & \textbf{16.35} \\
		2013  & 8.13 & 6.75  & 10.55 & 15.30 & 17.32 & 21.30 & 25.35 & 25.81 & 21.50 & 19.18 & 12.48 & 8.82  & \textbf{16.04} \\
		2012  & 7.53 & 2.98  & 13.15 & 14.13 & 17.28 & 23.78 & 26.39 & 27.47 & 21.67 & 18.19 & 14.42 & 7.74  & \textbf{16.23} \\
		2011  & 8.19 & 8.45  & 10.45 & 14.77 & 18.25 & 22.60 & 24.06 & 26.23 & 23.43 & 16.76 & 12.43 & 9.82  & \textbf{16.29} \\
		2010  & 6.47 & 8.13  & 9.92  & 13.80 & 16.82 & 21.57 & 26.71 & 25.05 & 20.70 & 16.65 & 13.21 & 8.37  & \textbf{15.62} \\
		2009  & 8.08 & 7.75  & 11.27 & 14.87 & 20.27 & 22.45 & 25.84 & 27.23 & 22.93 & 15.44 & 11.80 & 8.66  & \textbf{16.38} \\
		2008  & 8.59 & 8.49  & 10.72 & 13.88 & 18.28 & 22.48 & 25.29 & 25.66 & 20.73 & 17.81 & 12.87 & 8.69  & \textbf{16.12} \\
		2007  & 9.62 & 9.94  & 10.90 & 15.63 & 18.50 & 22.75 & 25.17 & 25.14 & 20.66 & 16.70 & 10.70 & 7.13  & \textbf{16.07} \\
		2006  & 5.85 & 7.93  & 9.87  & 14.42 & 18.48 & 22.10 & 26.93 & 24.40 & 22.20 & 18.15 & 12.22 & 9.92  & \textbf{16.04} \\
		2005  & 5.47 & 5.76  & 9.87  & 12.70 & 18.91 & 23.06 & 25.38 & 23.51 & 21.39 & 16.38 & 14.61 & 7.15  & \textbf{15.35} \\
		2004  & 5.45 & 8.15  & 9.41  & 13.35 & 16.19 & 21.94 & 24.83 & 24.63 & 21.68 & 18.81 & 12.40 & 10.54 & \textbf{15.61} \\
		2003  & 8.59 & 5.23  & 10.68 & 14.63 & 20.63 & 26.10 & 26.75 & 27.94 & 21.18 & 16.59 & 14.12 & 8.61  & \textbf{16.75} \\
		2002  & 6.00 & 9.80  & 12.90 & 14.50 & 18.20 & 23.40 & 25.30 & 23.60 & 18.00 & 16.50 & 12.00 & 11.40 & \textbf{15.97} \\
		2001  & 9.71 & 8.99  & 13.32 & 12.57 & 18.80 & 21.67 & 24.89 & 25.98 & 19.58 & 19.71 & 13.01 & 6.43  & \textbf{16.22} \\
		2000  & 5.20 & 7.91  & 10.48 & 14.22 & 19.15 & 22.29 & 23.32 & 25.19 & 21.25 & 17.43 & 13.08 & 10.12 & \textbf{15.81} \\
		1999  & 7.58 & 6.51  & 10.56 & 12.82 & 18.42 & 21.83 & 23.60 & 24.90 & 21.03 & 17.50 & 12.29 & 8.67  & \textbf{15.48} \\
		1998  & 8.13 & 9.45  & 9.67  & 13.73 & 17.39 & 22.70 & 24.79 & 26.06 & 20.37 & 16.40 & 10.86 & 7.45  & \textbf{15.58} \\
		1997  & 8.76 & 8.18  & 10.61 & 10.48 & 18.07 & 22.50 & 23.77 & 24.24 & 21.93 & 16.98 & 12.75 & 9.00  & \textbf{15.61} \\
		1996  & 9.81 & 7.37  & 9.41  & 13.85 & 17.89 & 21.93 & 23.58 & 23.66 & 18.03 & 15.11 & 13.00 & 7.98  & \textbf{15.14} \\
		1995  & 5.97 & 9.30  & 9.53  & 12.48 & 16.15 & 19.75 & 25.58 & 23.73 & 19.78 & 17.50 & 11.20 & 10.46 & \textbf{15.12} \\
		1994  & 8.76 & 8.24  & 11.85 & 12.43 & 17.08 & 21.17 & 26.32 & 27.20 & 22.23 & 16.74 & 13.30 & 9.58  & \textbf{16.24} \\
		1993  & 6.60 & 6.89  & 9.35  & 12.92 & 17.92 & 22.33 & 23.58 & 26.15 & 21.17 & 16.90 & 10.36 & 8.81  & \textbf{15.25} \\
		1992  & 7.06 & 7.90  & 10.58 & 13.17 & 18.10 & 20.25 & 23.55 & 26.10 & 21.60 & 17.37 & 13.10 & 8.47  & \textbf{15.60} \\
		1991  & 7.10 & 7.81  & 12.60 & 11.18 & 13.76 & 20.45 & 25.08 & 25.66 & 22.12 & 15.55 & 11.02 & 4.82  & \textbf{14.76} \\
		1990  & 6.98 & 10.20 & 11.15 & 12.13 & 17.87 & 21.78 & 24.84 & 24.27 & 20.83 & 18.40 & 11.52 & 6.34  & \textbf{15.53} \\
		\midrule
		\textbf{Average} & \textbf{7.56} & \textbf{7.96} & \textbf{10.79} & \textbf{13.53} & \textbf{17.86} & \textbf{22.16} & \textbf{24.97} & \textbf{25.36} & \textbf{21.09} & \textbf{17.25} & \textbf{12.56} & \textbf{8.59}  & \textbf{15.81} \\
		\bottomrule
	\end{tabular}
\end{table*}
\clearpage
\begin{sidewaystable*}[p]
	\centering
	\caption{Comprehensive Chronological Matrix of Temperature Extremes (${^{\circ}\text{C}}$) on a Decadal Scale from January to December (1990--2014).}
	\label{tab:decadal_temperatures_landscape}
	\resizebox{\linewidth}{!}{
		\begin{tabular}{lcccccccccccccccccccccccr}
			\toprule
			\textbf{Year} & \multicolumn{2}{c}{\textbf{Jan}} & \multicolumn{2}{c}{\textbf{Feb}} & \multicolumn{2}{c}{\textbf{Mar}} & \multicolumn{2}{c}{\textbf{Apr}} & \multicolumn{2}{c}{\textbf{May}} & \multicolumn{2}{c}{\textbf{Jun}} & \multicolumn{2}{c}{\textbf{Jul}} & \multicolumn{2}{c}{\textbf{Aug}} & \multicolumn{2}{c}{\textbf{Sep}} & \multicolumn{2}{c}{\textbf{Oct}} & \multicolumn{2}{c}{\textbf{Nov}} & \multicolumn{2}{c}{\textbf{Dec}} \\
			\cmidrule(lr){2-3} \cmidrule(lr){4-5} \cmidrule(lr){6-7} \cmidrule(lr){8-9} \cmidrule(lr){10-11} \cmidrule(lr){12-13} \cmidrule(lr){14-15} \cmidrule(lr){16-17} \cmidrule(lr){18-19} \cmidrule(lr){20-21} \cmidrule(lr){22-23} \cmidrule(lr){24-25}
			& Min & Max & Min & Max & Min & Max & Min & Max & Min & Max & Min & Max & Min & Max & Min & Max & Min & Max & Min & Max & Min & Max & Min & Max \\
			\midrule
			2014 & -2.0 & 16.0 &  1.0 & 20.0 &  0.0 & 21.0 &  4.0 & 22.0 &  6.0 & 26.0 & 10.0 & 34.0 & 14.0 & 33.0 & 14.0 & 31.0 & 12.0 & 28.0 &  6.0 & 27.0 &  2.0 & 20.0 & -2.0 & 19.0 \\
			2013 & -2.0 & 16.0 & -3.0 & 15.0 & -3.0 & 19.0 &  4.0 & 26.0 &  6.0 & 24.0 & 11.0 & 34.0 & 14.0 & 35.0 & 16.0 & 38.0 & 10.0 & 32.0 & 10.0 & 24.0 & -1.0 & 22.0 & -2.0 & 14.0 \\
			2012 & -5.0 & 15.0 & -7.0 & 18.0 &  2.0 & 24.0 &  1.0 & 26.0 &  6.0 & 28.0 & 12.0 & 35.0 & 17.0 & 37.0 & 16.0 & 38.0 & 10.0 & 32.0 &  1.0 & 27.0 &  3.0 & 19.0 & -5.0 & 14.0 \\
			2011 & -4.0 & 16.0 & -1.0 & 19.0 & -4.0 & 19.0 &  2.0 & 25.0 &  7.0 & 30.0 & 14.0 & 33.0 & 12.0 & 35.0 & 14.0 & 36.0 &  0.0 & 33.0 &  5.0 & 28.0 &  1.0 & 20.0 & -1.0 & 19.0 \\
			2010 & -3.0 & 13.0 & -3.0 & 16.0 & -2.0 & 18.0 &  3.0 & 23.0 &  6.0 & 25.0 & 10.0 & 30.0 &  0.0 & 35.0 & 13.0 & 34.0 &  9.0 & 28.0 &  4.0 & 25.0 & -1.0 & 23.0 & -7.0 & 17.0 \\
			2009 & -3.0 & 13.0 & -4.0 & 15.0 &  0.0 & 19.0 &  6.0 & 23.0 & 10.0 & 32.0 & 13.0 & 32.0 & 16.0 & 35.0 & 17.0 & 35.0 & 12.0 & 31.0 &  1.0 & 26.0 &  3.0 & 21.0 & -5.0 & 17.0 \\
			2008 & -2.0 & 19.1 & -6.0 &  0.0 & -0.8 & 19.5 &  1.2 & 21.1 &  6.4 & 29.5 & 12.0 & 32.0 & 16.0 & 34.0 & 16.0 & 35.0 &  7.0 & 35.0 &  4.0 & 26.0 &  1.0 & 22.0 & -1.0 & 14.0 \\
			2007 & -2.0 & 18.5 & -1.4 & 21.0 &  0.8 & 21.2 &  5.0 & 26.7 &  8.2 & 31.9 & 12.0 & 35.1 & 11.8 & 36.9 & 13.6 & 36.9 &  9.0 & 30.1 &  4.2 & 30.0 & -0.8 & 19.7 & -3.4 & 15.3 \\
			2006 & -5.3 & 15.5 & -4.4 & 18.1 & -2.4 & 20.0 &  3.0 & 23.7 &  6.2 & 29.3 &  7.6 & 34.7 & 13.6 & 38.0 & 13.8 & 33.7 & 11.6 & 32.7 &  5.8 & 28.5 & -1.0 & 20.3 &  1.0 & 19.5 \\
			2005 & -3.8 & 14.9 & -4.0 & 15.7 & -6.6 & 20.7 &  2.4 & 26.5 &  7.4 & 31.3 &  5.2 & 36.9 & 12.4 & 39.0 & 10.8 & 33.7 & 11.0 & 32.7 &  6.8 & 23.1 &  3.5 & 20.0 & -4.8 & 14.1 \\
			2004 & -5.4 & 16.1 & -2.0 & 17.5 & -2.3 & 20.7 &  2.8 & 23.0 &  3.8 & 25.0 &  9.4 & 33.1 & 11.4 & 37.7 & 12.8 & 34.7 &  9.4 & 33.7 &  8.8 & 26.5 & -2.0 & 25.3 & -2.4 & 18.9 \\
			2003 & -4.0 & 17.9 & -4.2 & 16.7 & -1.4 & 20.1 & -5.4 & 27.9 &  8.4 & 32.3 & 14.4 & 36.7 & 15.0 & 38.1 & 16.0 & 39.1 & 10.4 & 31.0 &  3.6 & 26.9 &  0.6 & 19.1 & -3.8 & 17.0 \\
			2002 & -7.0 & 15.0 & -1.0 & 16.3 & -0.8 & 25.3 &  1.0 & 26.8 &  6.4 & 27.3 & 10.6 & 37.1 & 11.8 & 34.5 & 10.4 & 32.5 &  7.8 & 26.5 &  6.2 & 23.0 &  0.8 & 21.7 & -0.2 & 14.9 \\
			2001 &  1.2 & 15.7 & -4.8 & 16.3 &  1.0 & 22.1 & -2.0 & 24.3 &  9.0 & 28.5 &  8.0 & 34.3 & 13.6 & 34.5 & 15.2 & 35.7 &  7.8 & 28.5 &  5.4 & 26.7 & -1.4 & 21.1 & -4.2 & 15.3 \\
			2000 & -5.0 & 13.0 &  1.0 & 16.0 & -1.0 & 18.0 &  3.0 & 25.0 & 10.6 & 27.0 & 11.6 & 31.7 & 10.0 & 36.1 & 12.8 & 38.1 &  9.0 & 29.9 &  8.0 & 27.9 &  3.0 & 20.3 & -1.0 & 17.3 \\
			1999 & -5.0 & 16.0 & -4.0 & 15.0 &  1.0 & 21.0 &  2.0 & 21.0 & 10.0 & 29.0 & 15.0 & 29.0 & 14.0 & 32.0 & 16.0 & 36.0 & 14.0 & 29.0 &  6.0 & 25.0 &  1.0 & 21.0 & -3.0 & 17.0 \\
			1998 & -3.0 & 15.0 &  0.8 & 18.0 & -2.0 & 19.0 &  2.0 & 23.0 &  9.0 & 27.0 & 10.0 & 34.0 & 15.0 & 36.0 & 11.0 & 37.0 & 10.0 & 30.0 &  6.0 & 21.0 & -1.0 & 21.0 & -5.0 & 13.0 \\
			1997 & -2.2 & 15.0 & -2.2 & 15.0 &  0.2 & 20.0 & -1.4 & 20.0 &  4.0 & 27.0 & 11.0 & 30.0 & 11.0 & 33.0 & 12.0 & 33.0 &  9.0 & 32.0 &  3.0 & 26.0 &  2.0 & 18.0 &  1.0 & 16.0 \\
			1996 & -3.0 & 14.0 & -4.0 & 13.0 & -6.0 & 15.0 &  1.0 & 21.0 &  6.0 & 26.0 &  0.0 & 33.0 & 11.0 & 35.0 & 13.0 & 33.0 &  7.0 & 24.0 &  4.0 & 21.0 &  1.0 & 20.0 & -10.0 & 15.0 \\
			1995 & -4.0 & 15.0 & -2.6 & 16.0 & -1.2 & 17.0 & -3.0 & 21.0 &  4.0 & 26.0 &  7.0 & 30.0 & 15.0 & 35.0 &  9.0 & 32.0 &  6.0 & 27.0 &  4.0 & 24.0 & -5.0 & 20.0 & -1.0 & 15.0 \\
			1994 & -1.0 & 18.0 & -3.6 & 16.0 &  0.4 & 24.0 & -1.0 & 23.0 &  4.0 & 27.0 & 10.0 & 31.0 & 15.0 & 34.0 & 15.0 & 38.0 &  9.0 & 33.0 &  2.0 & 23.0 &  2.0 & 23.0 &  3.0 & 21.0 \\
			1993 & -5.0 & 14.0 & -5.0 & 16.0 & -2.2 & 22.0 &  3.0 & 25.0 &  8.0 & 28.0 & 10.0 & 31.0 & 10.0 & 33.0 & 10.0 & 37.0 &  8.0 & 30.0 &  5.0 & 26.0 & -3.0 & 16.0 & -2.0 & 17.0 \\
			1992 & -3.8 & 14.0 & -5.0 & 16.0 & -1.2 & 20.0 &  2.0 & 23.0 &  5.0 & 28.0 & 10.0 & 29.0 & 12.0 & 35.0 & 14.0 & 36.0 &  8.0 & 30.0 &  5.0 & 25.0 &  2.0 & 20.0 & -1.8 & 15.0 \\
			1991 & -5.0 & 13.0 & -10.0 & 19.0 &  0.8 & 25.0 & -1.0 & 19.0 &  1.0 & 23.0 &  8.0 & 30.0 & 10.0 & 34.0 & 12.0 & 33.0 & 11.0 & 28.0 &  3.0 & 24.0 & -3.2 & 17.0 & -4.0 & 17.0 \\
			1990 & -4.0 & 15.0 & -1.6 & 17.0 & -2.6 & 19.0 &  1.0 & 17.0 &  6.0 & 26.0 &  7.0 & 32.0 & 13.0 & 34.0 & 13.0 & 32.0 &  8.0 & 27.0 &  5.0 & 27.0 & -2.2 & 21.0 & -5.0 & 13.0 \\
			\midrule
			\textbf{Average} & \textbf{-3.5} & \textbf{15.4} & \textbf{-3.2} & \textbf{16.1} & \textbf{-1.3} & \textbf{20.4} & \textbf{1.4} & \textbf{23.3} & \textbf{6.6} & \textbf{27.8} & \textbf{10.0} & \textbf{32.7} & \textbf{12.6} & \textbf{35.2} & \textbf{13.5} & \textbf{35.1} & \textbf{9.0} & \textbf{30.2} & \textbf{4.9} & \textbf{25.5} & \textbf{0.2} & \textbf{20.5} & \textbf{-2.8} & \textbf{16.2} \\
			\bottomrule
		\end{tabular}
	}
\end{sidewaystable*}
\clearpage
\centering

\end{document}